\documentclass[useAMS,usenatbib]{mn2e}

\usepackage{float}
\usepackage{epstopdf}
\usepackage[caption = false]{subfig}
\usepackage{graphicx}
\usepackage{multicol, blindtext}

\title[Photometric metallicity map of the LMC]{Photometric metallicity map of the Large Magellanic Cloud}
\author[Samyaday Choudhury et al.]{Samyaday Choudhury$^{1,2}$\thanks{E-mail:samyaday@iiap.res.in}, Annapurni Subramaniam$^1$, and Andrew A. Cole$^3$\\
 $^1$ Indian Institute of Astrophysics, 2B Koramangala, Bangalore, India 560034\\
 $^2$ Indian Institute of Science, Bangalore, India 560012\\
 $^3$ School of Physical Sciences, University of Tasmania, Private Bag 37, Hobart, Tasmania 7001, Australia
}
\begin{document}



\maketitle

\label{firstpage}
\begin{abstract}
We have estimated a metallicity map of the Large Magellanic Cloud (LMC) using the Magellanic Cloud Photometric Survey (MCPS) and Optical Gravitational Lensing Experiment (OGLE III) photometric data. This is a first of its kind map of metallicity up to a radius of 4 - 5 degrees, derived using photometric data and calibrated using spectroscopic data of Red Giant Branch (RGB) stars. We identify the RGB in the V, (V$-$I) colour magnitude diagrams of small subregions of varying sizes in both data sets. We use the slope of the RGB as an indicator of the average metallicity of a subregion, and calibrate the RGB slope to metallicity using spectroscopic data for field and cluster red giants in selected subregions. The average metallicity of the LMC is found to be [Fe/H] = $-$0.37 dex ($\sigma$[Fe/H] = 0.12) from MCPS data, and [Fe/H] = $-$0.39 dex ($\sigma$[Fe/H] = 0.10) from OGLE III data. The bar is found be the most metal-rich region of the LMC. Both the data sets suggest a shallow radial metallicity gradient up to a radius of 4~kpc ($-$0.049$\pm$0.002 dex kpc$^{-1}$ to $-$0.066$\pm$0.006 dex kpc$^{-1}$). Subregions in which the mean metallicity differs from the surrounding areas do not appear to correlate with previously known features; spectroscopic studies are required in order to assess their physical significance.
\end{abstract}
\section{Introduction}

The stars and gas in galactic disks have a mean metallicity which depends on the luminosity of the galaxy (e.g. \citealt{Tremonti+2004ApJtheorigin}) and often shows a radial gradient. A rich literature now exists which confirms these radial abundance trends in spirals (e.g. \citealt{Simpson+1995ApJfarIR,Afflerbach+1997ApJgalactic,Molla+1999ApJthestellar,Kewley+2010ApJmetal-grad,SanchezB+2011MNRASstarform} ).  Observations of nearby spiral galaxies show that the inner disks have higher metallicities than their associated outer disk regions; at the present day, typical gradients of $\sim$ $-$0.05 dex kpc$^{-1}$ are encountered \citep{Pilkington+2012A&Ametal-grad}. A study of a large sample of SDSS galaxies by \cite{Tremonti+2004ApJtheorigin} demonstrates a tight correlation between the stellar mass and the gas-phase oxygen abundance extending over 3 orders of magnitude in stellar mass and a factor of 10 in oxygen abundance. \cite{Kirby+2008ApJuncover} show that the relation between luminosity and metallicity continues down to the faintest known dwarf spheroidal galaxies. In this paper we have studied the average metallicity and metallicity gradient of the Large Magellanic Cloud (LMC), which is a gas rich, metal poor, actively star-forming, irregular type galaxy with a stellar mass of $\sim$ 10$^{10}$ M$_\odot$.

The abundance gradient for relatively young stars in the disk of the Milky Way (MW) is nicely delineated by the Cepheids \citep{Luck+2006AJthedistri}. The gradient is about $-$0.06 dex kpc$^{-1}$, in good agreement with the gas-phase gradient derived by \cite{Shaver+1983MNRASthegalactic}. \cite{Cioni2009A&Athemetallicity} used the change of the ratio of C and M-type Asymptotic Giant Branch (AGB) stars in the LMC and M33 to evaluate their stellar abundance gradients, finding a radial gradient in both galaxies; the shallower gradient in the LMC possibly attributable to the existence of its prominent central bar.

Bars are a very common feature in spiral galaxies. Analytical studies and numerical simulations have shown that the presence of a bar can greatly affect the evolution of the host galaxy. The presence of a bar appears to flatten or even erase the abundance gradient \citep{Alloin+1981A&Athemild}, probably due to the non-circular motions which the bar induces in the gas of the disk. We would expect barred galaxies to have higher central star formation rate (SFR), higher central metallicities, and possibly a flatter metallicity gradient, depending at what stage of their evolution they are being observed. This basic scenario is indeed supported by many observations (\citealt{Pagel+1979MNRASonthe, Martin&Roy1994theinfu}). The off-centre bar of the LMC is a stellar bar, but presently there is no activity as found in HI gas. Nevertheless, it will be interesting to understand the effect of the LMC bar on the metallicity of the central regions, if any. On the other hand, the bar in the LMC significantly differs from the bars formed in N-body models: there is little to no evidence for a change in stellar velocity dispersion \citep{Cole+2005AJspectroOfRGs}, or sign of non circular gas motions \citep{Staveley-Smith2003MNRASanewlook}. It is unclear whether the predictions based on bars in simulations of large spirals are applicable to the LMC. In this study, we plan to map the metallicity of the bar and estimate the gradient with respect to the disk. We expect to identify whether the bar feature has its expected imprint on the chemical enrichment in the LMC.

There are studies that estimated the age-metallicity relation (AMR) of the LMC using star clusters (\citealt{Olszewski+1991AJspectro,Geisler+1997AJasearch,Bica+1998AJages&metal,Dirsch+2000A&Aage&metal}), directed towards understanding the chemical enrichment history (CEH). \cite{Olszewski+1991AJspectro} used Ca II triplet (CaT) spectroscopy for RGB stars, an important metallicity indicator, to estimate the abundance for $\sim$ 80 star clusters. An analysis of the metallicity distribution of their cluster sample showed that the mean [Fe/H] values for all clusters in the inner (radius $<$ 5$^{\circ}$) and outer (radius $>$ 5$^{\circ}$) LMC are almost the same. This suggested the presence of very shallow, if any, radial metallicity gradient in the LMC disk. Later on, \cite{Grocholski+2006AJCaIItriplet} used CaT spectroscopy to derive updated metallicities for 28 populous LMC star clusters that had a good area coverage of the LMC. According to these authors, there was hardly any evidence for metallicity gradient in the LMC cluster system. This was in sharp contrast to what was seen in our own galaxy, the MW \citep{Friel+2002AJmetal}, and M33 \citep{Tiede+2004AJthestellar}.

\cite{Cole+2005AJspectroOfRGs} estimated the [Fe/H] of about $\sim$ 400 red giant (RGs) stars in an area of about 200 arcmin sq. around the central LMC, using CaT spectroscopy.  They found a distribution peaked at [Fe/H] $\sim$ $-$0.4 dex with a low metallicity tail reaching [Fe/H] $\sim$ $-$2.1 dex. \cite{Carrera+2008AJ-CEH-LMC} found a nearly constant metallicity [Fe/H] $\sim$ $-$0.5~dex out to a radius of 6$^{\circ}$, with a suggestion of a decrease at larger radii, likely driven by a change in the mean magnitude of stars selected for analysis.
A distribution peaked at a slightly lower value ([Fe/H] = $-$0.75 dex, $\sigma$[Fe/H] = 0.23 dex) was found by \cite{Pompeia+2008A&Achemi}, for a region at 1.2 kpc from the centre. \cite{Lapenna+2012ApJtagging} estimated a [Fe/H] = $-$0.48 dex with a quite small dispersion ($\sigma$[Fe/H] = 0.13 dex), using 91 stars in the region surrounding NGC 1786, 2 degree north west from the LMC centre. The results from the above studies are compared by \cite{Lapenna+2012ApJtagging}. \cite{Olsen+2011ApJapopulation} found a metallicity distribution with a peak at [Fe/H] = $-$0.45, a median of [Fe/H] = $-$0.56$\pm$0.02, and a dispersion of 0.5 dex compared to a median error of 0.15 dex, using $\sim$ 1000 LMC field stars. 
\cite{VanderSwaelmen+2013A&Achemiabun} performed a detailed chemical analysis of a sample of 106 and 58 LMC field red giant stars, located in the bar and the disk of the LMC respectively. The authors found the bar to be chemically enriched (primarily the $\alpha$-elements) compared to the disk, which was possibly due to the formation of a new stellar population in the central part of the LMC.

As the LMC presents a large area in the sky, the estimation of metallicity using spectroscopy of individual stars is a laborious and time consuming process. One can also see that all the above spectroscopic studies of star clusters and stellar populations have been carried out only in small pockets of the LMC, and have small sample size. Also, the spatial variation of metallicity has been estimated using studies which do not cover substantial area of the LMC. As a result, there has been little progress or consensus in understanding the effect of the bar on the LMC metallicity or the possibility of large scale inhomogeneities due to interactions with the SMC or MW. These require a metallicity map of the LMC with good spatial resolution and coverage.

\cite{Cioni2009A&Athemetallicity} used a photometric technique to estimate the metallicity gradient of field asymptotic giant branch (AGB) stars as tracers. They calculated the C/M ratio, which is an indicator of metallicity. The plot of [Fe/H] against galactocentric distance out to about 8 kpc,  showed a small metallicity gradient with a high degree of scatter. Although these authors could cover a large area of the LMC and estimate the metallicity gradient, their indicators (AGB) and calibrators (RGB) were different, and the C/M ratio is potentially susceptible to age effects. \cite{Piatti&Geisler2013AJtheage} used Washington photometry of over 5 million field stars and estimated the mean metallicity, radial variation and age-metallicity relation for the LMC; they were able to explain the differences between their inner and outer disk as due to a consistent age-metallicity relation but with an increasing concentration of star formation towards the inner disk at young ages. However, their study avoided the densely concentrated regions of the central bar.

Studies using photometric data of giants can cover a sufficiently large area and estimate the metallicity variation, suggesting that it is an efficient technique. In this study, we have created a metallicity map of the LMC using the optical photometric data of the Magellanic Cloud Photometric Survey (MCPS, \citealt{Zaritsky+2004AJmcps}) and the Optical Gravitational Lensing Experiment (OGLE III, \citealt{Udalski+2008AcAogleIII}), two of the most recent, high-resolution, large area surveys of the LMC. Using the red giant stars, we estimated the average metallicity of the LMC, and its radial variation. This is a first of its kind map of mean metallicity up to a radius of 4$^{\circ}$-5$^{\circ}$. The paper is organised in the following way: In Section 2 we describe the data (OGLE III and MCPS) used in this study . The OGLE III analysis, metallicity maps and results are presented in Section 3, whereas those corresponding to the MCPS data are presented in Section 4. Section 5 describes the error analysis corresponding to our estimations. The discussion related to our study are presented in Section 6. We summarise the conclusions from this study in Section 7.

\section{Data}

In this study we have made use of two large photometric surveys of the LMC, the OGLE III and the MCPS. The OGLE III survey covered 39.7 square degrees of the central LMC, and presented the mean, calibrated VI photometry of about 35 million stars. The photometric calibration of OGLE III is based on \cite{Landolt1992AJ-BroadUBVRI} standard stars \citep{Udalski+2008AcAOGLEIIIreduc}. OGLE III covers the bar region as well as the eastern and western wing of the inner LMC till a radius of $\sim$ 4 degrees from the LMC centre. We divided the OGLE III data into 1854 regions, each of dimension (8.88 $\times$ 8.88) sq. arcmin in RA and DEC. This was done to ensure that each area bin has a well populated RGB, in their corresponding V, (V$-$I) colour magnitude diagram (CMD). The MCPS survey covered an area of about 64.0 square degrees of the central LMC, and provided the optical photometric data of about 24 million stars in the U, B, V, and I pass bands. The MCPS data are placed on the Johnson-Kron-Cousins photometric system, using \citealt{Landolt1983AJ-UBVRIphoto,Landolt1992AJ-UBVRIphoto} \citep{Zaritsky+2004AJmcps}. The survey is of comparatively lower resolution (0$^{''}$.70 pixel$^{-1}$) than the OGLE III (0$^{''}$.26 pixel$^{-1}$) survey. For our analysis, we have divided the whole MCPS survey region into 1512 regions of equal areas, (10.53 $\times$ 15) sq. arcmin, so as to have a well populated RGB in the V, (V$-$I) CMD. We only consider stars with photometric error less than 0.15 mag in the V and I pass bands in both the data sets. The MCPS has more coverage of the northern and southern regions of the central LMC, whereas the OGLE III survey has more coverage of the eastern and western regions. The two surveys thus complement each other in terms of covered area of the LMC.

\section{Analysis of OGLE III data}

The slope of the red giant branch (RGB) in the CMD of a small region in the galaxy is used as an indicator of the mean metallicity of the region. Metal-rich RGB stars are redder and fainter in the visible band-passes than their metal-poor counterparts. A region in a galaxy with metal-rich stars is expected to have a shallower RGB slope when compared to a relatively metal-poor region. The dependence of slope of the RGB on metallicity is well known \citep{DaCosta&Armandroff1990AJstandard, Kuchinski+1995AJ-IRarray}. In the case of field stars, the stellar population is heterogeneous with respect to age and metallicity. The RGB will consist of many populations, but the dominant population dictates the shape as well as the slope of the RGB. Thus, the RGB slope of a field region will correspond to the metallicity of the dominant RGB population of the region. Here we describe the steps carried out to identify the RGB consistently in all regions and estimate the RGB slope in the OGLE III data.

\begin{figure*} 
\begin{center} 
\includegraphics[height=5.0in,width=6.0in]{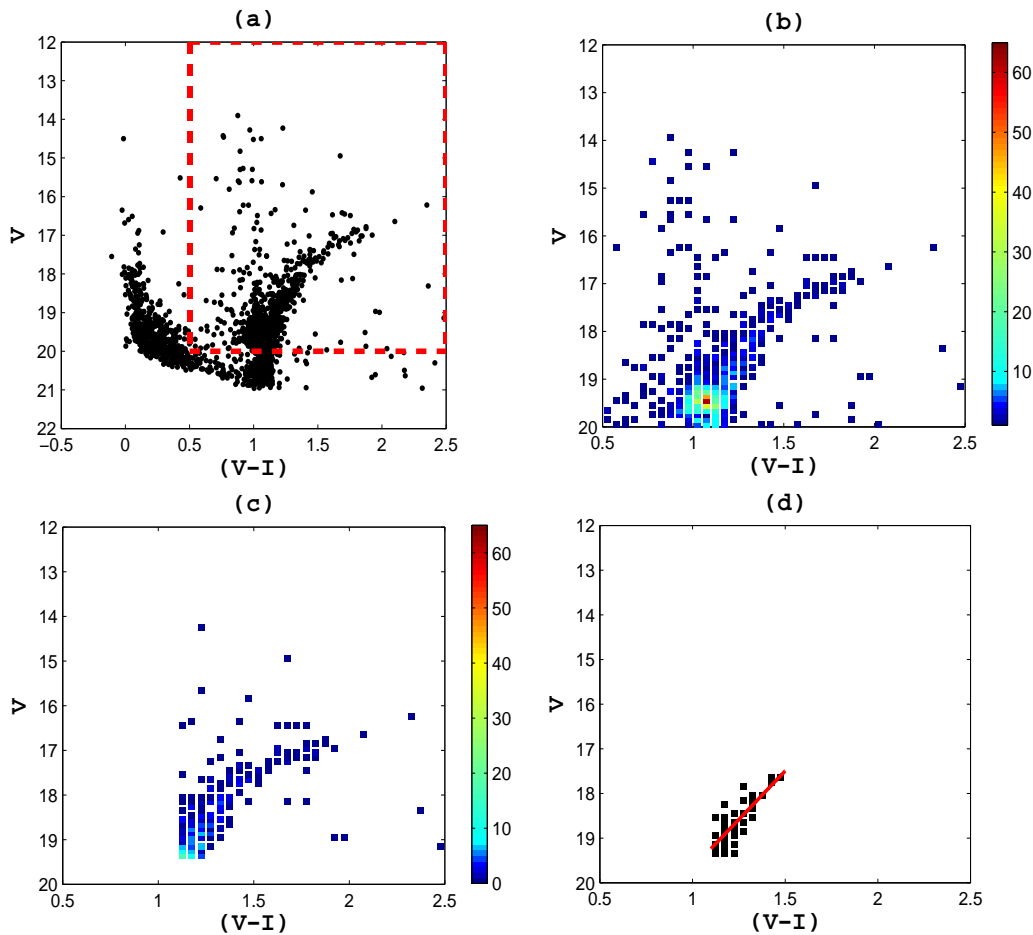}
\caption {\small (a) The V versus (V$-$I) CMD of an OGLE III region at (70.78$^\circ$, $-$70.07$^\circ$), of size (8.88$\times$8.88) sq. arcmin, with N=1762 stars (black filled circles). The stars within the rectangle (red dashed line) belongs to the evolved part of the CMD. (b) Density diagram of the evolved part of CMD, where the CMD bins are colour coded based on the number of stars contained in them, as denoted in the colour bar. (c) Density diagram after giving a colour-magnitude cut at the peak value of RC distribution. (d) Density diagram showing CMD bins that have $\ge$ 3 stars (black) . Straight line fit to these bins representing the RGB, after 3-sigma clipping, is shown as a red solid line. The estimated parameters are: $|$slope$|$= 4.34$\pm$0.56, $r$=0.80, and $N_p$= 36.}
\label{fig:lsqfit}
\end{center} 
\end{figure*}

\begin{figure*}
\centering
\begin{minipage}[b]{0.45\linewidth}
\includegraphics[height=3.0in,width=3.0in]{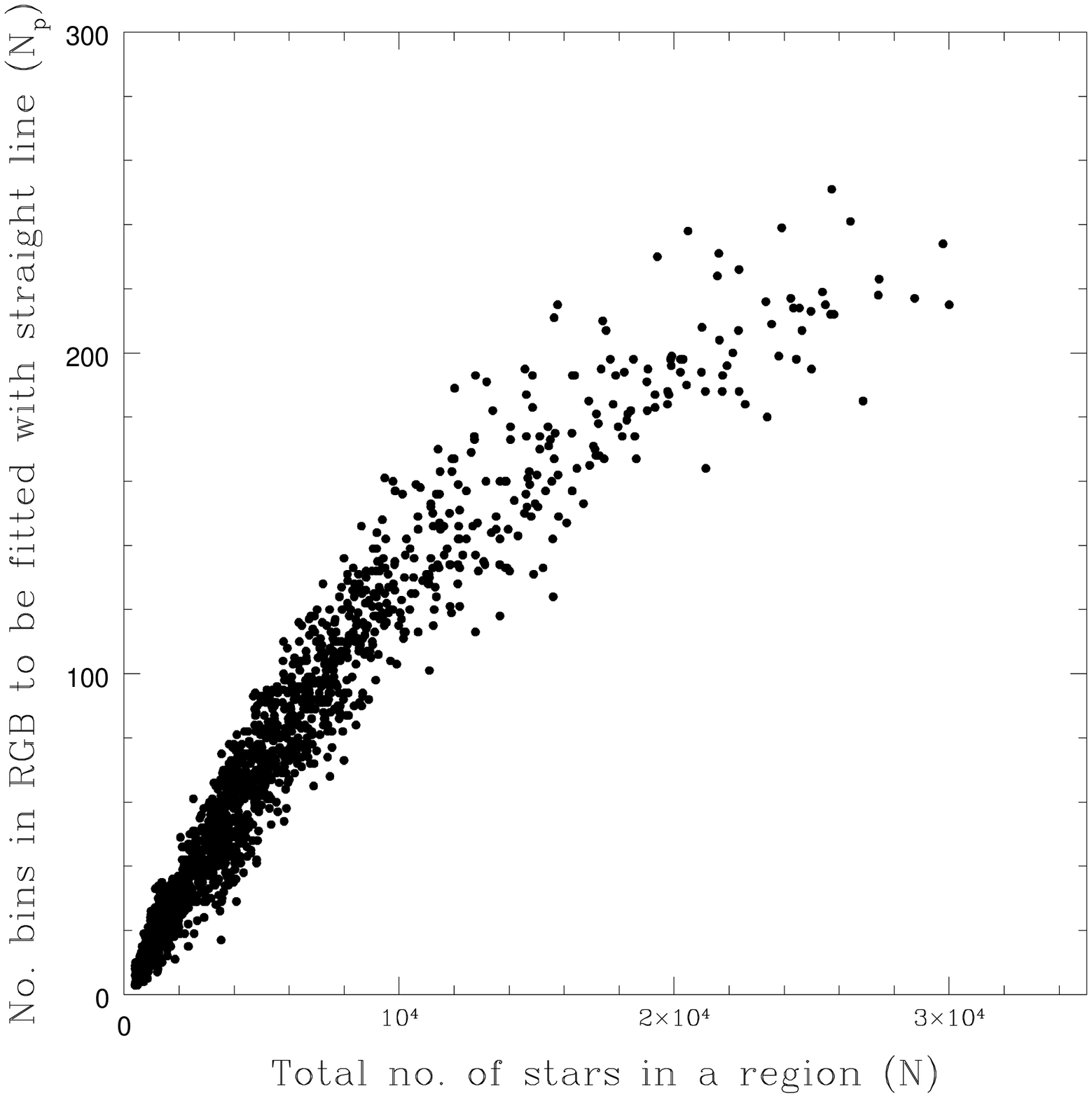}
\caption{\small Plot of number of bins in RGB to be fitted with straight line ($N_p$) versus the total number of stars ($N$) for OGLE III subregions, after initial area binning.
\label{fig:ogle3_np_vs_isum_undiv} } 
\end{minipage}
\quad
\begin{minipage}[b]{0.45\linewidth}
\includegraphics[height=3.0in,width=3.0in]{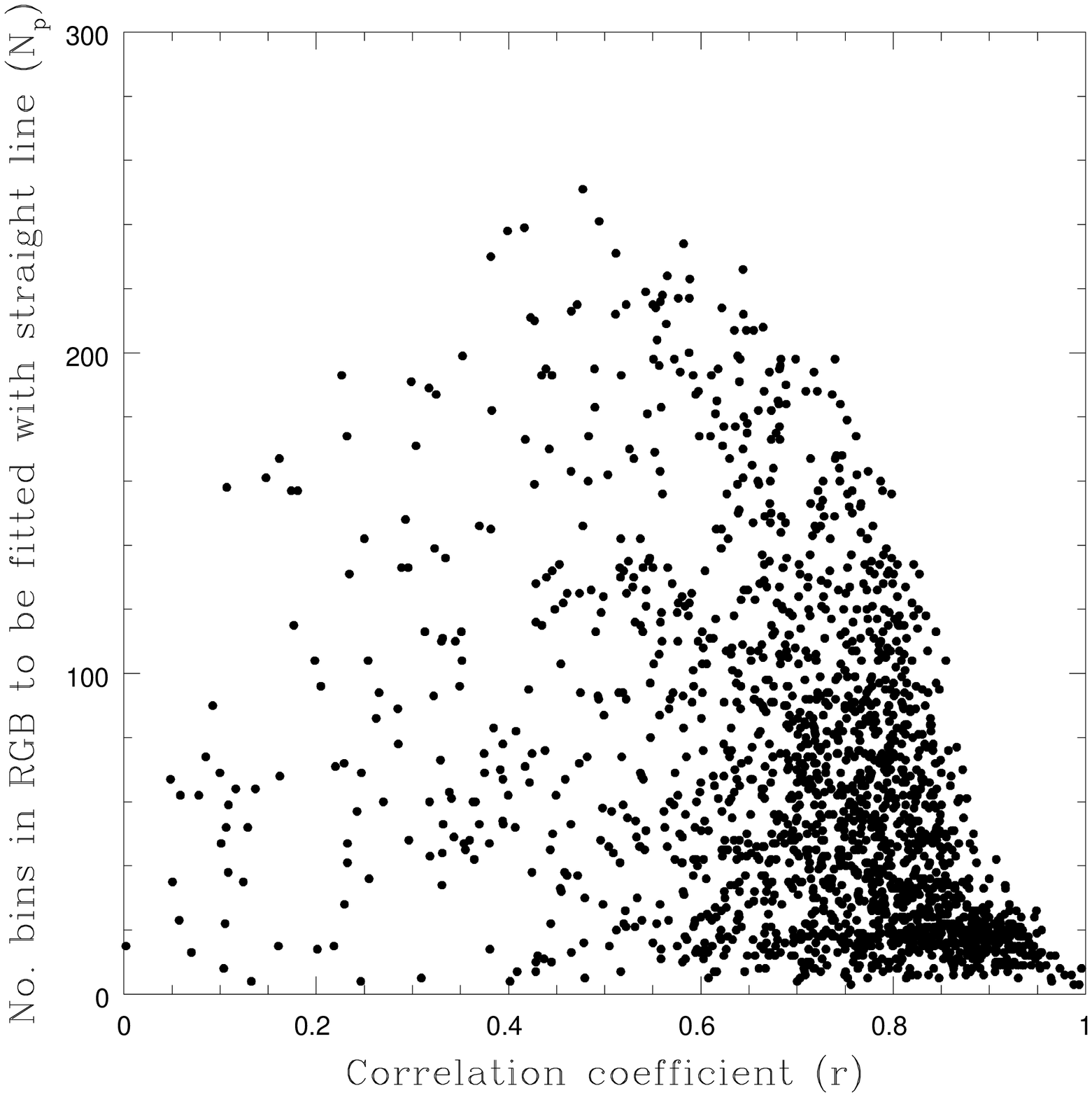}
\caption{\small Plot of  number of bins in RGB to be fitted with straight line ($N_p$) versus correlation coefficient ($r$) for OGLE III subregions, after initial area binning.
\label{fig:ogle3_np_vs_r_undiv} }  
\end{minipage}
\end{figure*} 


Our aim is to robustly identify the RGB in the CMD of each region, independent of the reddening, as location of the RGB varies in each region. We also need to identify a feature that can be used to fix the location of the base of the RGB and trace it in CMDs of all the regions.

The OGLE III data for the LMC is spatially binned into 1854 small regions, the dimension of each bin being (8.88 $\times$ 8.88) sq. arcmin in RA and DEC. The CMDs of each region clearly shows the presence of the main sequence (MS), RGB, red clump (RC) stars, as well as other evolutionary stages. We used the location of RC stars to trace the base of the RGB. \cite{Smitha&Purni2009A&Adepth} found that most of the LMC regions have large number of RC stars and their distribution can be identified well in the CMDs of the regions. We identified the location of RC stars and assume the peak (V$-$I) and V of the RC distribution as the base of the RGB. Because the RGB and RC colours are similar, they are similarly affected by reddening. That is, we do not expect any differential shift between RC and RGB stars. As the RC and RGB stars are identified together, this reduces the effect of reddening in locating the RGB similarly across the LMC. The steps involved in this method are described below:

\begin{enumerate}
\item Excluding the MS: We studied the MS in the CMDs of a few tens of regions located in different parts of the LMC, in order to decide a cut in colour and magnitude that would exclude most of the MS stars.  We find that if we include only stars with 0.5 $<$ (V$-$I) $\le$ 2.5 mag and 12.0 $\le$ V $<$ 20.0 mag, we can isolate the evolved part of the CMD. Figure \ref{fig:lsqfit}(a) shows the CMD of such a region (black filled circles), where the total number of stars in a particular region is given by N. The stars which remain after the above exclusion are shown within the rectangle.

\item Identifying the base of the RGB by constructing the density diagram: In order to identify the densest part of this portion of the CMD, we constructed a density diagram, where the CMD is binned in magnitude and colour with a bin width of 0.10 mag in V magnitude and 0.05 mag in (V$-$I) colour, and count the number of stars in each bin. Figure \ref{fig:lsqfit}(b) shows such a plot. The bins are colour coded based on the number of stars in each bin, as denoted in the colour bar. The most populated bin falls near the centre of the RC region, and is chosen as the base of the RGB. This location changes with the reddening, along with the location of the RGB, and is used to identify the RGB similarly in all the regions. 

\item Identifying the RGB: We remove the bluer and fainter bins with respect to the RC peak. That is, we give a cut in colour and magnitude corresponding to the densest point in the density CMD and assume that the RGB in the CMD consists of redder and brighter bins. This part of the CMD is dominated by RGB, but contaminated with some other evolutionary phases (AGBs etc.) to a lesser extent. This is shown in Figure \ref{fig:lsqfit}(c). The use of the densest bin at the magnitude of the RC allows us to uniquely and consistently define a location in the RGB and identify it uniformly in CMDs of all the regions. Thus, even if this is not the actual base of the RGB, the part of RGB used for slope estimation is made uniform for all location.

\item Estimating Slope: In order to identify the RGB unambiguously, and to reduce the scatter in the RGB, we consider only those colour-magnitude bins that contain a minimum number of stars. After inspecting the density CMD of various regions located at different parts of the LMC, we choose this minimum number to be 3. We also tried 5 stars as the minimum number, but this limits the extent of the RGB considered, as the brightest part of the RGB is, in general, poorly populated. The chosen criterion eliminates the brighter part of the RGB, typically sampling the RGB from the RC peak up to 2 magnitude brighter in most of regions. As shown in Figure \ref{fig:lsqfit}(d), the selected part of the RGB thus appears to be more or less a straight line, without the curved brighter part. These bins, thus representing the RGB, are fitted with a straight line and the slope is estimated using the method of least square fit. We carry out a 3 sigma clipping with a single iteration to refine our estimates. We have inspected many CMDs and the RGB fits using this method, and find that the straight line fit is satisfactory in the vast majority of regions. The fit is shown in Figure \ref{fig:lsqfit}(d). We define $N_p$, as the number of  CMD bins (with number of stars in each bin $\ge$ 3) representing the RGB, to which straight line is fitted after 3 sigma clipping, to estimate the slope. We denote the estimated slope of RGB by its absolute value, as $|$slope$|$, and the error in slope as $\sigma_{slope}$. We also express the correlation coefficient by its absolute value, as $r$.

\end{enumerate}

Thus, we have calculated $N_p$, $|$slope$|$, $\sigma_{slope}$, and $r$ for each of the regions. While doing so we have excluded the regions that have $N_p \sim 0$, thus making estimations for 1849 regions (out of 1854). The above method is found to work consistently for most of the regions, except when there is a large variation in reddening. In some regions, we identify the RGB to be very broad resulting in poor straight line fit. The number density of stars across the region observed by OGLE III varies significantly, with the maximum stellar density in the bar regions. If we use bins of equal area across the observed region, then CMDs of some of the central regions are found to be very dense and broad, resulting in poor fits. This suggests that large number of stars in the RGB might have an effect on the fit and estimation of slope. In order to probe the effect on the fit due to over-populated  CMDs, Figure \ref{fig:ogle3_np_vs_isum_undiv} shows a plot of $N_p$ versus N. The figure shows that as N increases, $N_p$ also increases, suggesting that, in general, the denser areas have well populated RGB. Now, we need to check the correlation between $N_p$ and $r$. A plot between $r$ and $N_p$ is shown in Figure \ref{fig:ogle3_np_vs_r_undiv}. It is seen that the regions with high $N_p$ (or correspondingly higher N) have lower $r$ ($<$ 0.50), suggesting a poor fit/estimation of slope. We identified the locations of such regions and inspected a few of them. It turns out that, the CMDs of these regions show a broad RGB. The broad RGB is likely to be due to small scale variation in reddening and/or multiple dominant population.

\begin{figure*}
\centering
\begin{minipage}[b]{0.45\linewidth}
\includegraphics[height=3.0in,width=3.0in]{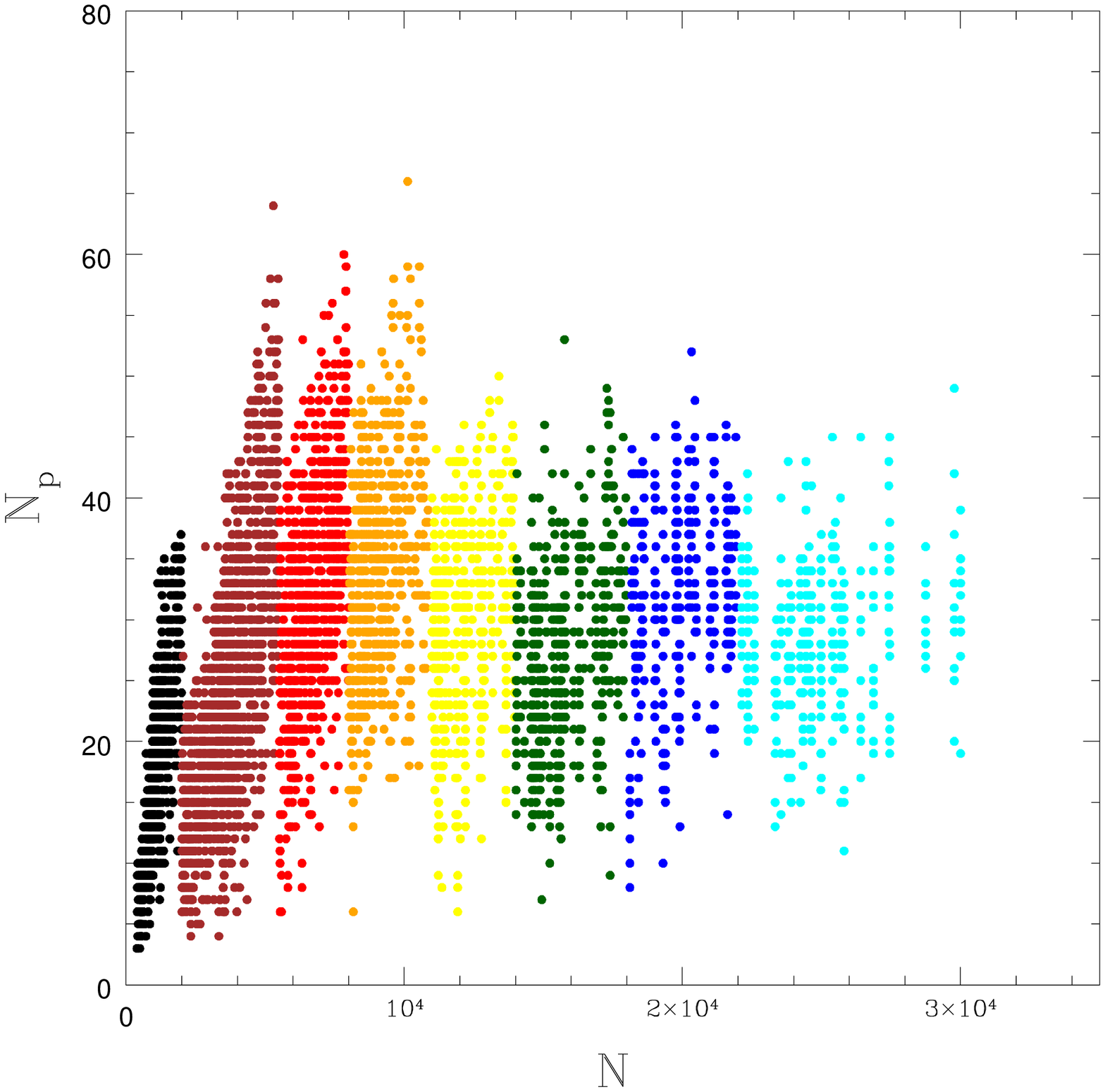}
\caption{\small Plot of $N_p$ versus $N$ for OGLE III subregions, after finer area binning. The colours correspond to the eight different bin areas, as mentioned in the eighth column of Table \ref{table:tab1}. 
\label{fig:ogle3_np_vs_isum_div}}   
\end{minipage}
\quad
\begin{minipage}[b]{0.45\linewidth}
\includegraphics[height=3.0in,width=3.0in]{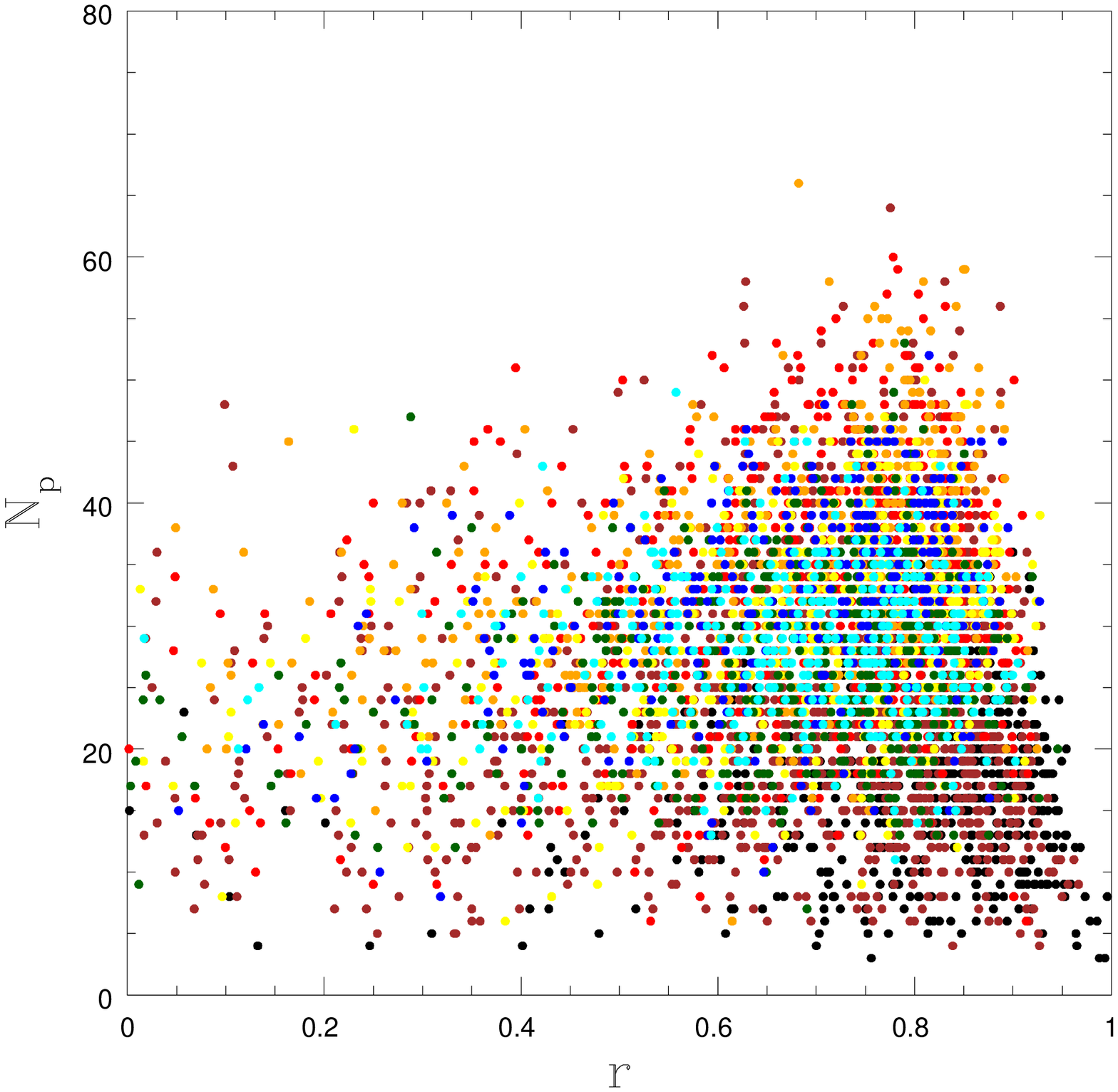}
\caption{\small Plot of $N_p$ versus $r$ for OGLE III subregions, after finer area binning. The colours correspond to the eight different bin areas, as mentioned in the eighth column of Table \ref{table:tab1}. 
\label{fig:ogle3_np_vs_r_div} }  
\end{minipage}
\end{figure*}

A way out from this particular issue is to further sub-divide the denser OGLE III regions spatially. To achieve this, we perform 8 types of area binning for OGLE III observed regions, where the binning criteria is solely based on the stellar density. We tried to keep a similar but sufficient number of density bins in the RGB, in order to get a good estimation of slope with high $r$ ($>$0.50). Table \ref{table:tab1} lists the 8 division criteria adopted based on the number of stars in a region. It also lists the total number of subregions extracted under each division criteria, along with their corresponding areas. The area of largest subregion is (8.88 $\times$ 8.88) sq. arcmin and the smallest area being (2.22 $\times$ 2.96) sq. arcmin. Excluding subregions with $N_p \sim 0$, the number of subregions analysed is 4777 (out of 4779). Figures \ref{fig:ogle3_np_vs_isum_div} and \ref{fig:ogle3_np_vs_r_div} show the plots of $N_p$ versus N and $N_p$ versus $r$ respectively, after the above area binning is performed. It is seen that the upper limit of $N_p$ is now confined to a lower value, and remains similar for almost all the 8 cases of sub-divisions. Also, the number of regions with higher $r$ value is found to increase, as compared to the previous case. Thus, we find that the sub-division has improved the fit to the RGB for a large number of regions. On the other hand, we still have some regions with low value of $r$. These regions were found to have either poorly  defined RGB due to scattered RGB stars, or very broad RGB. 

\begin{figure*}
\centering
\begin{minipage}[b]{0.45\linewidth}
\includegraphics[height=3.0in,width=3.0in]{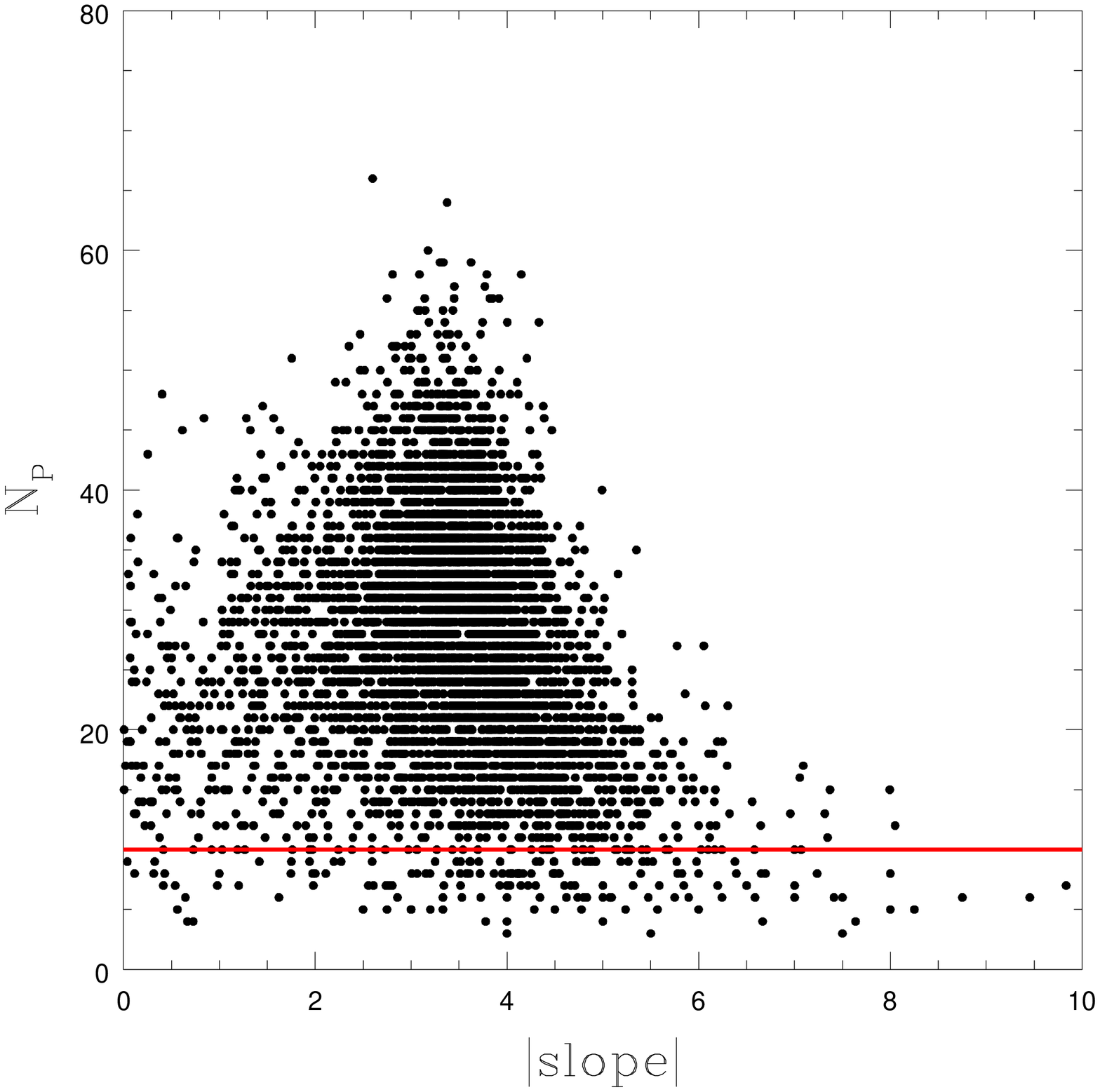}
\caption{\small Plot of $N_p$ versus $|$slope$|$ for OGLE III subregions. The red line at $N_p$ = 10 denotes the cut-off decided to exclude regions with poorly populated RGB.
\label{fig:ogle3_np_vs_slope} } 
\end{minipage}
\quad
\begin{minipage}[b]{0.45\linewidth}
\includegraphics[height=3.0in,width=3.0in]{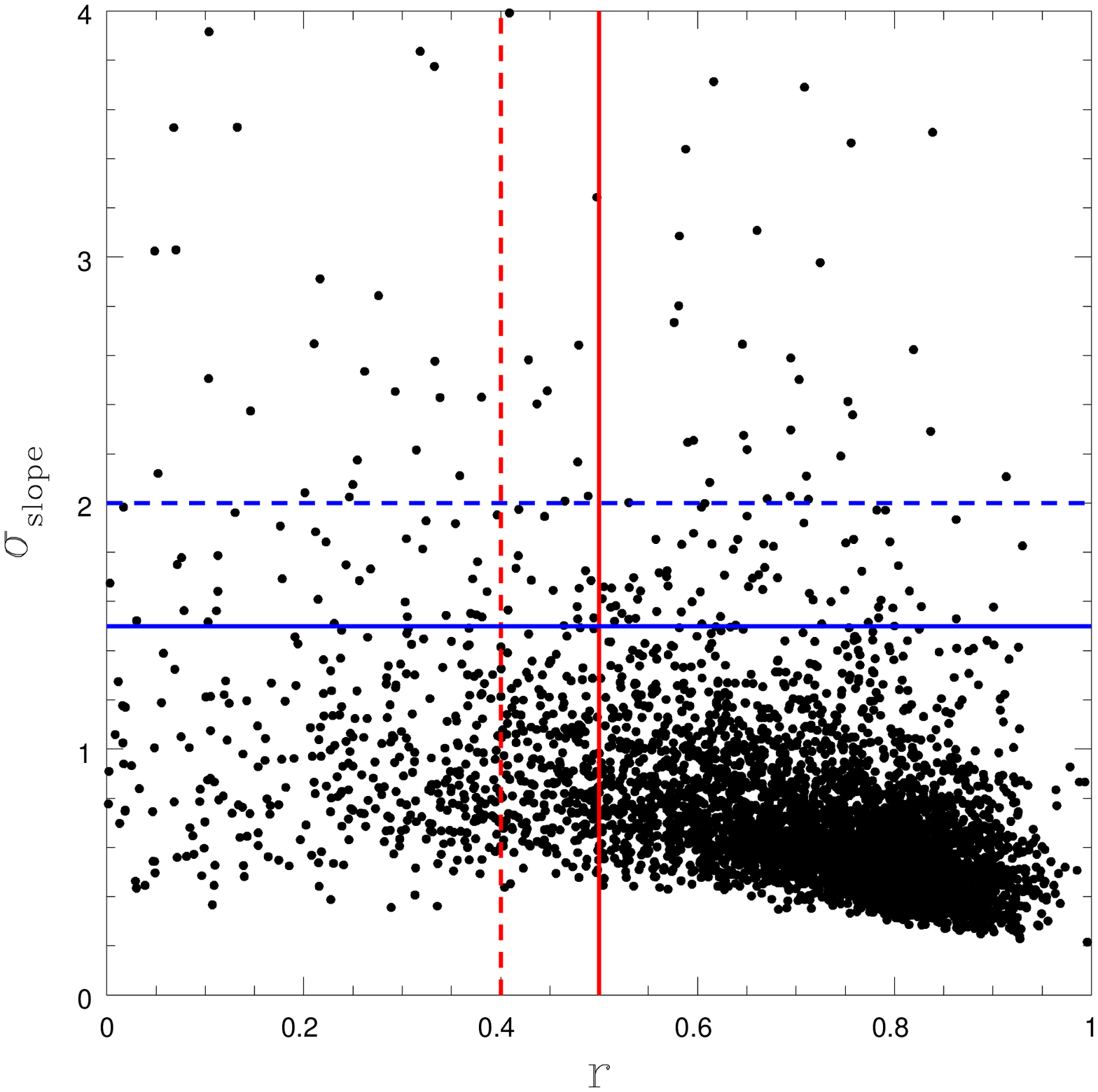}
\caption{\small Plot of $\sigma_{slope}$ versus $r$ for OGLE III subregions. The blue dashed and solid lines corresponds to the cut-off criteria on $\sigma_{slope}$ at 2.0 and 1.5 respectively. The red dashed and solid lines denote the cut-off corresponding to $r$ at 0.4 and 0.5 respectively. 
\label{fig:ogle3_errslope_vs_r} }  
\end{minipage}
\end{figure*}

\begin{figure*} 
\begin{center} 
\includegraphics[height=4.0in,width=4.0in]{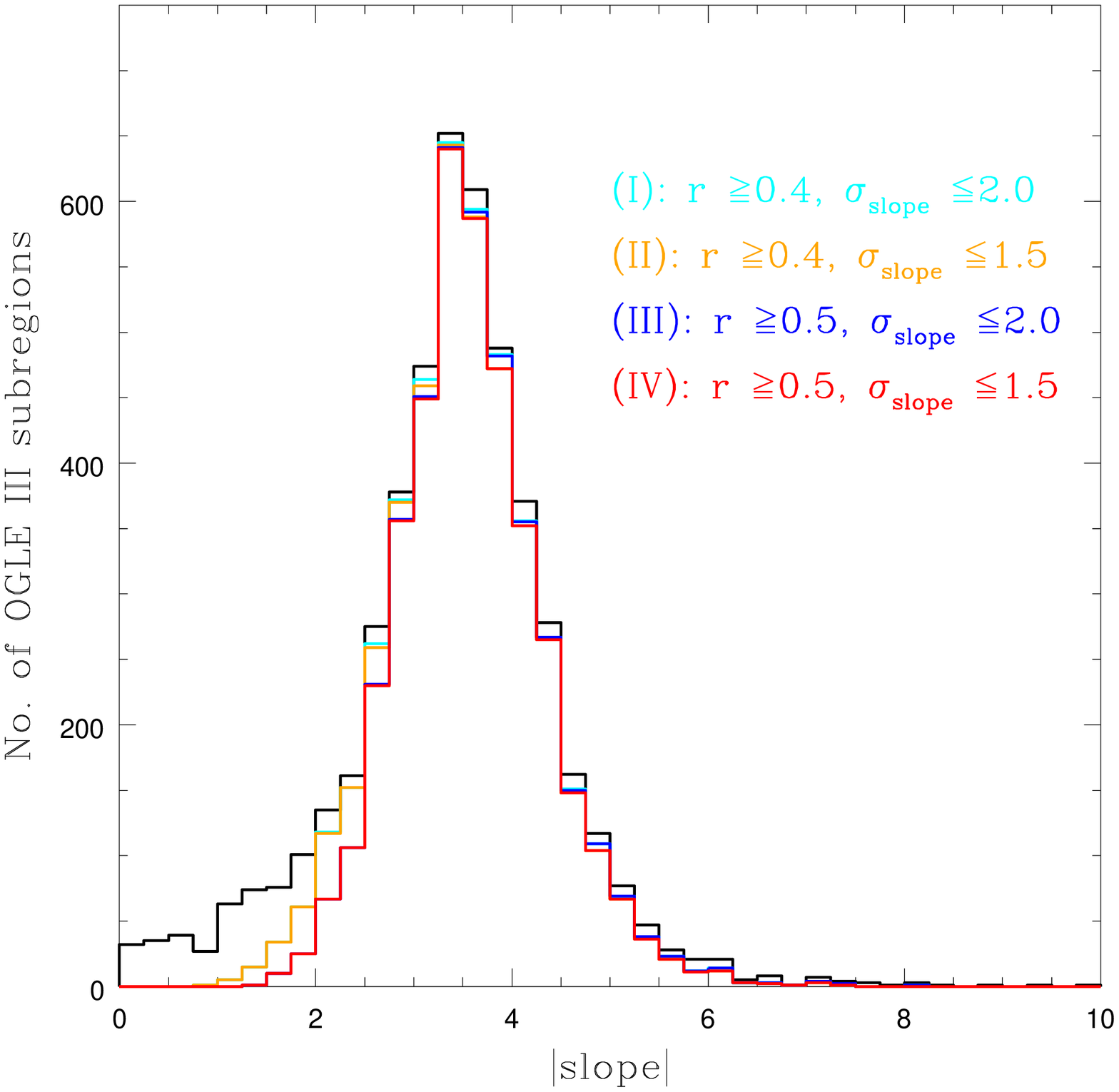}
\caption{\small Histogram of $|$slope$|$ for OGLE III subregions estimated for all the four cut-off criteria ($(I)$ in cyan, $(II)$ in orange, $(III)$ in blue, and $(IV)$ in red). $N_p$ $\ge$ 10 for all these four cases. The black solid line shows the distribution of $|$slope$|$ with no cut-off corresponding to all OGLE III subregions. 
\label{fig:ogle3_histslope}}  
\end{center} 
\end{figure*}

To carry forward our analysis, we need to remove regions that have a poor estimation of slope. Although the goodness of fit is best described  by $r$, a few more  parameters may also be considered. Figure \ref{fig:ogle3_np_vs_slope} shows a plot between $N_p$ and estimated RGB slope ($|$slope$|$). It is seen from the plot that for regions with $N_p$ $<$10, there is a large range in slope value, along with a few regions estimated with a very large value for $|$slope$|$. A low value of $N_p$ signifies a sparsely populated RGB, hence the large slope value may be an artefact. To exclude regions with poorly populated RGB, we select only those regions that have $N_p$ $\ge$ 10, implying that the fitted RGB should at least have 30 stars. We need to decide the optimal cut-off for $r$, to remove regions with poor fit to the RGB slope. It is also necessary to exclude regions that have large error in $|$slope$|$, i.e. $\sigma_{slope}$. Figure \ref{fig:ogle3_errslope_vs_r} shows a plot between $\sigma_{slope}$ versus $r$. The figure shows that most of the regions have $r$ in the range 0.4-0.95, and $\sigma_{slope}$ in the range 0.5-2.0. Large scatter is observed for regions with $\sigma_{slope}$ $>$ 2.0 and $r$ $<$ 0.4. It is also seen that the densest part of the diagram is found for $\sigma_{slope}$ $<$ 1.5 and $r$ $>$ 0.5. Thus, to remove regions with poor fit from our analysis, we have considered four different cases based on the cut off for $r$ and $\sigma_{slope}$ (with $N_p$ $\ge$ 10 for all cases). They are as follows: 
\begin{itemize}
\item criteria (I): $r$ $\ge$ 0.4 and $\sigma_{slope}$ $\le$ 2.0.
\item criteria (II): $r$ $\ge$ 0.4 and $\sigma_{slope}$ $\le$ 1.5.
\item criteria (III): $r$ $\ge$ 0.5 and $\sigma_{slope}$ $\le$ 2.0.
\item criteria (IV): $r$ $\ge$ 0.5 and $\sigma_{slope}$ $\le$ 1.5.
\end{itemize}
Figure \ref{fig:ogle3_histslope} shows the histogram of the RGB slope, estimated using OGLE III data. The slope distribution for all the 4 cut-off cases mentioned above are plotted with respect to the original distribution. It is observed from the plot that most of the regions with slopes less than 2.0 get removed after the cut-off criteria are implemented. A decrease in $\sigma_{slope}$ for same value of $r$ does not produce significant change in the width of the distribution. However, with increasing value of the cut-off in $r$, the width of the distribution reduces mostly for the lower value of slope. Thus, out of these 4 cut-off criteria listed above, the last one is the most strict criteria for selecting the regions with best fits, and the first one is the most relaxed criteria. 

\begin{table*}
{\small
\caption{Sub-division of OGLE III regions:}
\label{table:tab1}
\begin{tabular}{|c|c|c|c|c|c|c|c|}
\hline \hline
Sl. no. & No. of  & No. of  & No. of     & No. of     & No. of             & Area of        &  Number of       \\
        & stars   & regions & division   & division   & sub-divisions      & a sub-division &  subregions      \\
        &         & (a)     & along RA   & along Dec  & (d=b$\times$c)     & (arcmin sq.)   &  (a$\times$d)    \\
        &         &         & (b)        & (c)        &                    &                &                  \\
\hline\hline
1  & 0 $<$ N $\le$ 2000      & 637 & 1 & 1 & 1  & (8.88$\times$8.88) & 637 (black)     \\
2  & 2000 $<$ N $\le$ 5500   & 628 & 2 & 1 & 2  & (4.44$\times$8.88) & 1256 (brown)    \\
3  & 5500 $<$ N $\le$ 8000   & 245 & 3 & 1 & 3  & (2.96$\times$8.88) & 735 (red)       \\
4  & 8000 $<$ N $\le$ 11000  & 141 & 2 & 2 & 4  & (4.44$\times$4.44) & 564 (orange)    \\
5  & 11000 $<$ N $\le$ 14000 & 74  & 3 & 2 & 6  & (2.96$\times$4.44) & 444 (yellow)    \\
6  & 14000 $<$ N $\le$ 18000 & 60  & 4 & 2 & 8  & (2.22$\times$4.44) & 480 (dark green)\\
7  & 18000 $<$ N $\le$ 22000 & 35  & 3 & 3 & 9  & (2.96$\times$2.96) & 315 (blue)      \\
8  & N $>$ 22000           & 29  & 4 & 3 & 12 & (2.22$\times$2.96) & 348 (cyan)      \\
\hline     
\end{tabular}
\vskip 1.0ex
\begin{minipage} {180mm}
{Note: The table describes the 8 binning criteria used to sub-divide OGLE III regions. For each criteria, the second column denotes the limit on total number of stars (N) within a region. The third column gives the number of regions, having N, within that specified limit. Column four and five specify the number by which a region is binned along RA and Dec respectively. Column six thus gives the total number of subregions, a single region is binned into. Whereas, the seventh column gives the area of each such subregion. The last (eighth) column denotes the total number of subregions corresponding to each of the 8 sub-division criteria. The colours adjacent to the numbers are used to denote them in Figure \ref{fig:ogle3_np_vs_isum_div} and \ref{fig:ogle3_np_vs_r_div}.}
\end{minipage}
}
\end{table*} 


\subsection{Calibration of slope to metallicity}

After estimating the RGB slope of subregions within the LMC, the next task is to convert the slope to metallicity. As the slope of the RGB is a measure of the average metallicity of the region, we use the spectroscopically estimated average metallicity of red giants in that region to build the required relation. In order to build this relation, we need to estimate the spectroscopically determined mean metallicity for several regions within the area studied. As there is no such single study, we identified three studies which cover the inner to outer regions. We used metallicities of field red giants \citep{Cole+2005AJspectroOfRGs}, star clusters \citep{Grocholski+2006AJCaIItriplet} and field red giants around these clusters (Cole et al. 2015, in preparation). We have chosen the above three studies for the following reasons: (1) These studies derive metallicities for the same population as our study, the red giants, (2) Our slope map covers a fairly large range in slope and hence metallicity. As the map also covers a large area of the LMC, the slope to metallicity relation should be built from a range in metallicity as well as location in the LMC. These three studies put together cover a range in metallicity as well as location in the LMC. (3) All the three studies have used the Ca II triplet lines to derive the metallicity, including the calibration of CaT strength to [Fe/H], thus there is no inconsistency or systematic offsets between these studies.

\cite{Cole+2005AJspectroOfRGs} calculated abundance of 373 field RGs within a 200 arcmin$^2$ area at the optical centre of the LMC bar. We calculate the mean metallicity for a subregion by averaging over the Cole et al.\ metallicities within its area. While doing so we consider stars lying within twice the standard deviation about the mean metallicity. To ensure good calibration, we consider those subregions that have $r$ higher than 0.70, $\sigma_{slope}$ lower than 1.0, and contain  spectroscopic metallicity estimate of at least 5 RGs. Under this criteria, we could identify 12 subregions which can be used for calibrating the RGB slope-metallicity relation. It is found that the range of slope value covered by all these sub regions, except for one, lies mostly within a range of $|$slope$|$ $\sim$ 3.2--3.8. One point is found near $|$slope$|$ $\sim$ 4.4. We are unable to find any subregion overlapping with the Cole et al.\ (2005) sample that has higher mean metallicity than that corresponding to $|$slope$|$ $\sim$ 3.2. The slope histogram the full LMC shows that the majority of regions have $|$slope$|$ $>$ 3.2.

To cover a larger range in slope, so as to formulate a good calibration relation, we used \cite{Grocholski+2006AJCaIItriplet}. The authors studied the abundance of 28 LMC clusters, and the metallicity range covered in their study is: $-$0.3 $\ga$ [Fe/H] $\ga$ $-$2.0. Out of  these 28 clusters, only 12 are located in the OGLE III field. Using the central co-ordinates and radii of these clusters, we extracted their corresponding OGLE III data to construct their V, (V$-$I) CMDs. We tried to estimate the slope of the RGB for all these clusters using the same technique as mentioned in Section 3. For most of the clusters, the RGB was either sparsely populated or had significant scatter about the mean RGB (either due to crowding or intermingling of cluster and field stars in the OGLE III data or, less likely, differential reddening). Therefore, we could make a good estimate of the cluster RGB slope only for NGC 2121. Examination of the remaining clusters showed that by manually cleaning the OGLE III sample in the subregion we could additionally make a good estimate of the RGB slope for NGC 1651, in which the cluster RGB was not very dense. The estimated slope values for these two clusters increased the range of slopes in our calibration relation, especially to lower value of metallicity. From the slope histogram, it can be seen that only a small number of regions have $|$slope$|$ $>$ 4.6.

The third set of data used for calibration is from Cole et al. (2015, in preparation). The authors have estimated the mean metallicity of field regions around the 28 clusters that were studied by \cite{Grocholski+2006AJCaIItriplet}. We repeated a similar process as described above to extract the OGLE III data corresponding to each such field, with the cluster area subtracted. The CMD of the resultant field was then used to estimate the RGB slope. For most of these fields our usual method did not work either due to scatter in RGB or due to differential reddening. The fields around NGC 2121 and NGC 1751 suffered least from these issues but had sparse RGB. For these two cases, we adopted a similar technique as for the cluster NGC 1651, to estimate the slope. The metallicity of these fields were found to be similar to that of the clusters, with the slopes found to be $|$slope$|$ $\sim$ 4.2. 

\begin{figure*} 
\begin{center} 
\includegraphics[height=4in,width=4in]{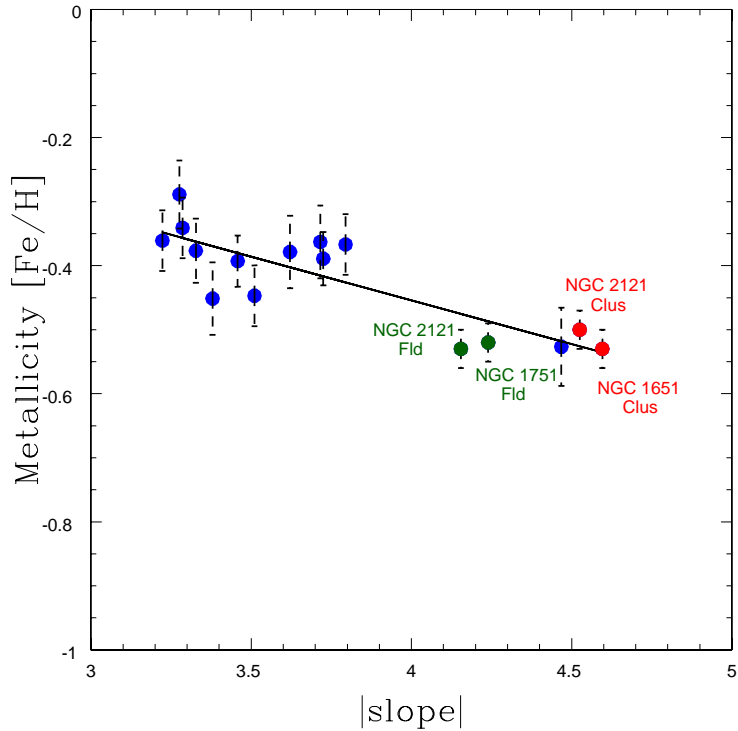}
\caption{Plot of metallicity ([Fe/H]) versus $|$slope$|$. Blue points denote our subregions whose mean [Fe/H] has been found using RGs from \protect\cite{Cole+2005AJspectroOfRGs}, red points denote clusters (NGC 1651 and NGC 2121) from \protect\cite{Grocholski+2006AJCaIItriplet}, 
and the dark green points correspond to field around clusters (NGC 2121 and NGC 1751) from Cole et al. (2015, in preparation). The straight line fit to all the calibrators is shown as a black solid line. The error bar (black dashed line) shown for each point is the standard error of mean [Fe/H].}
\label{fig:ogle_calib}
\end{center} 
\end{figure*}

In Figure \ref{fig:ogle_calib}  the final 16 subregions used for the slope-metallicity calibration are plotted in the metallicity versus $|$slope$|$ plane. These values are also tabulated in Table \ref{table:tab2}. The trend shows that with increasing value of $|$slope$|$, the metallicity decreases. We estimate a linear relation between them by fitting a straight line using least square fit. The slope-metallicity relation estimated is given by:
\begin{equation} \label{eq:1}
[Fe/H]=(-0.137\pm0.024)\times|slope|+ (0.092\pm0.091);
\end{equation}
with $r$=0.83.We need to extrapolate this relation to the whole range of slopes estimated for different subregions within the LMC and obtain the value of corresponding metallicity. Although the range of slope covered by the 16 calibration points (3.2 $\la$ $|$slope$|$ $\la$ 4.6) is small as compared to the range in slope we have obtained (1.5 -- 6.0), nearly all of the regions have 3.0 $\la$ $|$slope$|$ $\la$ 5.0. Thus, we have used the above relation for the entire range of the slope, to estimate metallicity.

\begin{table*}
{\small
\caption{Calibrators for OGLE III slope-metallicity relation:}
\label{table:tab2}
\begin{tabular}{|c|c|c|c|c|c|}
\hline \hline
Name           & (RA$^{\circ}$, Dec$^{\circ}$) & $|$slope$|$ $\pm$ $\sigma_{slope}$ & $r$ & Mean [Fe/H] & Standard error         \\
               &                               &                                    &     & (dex)       & of mean [Fe/H] (dex)   \\
\hline\hline
Subregion 1    & (81.48, $-$69.72) & 3.27$\pm$0.32 & 0.87 & $-$0.29 & 0.05 \\      
Subregion 2    & (81.48, $-$69.66) & 3.79$\pm$0.49 & 0.80 & $-$0.37 & 0.05 \\     
Subregion 3    & (81.48, $-$69.62) & 3.33$\pm$0.40 & 0.83 & $-$0.38 & 0.05 \\      
Subregion 4    & (80.63, $-$69.62) & 3.22$\pm$0.49 & 0.74 & $-$0.36 & 0.05 \\      
Subregion 5    & (80.62, $-$69.77) & 3.38$\pm$0.40 & 0.83 & $-$0.45 & 0.06 \\     
Subregion 6    & (80.73, $-$69.77) & 3.72$\pm$0.49 & 0.78 & $-$0.39 & 0.04 \\       
Subregion 7    & (81.16, $-$69.82) & 3.62$\pm$0.38 & 0.89 & $-$0.38 & 0.06 \\      
Subregion 8    & (81.16, $-$69.77) & 3.71$\pm$0.46 & 0.84 & $-$0.36 & 0.06 \\      
Subregion 9    & (81.37, $-$69.97) & 4.47$\pm$0.75 & 0.79 & $-$0.53 & 0.06 \\      
Subregion 10   & (80.78, $-$69.92) & 3.51$\pm$0.52 & 0.76 & $-$0.45 & 0.05 \\      
Subregion 11   & (81.21, $-$70.02) & 3.46$\pm$0.50 & 0.75 & $-$0.39 & 0.04 \\      
Subregion 12   & (81.36, $-$70.02) & 3.28$\pm$0.44 & 0.75 & $-$0.34 & 0.05 \\                   
NGC 1651       & (69.39, $-$70.58) & 4.59$\pm$0.37 & 0.84 & $-$0.53 & 0.03 \\
NGC 2121       & (87.05, $-$71.48) & 4.52$\pm$0.35 & 0.96 & $-$0.50 & 0.03 \\
NGC 1751 Field & (73.56, $-$69.83) & 4.24$\pm$0.20 & 0.82 & $-$0.52 & 0.03 \\
NGC 2121 Field & (87.04, $-$71.50) & 4.15$\pm$0.22 & 0.83 & $-$0.53 & 0.03 \\
\hline     
\end{tabular}
\begin{minipage} {180mm}
\vskip 1.0ex
{Note: The table lists out the 16 calibrators used to construct the slope-metallicity relation for OGLE III data. The first 12 entries in column number one corresponds to subregions lying near the central LMC, followed by two clusters (NGC 1651 and NGC 2121), and field around two clusters (NGC 1751 and NGC 2121). The central (RA, Dec) corresponding to each calibrator is listed down in second column. The third column gives the estimated slope and its associated error for each calibrator, whereas the correlation coefficient ($r$) for each case is specified in the fourth column. The fifth column denotes the mean [Fe/H] estimated for each calibrator using spectroscopic results of \cite{Cole+2005AJspectroOfRGs} (for 12 subregions), \cite{Grocholski+2006AJCaIItriplet} (for two clusters), and Cole et al. (2015, in preparation) (for field around two clusters). The standard error of mean [Fe/H] is  mentioned in the sixth column. It is calculated as 0.15/$\sqrt{n}$, where n is the number of RGs located within the area of the 
calibrator.}
\end{minipage}
}
\end{table*} 


\begin{figure*} 
\begin{center} 
\includegraphics[height=4.5in,width=6in]{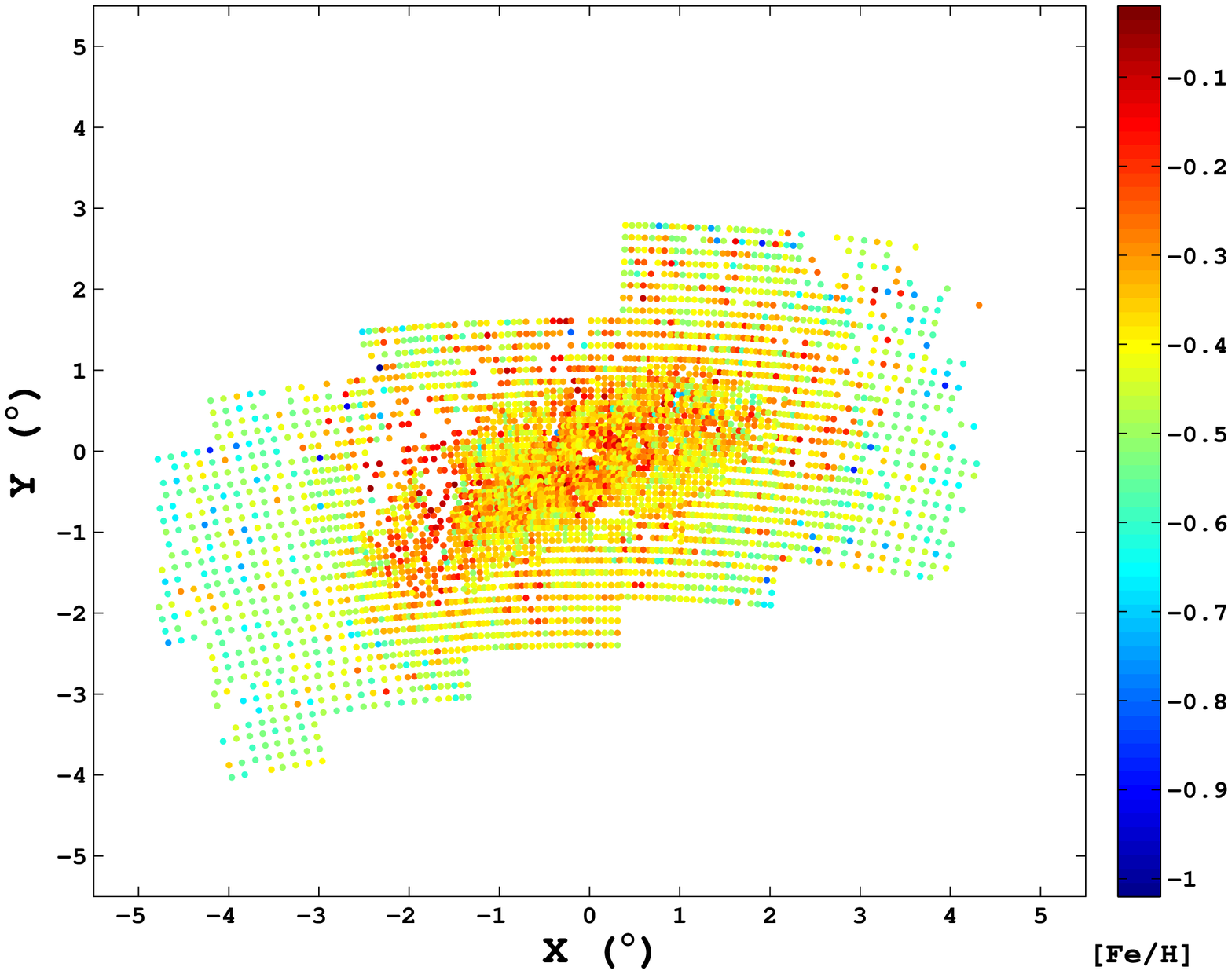}
\caption{\small OGLE III metallicity map with cut-off criteria $(I)$: $N_p$ $\ge$ 10, $r$ $\ge$ 0.4 and $\sigma_{slope}$ $\le$ 2.0.
\label{fig:ogle3map1}} 
\end{center} 
\end{figure*}

\begin{figure*} 
\begin{center} 
\includegraphics[height=4.5in,width=6in]{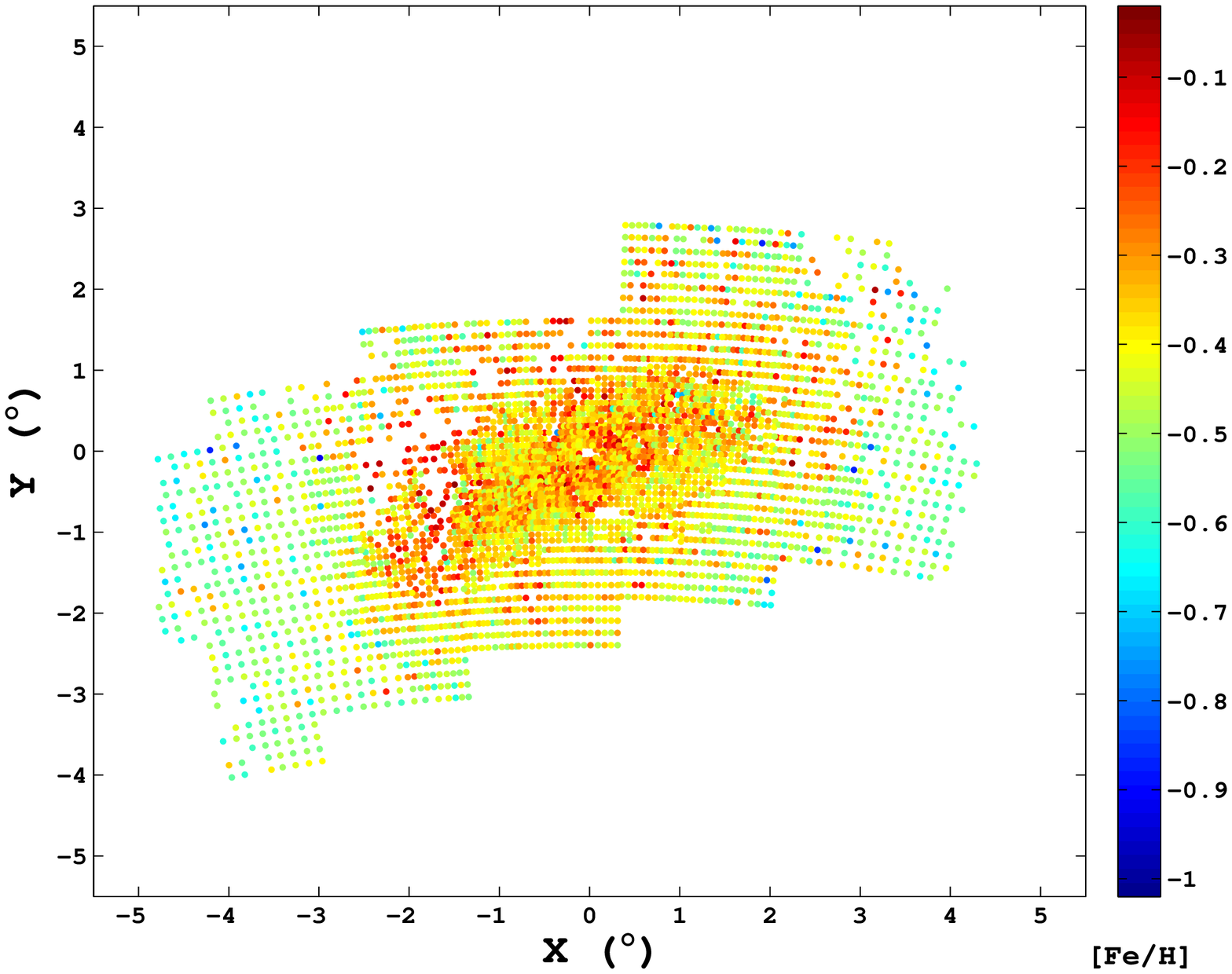}
\caption{\small OGLE III metallicity map with cut-off criteria $(II)$: $N_p$ $\ge$ 10, $r$ $\ge$ 0.4 and $\sigma_{slope}$ $\le$ 1.5.
\label{fig:ogle3map2}} 
\end{center} 
\end{figure*}

\begin{figure*} 
\begin{center} 
\includegraphics[height=4.5in,width=6in]{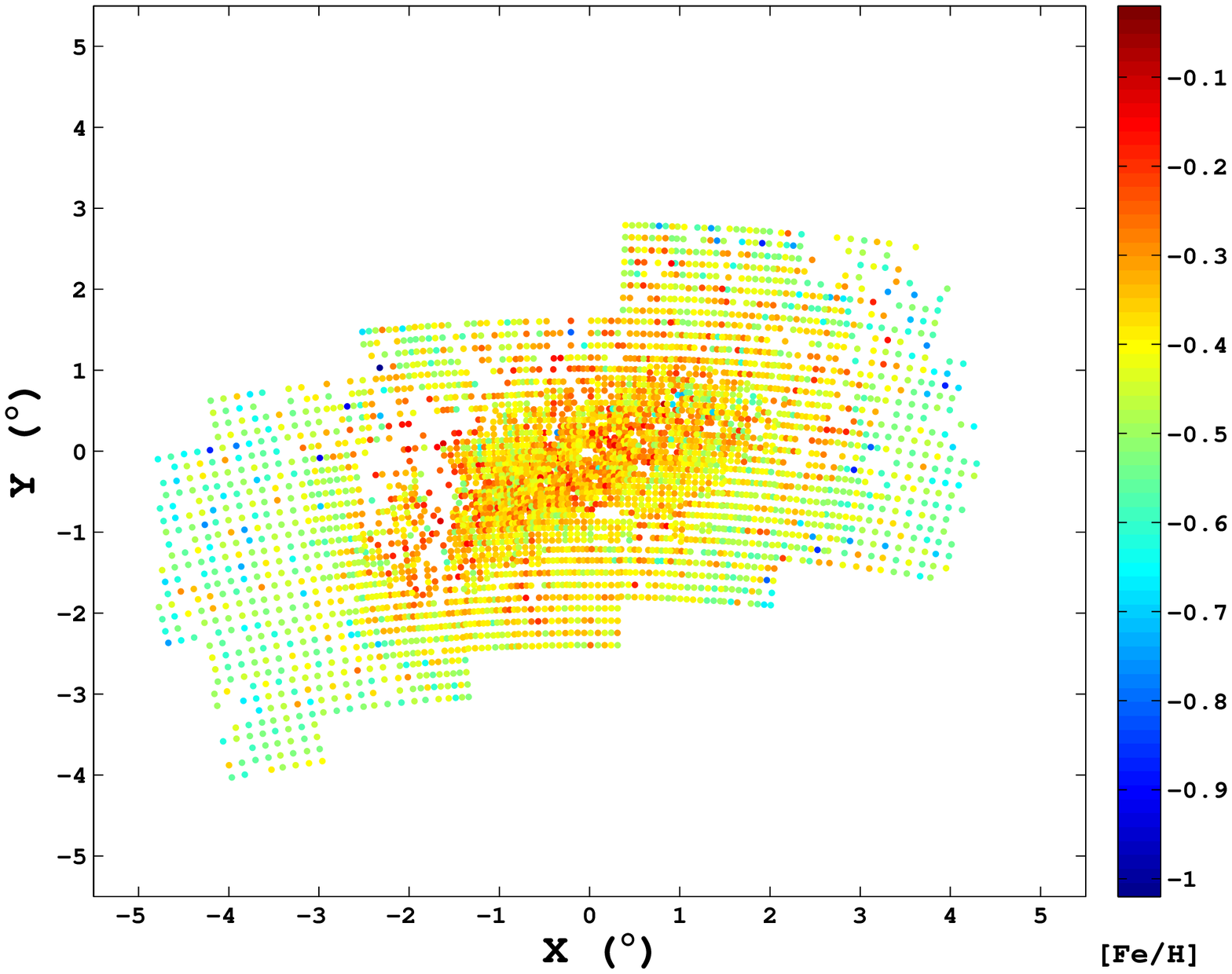}
\caption{\small OGLE III metallicity map with cut-off criteria $(III)$: $N_p$ $\ge$ 10, $r$ $\ge$ 0.5 and $\sigma_{slope}$ $\le$ 2.0.
\label{fig:ogle3map3}} 
\end{center} 
\end{figure*}

\begin{figure*} 
\begin{center} 
\includegraphics[height=4.5in,width=6in]{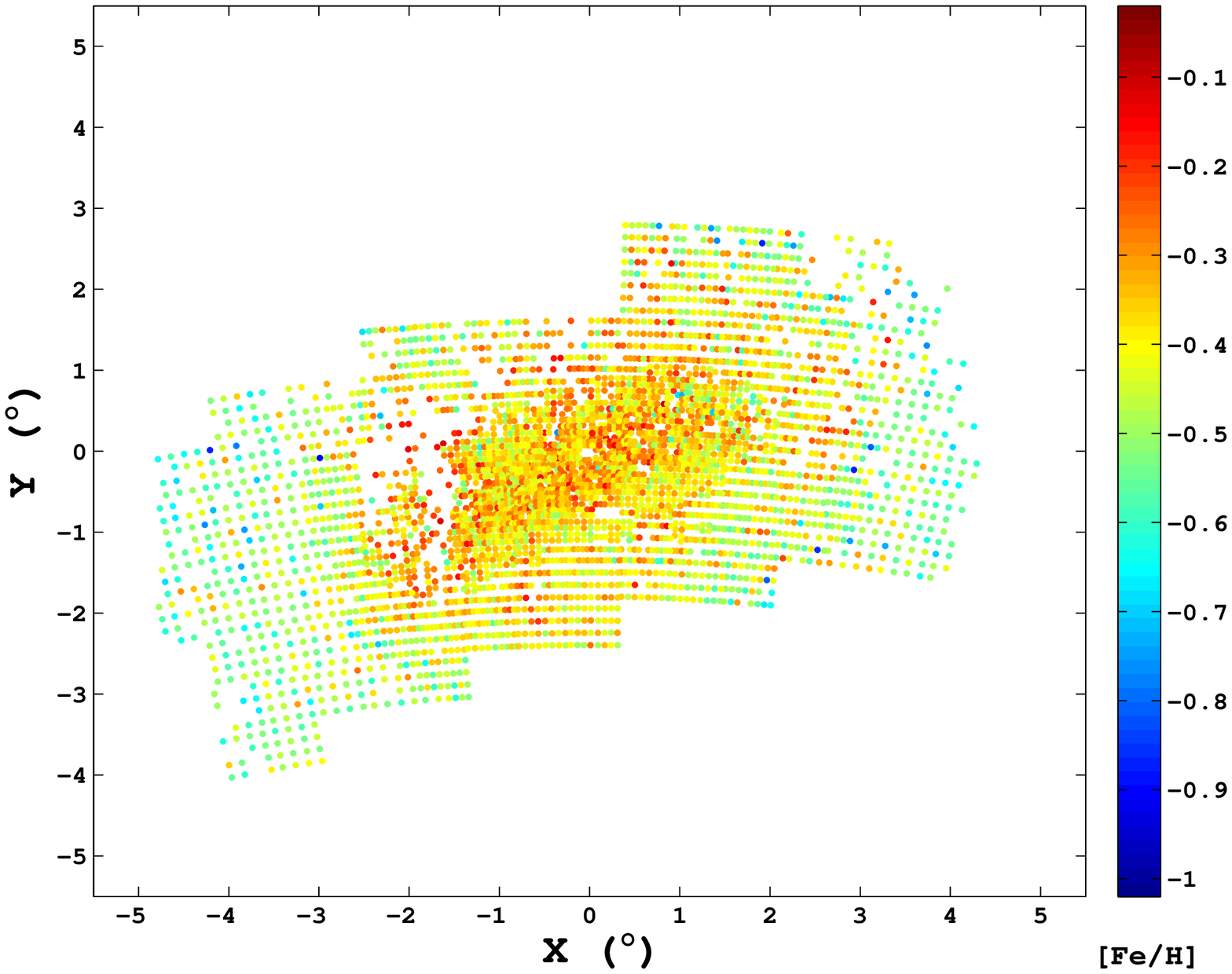}
\caption{\small OGLE III metallicity map with cut-off criteria $(IV)$: $N_p$ $\ge$ 10, $r$ $\ge$ 0.5 and $\sigma_{slope}$ $\le$ 1.5.
\label{fig:ogle3map4}} 
\end{center} 
\end{figure*}
\subsection{ The OGLE III metallicity map}

After converting the OGLE III slopes to metallicities using Equation \ref{eq:1}, we can make the average metallicity map for the inner LMC, derived from photometry. This is the first such average metallicity map of the LMC estimated using red giant photometry, with both high spatial resolution and complete spatial coverage of the bar and disk. To understand the variation of metallicity in the projected sky plane we convert the RA-Dec to the Cartesian coordinates (X, Y). The value of the LMC centre used is, RA = 5$^h$ 19$^m$ 38$^s$; Dec = $-$69$^{\circ}$ 27$^m$ 5.2$^s$ (J2000.0 \citealt{deVaucouleurs&Freeman1972VAstructure}). We created a metallicity map in the (X, Y) plane using all the 4 cut-off criteria described in Section 3 (Figures \ref{fig:ogle3map1}, \ref{fig:ogle3map2}, \ref{fig:ogle3map3} and \ref{fig:ogle3map4}). The map primarily reveals the metallicity trend in the bar, eastern and western regions of the LMC.  The four maps look more or less similar, with the last two being almost identical. All the maps show a gradient in the metallicity from the centre to the outer regions. The bar region is found to be more metal rich, with a small scale variation in metallicity. We detect a drop in the metallicity beyond a radial distance of 2--2.5 degrees. The maps also suggest that most of the regions with metallicity less then $-$0.6 dex are located in the outer regions, with a few of them located in the middle of the metal rich bar region. The eastern and the western LMC are found to be metal poor compared to the bar region. Due to our cut-off criteria, several regions get removed from our analysis due to poor estimation of slope and thus metallicity. Most of the regions which are excluded due to poor slope estimation are located close to the 30 Dor star forming region, with a few in the bar region. These regions are likely to suffer from large reddening or variation in reddening within the small subregion. 

The average values of metallicity of the complete LMC, the bar region and the outer region of LMC are calculated and tabulated in Table \ref{table:tab3}, for all the 4 cut-off criteria. By the complete LMC, we mean all the regions for which metallicity has been estimated. The bar region is defined according to \cite{Smitha&Purni2010A&Aanestimate}. By the outer LMC, we mean the regions that lie beyond a radial distance of 2.5 deg away from the LMC centre. It can be seen that the estimated average metallicity values do not change significantly between the criteria, though the number of regions considered change. Also, the difference between the average metallicity and the number of regions are not significantly different between criteria $(III)$ and $(IV)$. Nevertheless, we use criteria $(IV)$ and the last row as final values and discuss these. The errors shown along with the average metallicity values are the standard deviation about the mean and do not reflect error in the individual metallicity estimation. It can be seen that there is a metallicity difference between the bar region and the outer regions, though it is not very high. The average metallicity value of [Fe/H] = $-$0.40$\pm$0.10 suggests that the part of the LMC studied here, does not have much of variation in metallicity as suggested by the relatively low standard deviation. The last row in the table also suggests that, the number of subregions in the bar region and the outer region is similar, hence similarly sampled. Also, the outer regions sampled here are mainly the eastern and the western regions and have an average metallicity of [Fe/H] = $-$0.47$\pm$0.11. 

\begin{figure*}
\centering
\begin{minipage}[b]{0.45\linewidth}
\includegraphics[height=3.0in,width=3.0in]{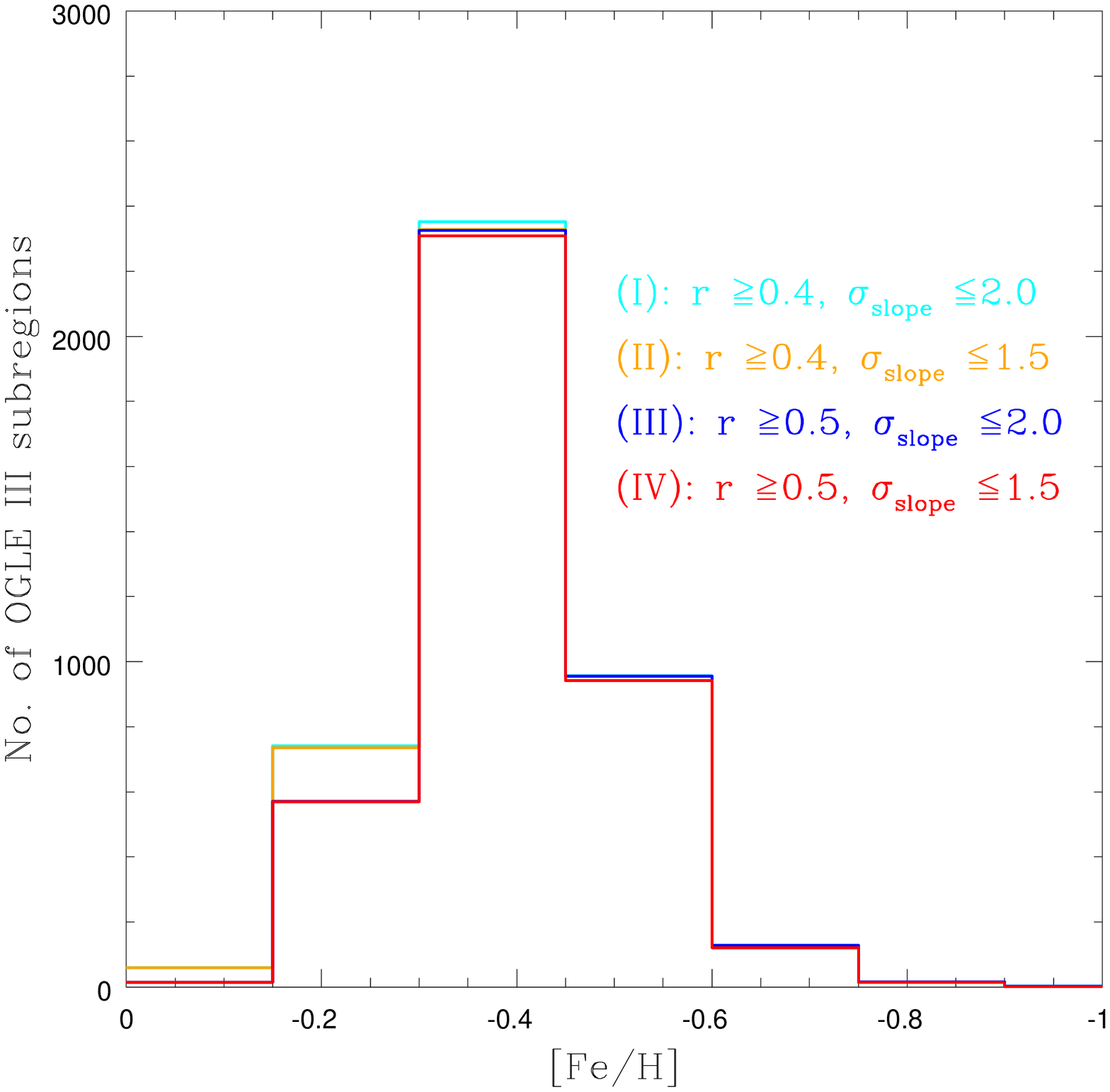}
\caption{\small Histogram of metallicity ($[Fe/H$]) for OGLE III data, estimated for all the four cut-off criteria ($(I)$ in cyan, $(II)$ in orange, $(III)$ in blue, and $(IV)$ in red). $N_p$ $\ge$ 10 for all these four cases.
\label{fig:ogle3_histabun}} 
\end{minipage}
\quad
\begin{minipage}[b]{0.45\linewidth}
\includegraphics[height=3.0in,width=3.0in]{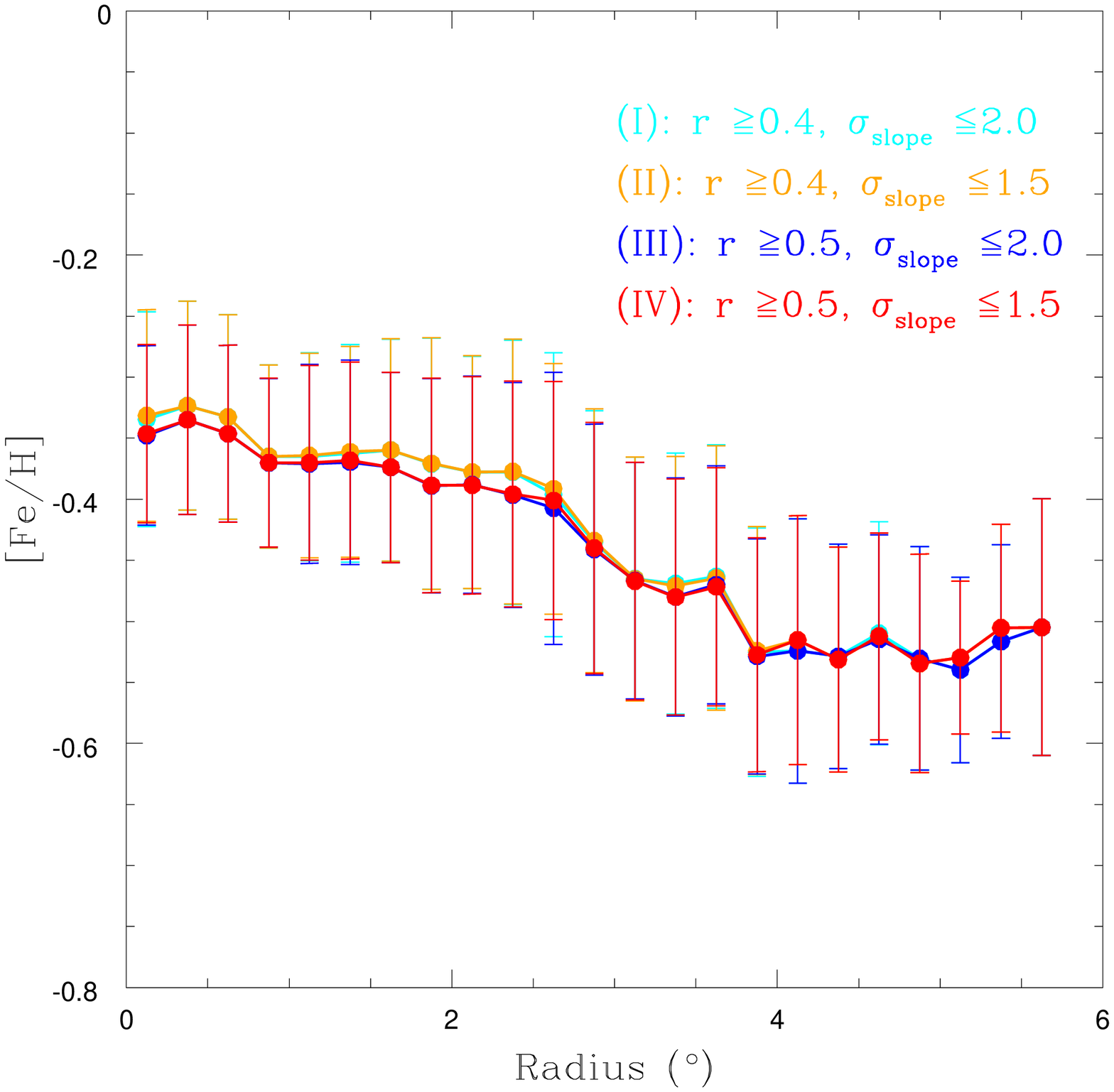}
\caption{\small Radial variation of metallicity ($[Fe/H$]) for OGLE III data, estimated for all the four cut-off criteria ($(I)$ in cyan, $(II)$ in orange, $(III)$ in blue, and $(IV)$ in red). $N_p$ $\ge$ 10 for all these four cases.
\label{fig:ogle3rad}}
\end{minipage}
\end{figure*}

The metallicity distribution is presented in Figure \ref{fig:ogle3_histabun} for all the 4 cut-off criteria. The distribution is binned with a width of 0.15 dex, which is of the order of one sigma error. The distribution contains a smaller number of regions with high metallicity as we progressively tighten the selection criteria, from $(I)$ to $(IV)$. On the other hand, the difference in distribution between criteria $(III)$ and $(IV)$ is marginal, suggesting that all estimates in the distribution are likely to be more or less reliable. The peak of the distribution is found be in the bin $-$0.30 to $-$0.45 dex, which is consistent with the values listed in Table \ref{table:tab3}. Almost no regions are found to have metallicity lower than $-$0.75 dex and higher than $-$0.15 dex. 

Figure \ref{fig:ogle3rad} shows a plot of the radial metallicity gradient for all 4 cut-off criteria. To construct the radial gradient, we binned the metallicity map into annular bins of width 0.25$^{\circ}$, and estimated the mean metallicity within each annular bin. It is seen from the figure that the mean metallicity is almost constant in the bar region, i.e. within a region of $\sim$2$^\circ$ around the optical centre. Also, the radial metallicity gradient is shallow. The radial profile shows the variation across the bar and off the bar, as the northern and the southern regions are not covered in OGLE III.  The profile thus suggests a homogeneous and metal rich bar region with a relatively metal poor eastern and western disk. A more complete analysis of the disk metallicity requires the northern and the southern regions to be included, using the MCPS data.

\begin{table*}
{\small
\caption{Mean metallicity for different regions of the LMC using OGLE III data:}
\label{table:tab3}
\begin{tabular}{|c|c|c|c|c|c|c|}
\hline \hline
Cut-off criteria & $r$  & $\sigma_{slope}$  & Region of the LMC & Number of subregions & Mean [Fe/H] (dex) \\
\hline\hline
I   & $\ge$ 0.40 & $\le$ 2.0 & COMPLETE & 4259 & $-$0.39$\pm$0.11\\      
    &            &           & BAR      & 1284 & $-$0.34$\pm$0.09\\      
    &            &           & OUTER    & 1248 & $-$0.46$\pm$0.11\\         
\hline    
II  & $\ge$ 0.40 & $\le$ 1.5 & COMPLETE & 4202 & $-$0.39$\pm$0.11\\      
    &            &           & BAR      & 1278 & $-$0.34$\pm$0.09\\      
    &            &           & OUTER    & 1210 & $-$0.46$\pm$0.11\\         
\hline    
III & $\ge$ 0.50 & $\le$ 2.0 & COMPLETE & 4014 & $-$0.40$\pm$0.10\\      
    &            &           & BAR      & 1191 & $-$0.35$\pm$0.08\\      
    &            &           & OUTER    & 1212 & $-$0.47$\pm$0.11\\         
\hline    
IV  & $\ge$ 0.50 & $\le$ 1.5 & COMPLETE & 3969 & $-$0.40$\pm$0.10\\      
    &            &           & BAR      & 1189 & $-$0.35$\pm$0.08\\      
    &            &           & OUTER    & 1180 & $-$0.47$\pm$0.11\\         
\hline       
\end{tabular}
\begin{minipage} {180mm}
\vskip 1.0ex
{Note: The first column denotes the four different cut-off criteria considered to filter out the LMC subregions. It is to be noted that we considered $N_p$ $\ge$ 10 for all four cut-off criteria. The second and third column specify the constraint on correlation coefficient ($r$) and $\sigma_{slope}$ respectively, corresponding to each cut-off criteria. The fourth column mentions three specific regions of the LMC: the complete coverage (for OGLE III data), bar region and outer region (as defined in Section 3.2). The number of subregions that satisfy the cut-off for each of these three specific regions, are mentioned in the fifth column. The mean metallicity and standard deviation for these three specific LMC regions, are mentioned in the last (sixth) column.}
\end{minipage}
}
\end{table*}


\section{Analysis of MCPS data}

The MCPS survey covers a comparatively larger area than the OGLE III survey in the optical bands. Although the resolution of MCPS survey is comparatively lower than the OGLE III survey, the MCPS provides more coverage on the northern and southern regions of the central LMC unlike OGLE III.  As mentioned previously, we have binned the MCPS data into 1512 regions of RA and DEC, each of dimension (10.53 x 15.00) sq. arcmin.

\subsection{Estimation of RGB slope}

\begin{figure*}
\centering
\begin{minipage}[b]{0.45\linewidth}
\includegraphics[height=3.0in,width=3.0in]{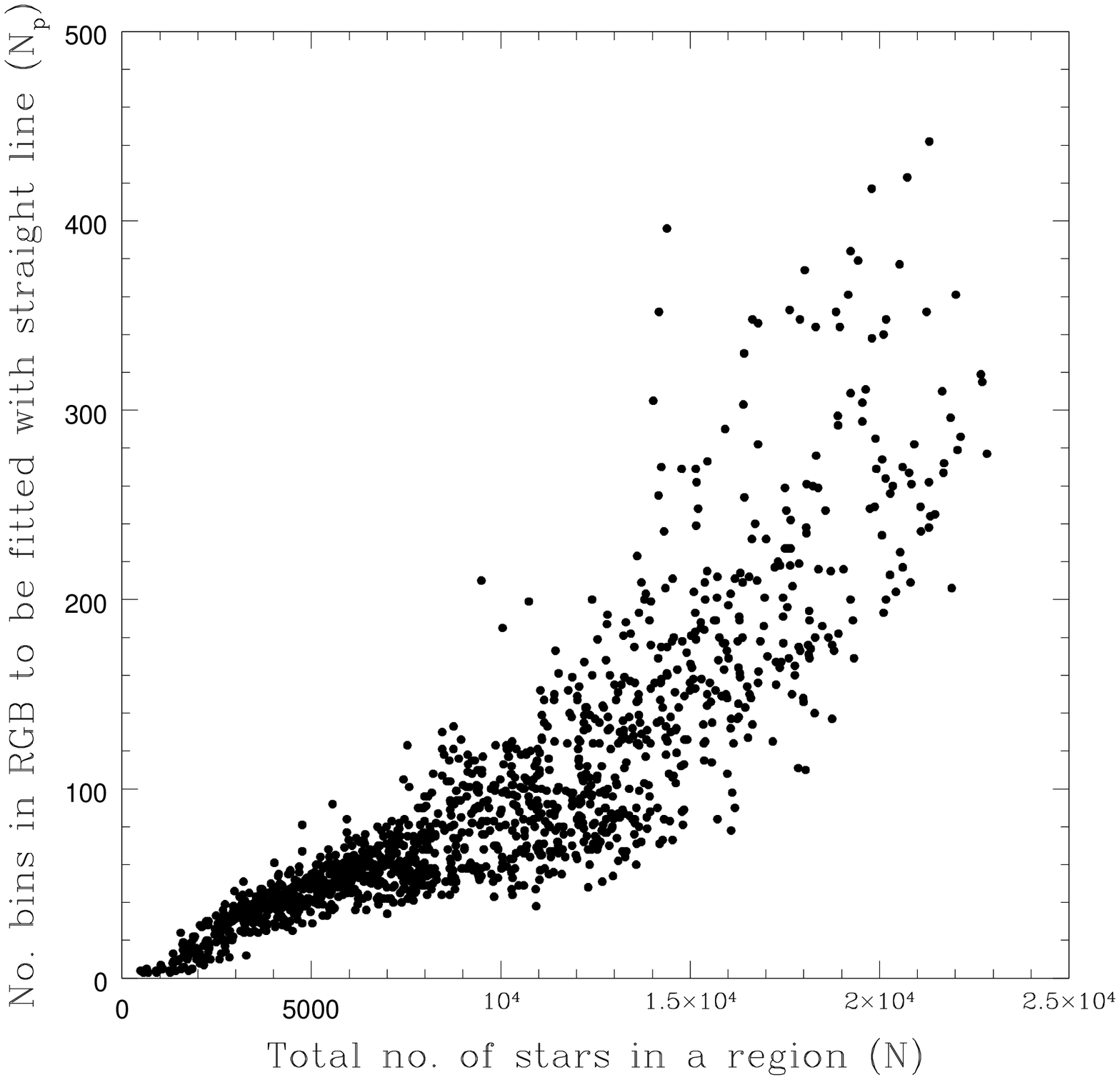}
\caption{\small Plot of number of bins in RGB to be fitted with straight line ($N_p$) versus the total number of stars ($N$) for MCPS subregions, after initial area binning.
\label{fig:mcps_np_vs_isum_undiv}}  
\end{minipage}
\quad
\begin{minipage}[b]{0.45\linewidth}
\includegraphics[height=3.0in,width=3.0in]{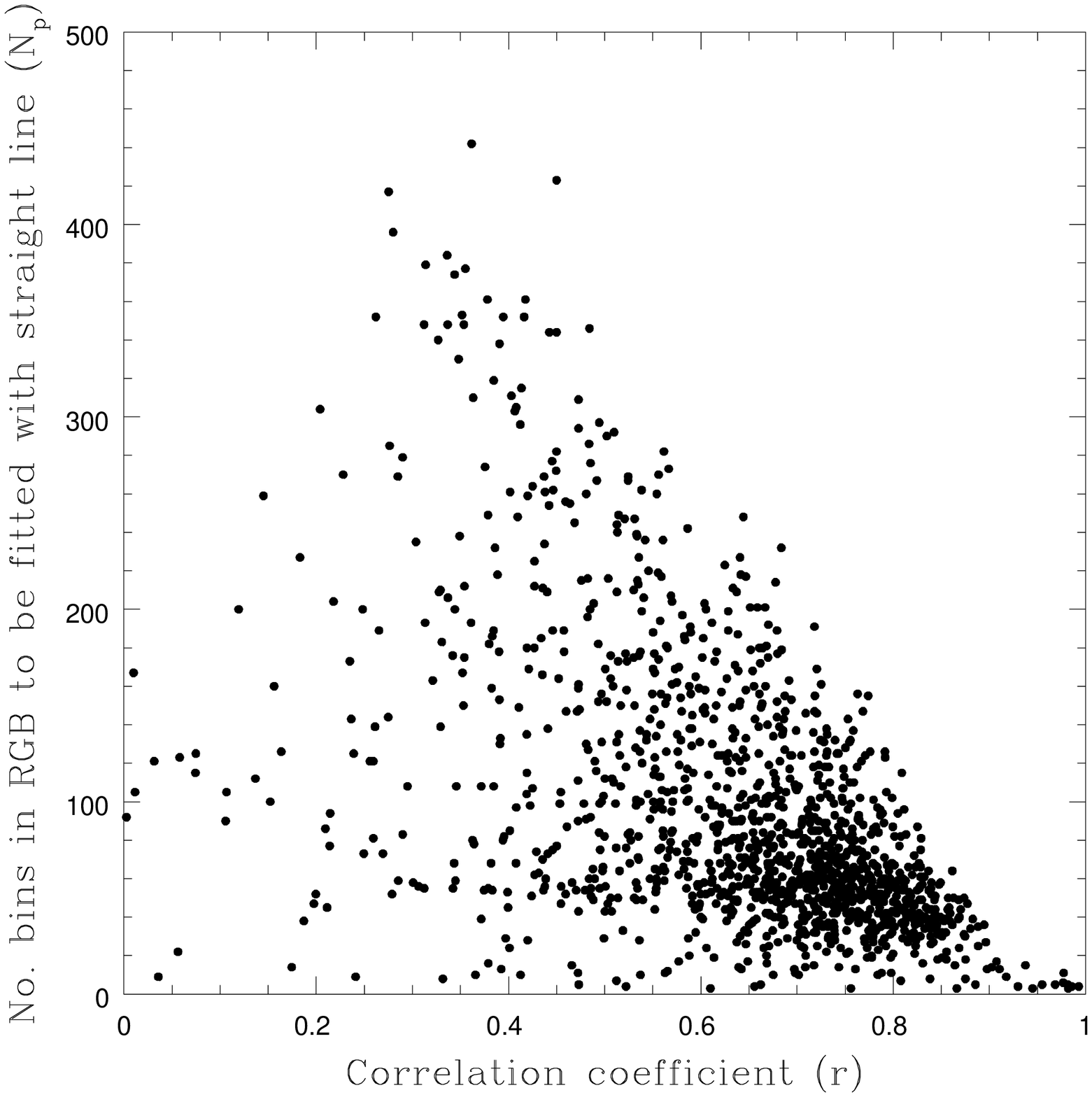}
\caption{\small Plot of number of bins on RGB to be fitted with straight line ($N_p$) versus correlation coefficient ($r$) for MCPS subregions, after initial area binning.
\label{fig:mcps_np_vs_r_undiv}}  
\end{minipage}
\end{figure*}

For the MCPS data, we used similar steps for estimation of slope of the RGB (Section 3), starting from isolating the RGB, locating the densest point of the RC and estimating the density distribution of stars on the RGB. We have excluded regions that have $N_p \sim 0$, thus making estimations for 1355 regions (out of 1512). We present the plot for $N_p$ versus N in Figure \ref{fig:mcps_np_vs_isum_undiv} for these regions of the MCPS data, which is similar to Figure \ref{fig:ogle3_np_vs_isum_undiv} for OGLE III data. The plot suggests that with increase in N, $N_p$ also increases. The plot for $N_p$ versus $r$ is shown in Figure \ref{fig:mcps_np_vs_r_undiv} resembles Figure \ref{fig:ogle3_np_vs_r_undiv} created for OGLE III, and suggests that as $N_p$ increases, $r$ decreases. Similar to  OGLE III, we performed 7 types of area binning of the MCPS regions in order to achieve similar value of $N_p$ and high $r$ ($>$ 0.50). The division criteria adopted based on the number of stars in a region is presented in Table \ref{table:tab4}. Excluding subregions with $N_p \sim 0$, the number of subregions analysed is 4742 (out of 4750). The area of largest bin being (10.53 $\times$ 15.00) sq. arcmin, whereas the area of smallest bin is (3.51 $\times$ 5.00) sq. arcmin. The slopes for each of these subregions are then re-estimated.

\begin{figure*}
\centering
\begin{minipage}[b]{0.45\linewidth}
\includegraphics[height=3.0in,width=3.0in]{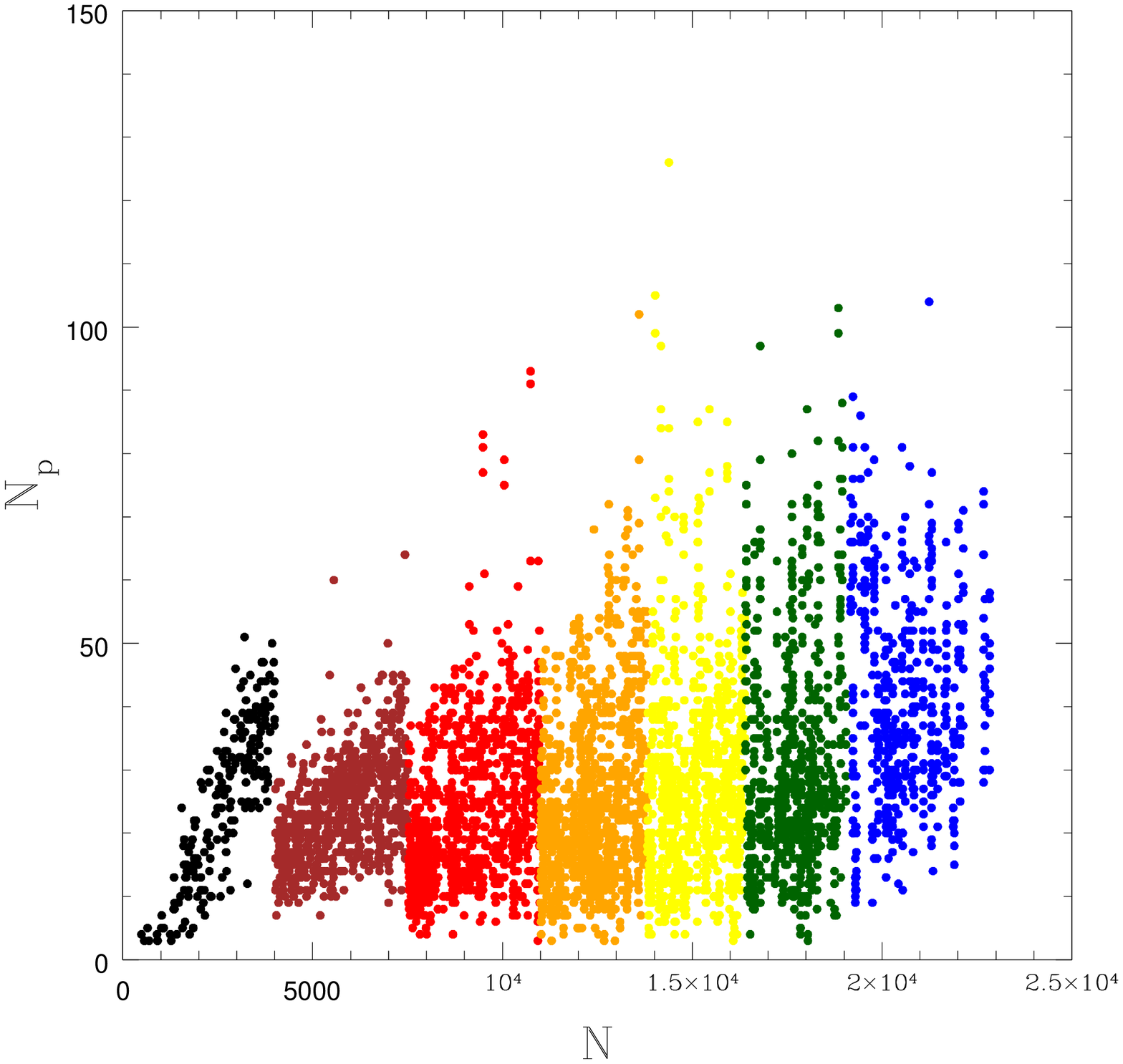}
\caption{\small Plot of $N_p$ versus $N$ for MCPS subregions, after finer area binning. The seven different colours correspond to the seven different binning criteria, as mentioned in the eighth column of Table \ref{table:tab4}. 
\label{fig:mcps_np_vs_isum_div}}  
\end{minipage}
\quad
\begin{minipage}[b]{0.45\linewidth}
\includegraphics[height=3.0in,width=3.0in]{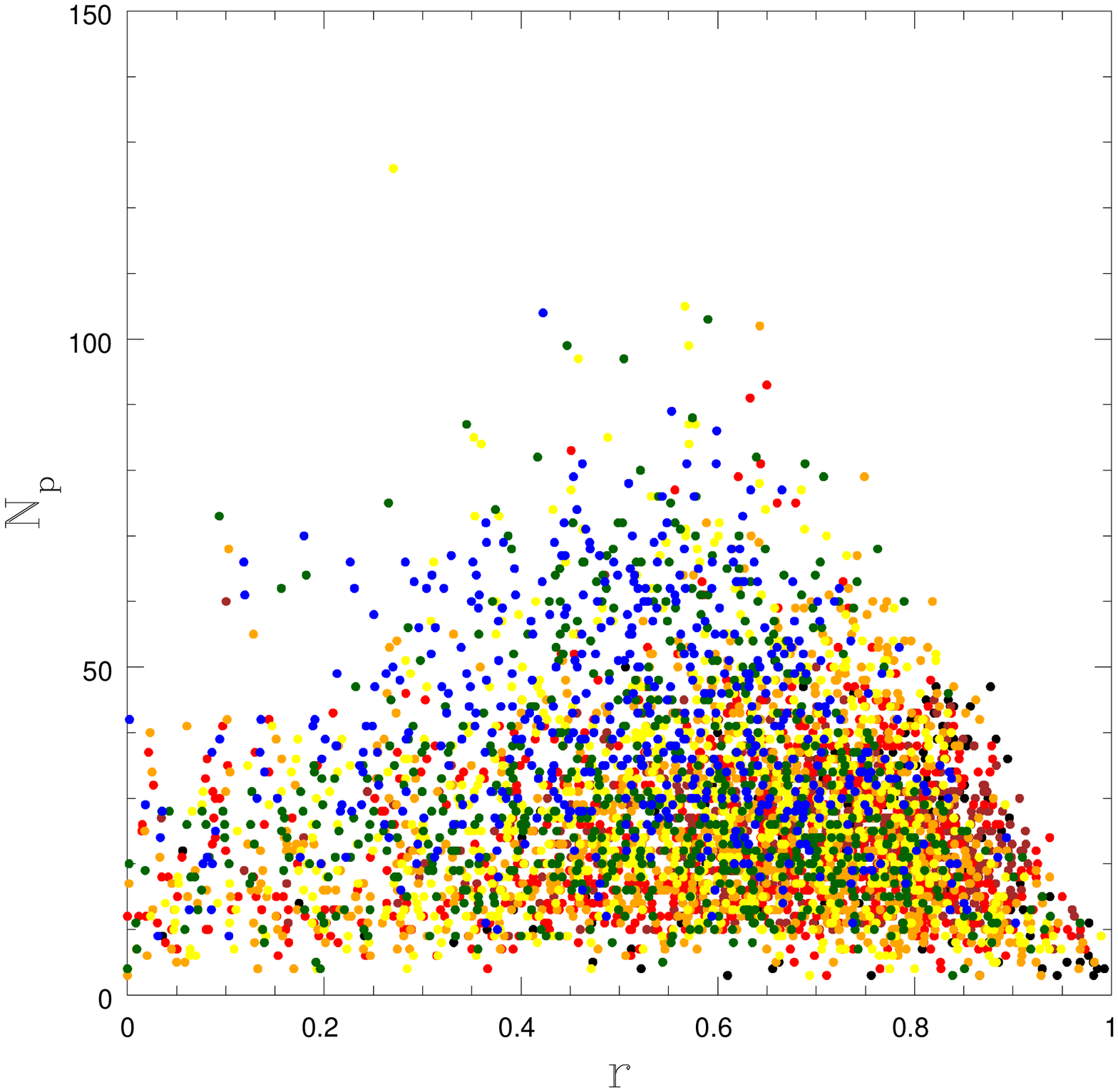}
\caption{\small Plot of $N_p$ versus $r$ for MCPS subregions, after finer area binning. The seven different colours correspond to the seven different binning criteria, as mentioned in the eighth column of Table \ref{table:tab4}. 
\label{fig:mcps_np_vs_r_div}}  
\end{minipage}
\end{figure*}

Figures \ref{fig:mcps_np_vs_isum_div} and \ref{fig:mcps_np_vs_r_div} show the plots of $N_p$ versus N and $N_p$ versus $r$ respectively, after the revised area binning is performed. It is seen that the upper limit of $N_p$ is now confined to a lower value, and remains almost same for all the 7 cases of division. Also, there are more regions with relatively high $r$ value. The regions with lower values of $r$ ($<$ 0.5) are likely to be those that suffer either from issues of multiple dominant population and/or small scale variation in reddening. 

\begin{figure*} 
\begin{center} 
\includegraphics[height=4in,width=4in]{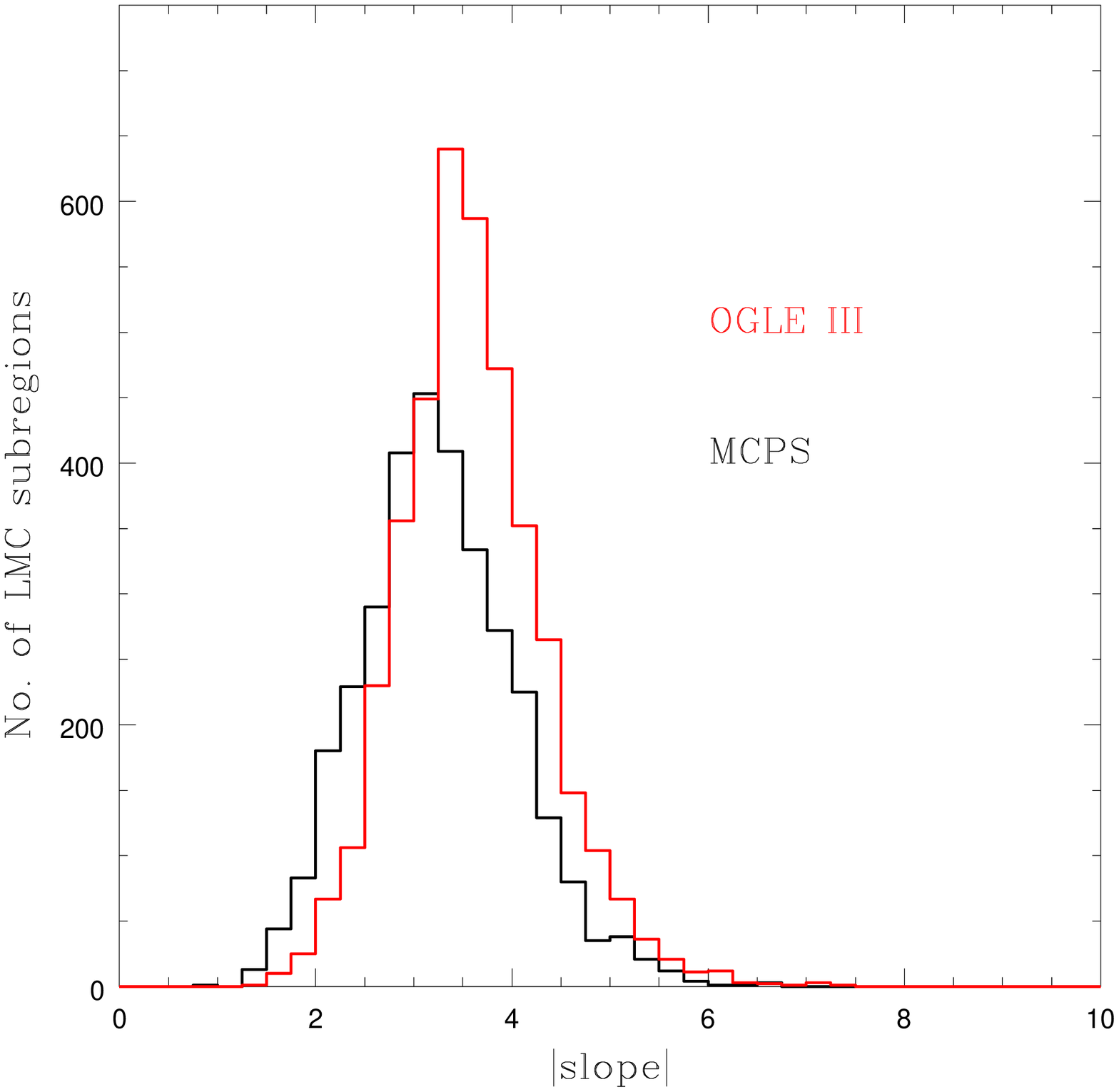}
\caption{\small Histogram for MCPS and OGLE III $|$slope$|$ values with cut-off criteria $(IV)$: $N_p$ $\ge$ 10, $r$ $\ge$ 0.5 and $\sigma_{slope}$ $\le$ 1.5.
\label {fig:histslope_mcps_ogle3}}
\end{center} 
\end{figure*}

To carry forward our analysis, we need to remove regions that have a poor estimation of slope. We consider the most stringent criteria, similar to that adopted for the OGLE III slope values, ($N_p$ $\ge$ 10, $r$ $\ge$ 0.5, and $\sigma_{slope}$ $\le$ 1.5) that filters out the best estimated slope. The slope distribution is compared with that of OGLE III, as shown in Figure \ref{fig:histslope_mcps_ogle3}. It is seen that that peak of the MCPS as well as the overall slope distribution for MCPS is shifted towards a lower value with respect to the OGLE III distribution. This shift could be a genuine shift or due to some other effect. One possibility is that the slope would depend on the V magnitude and (V$-$I) colour calibration. If there is any difference between the filter systems of OGLE III and MCPS, it might appear as a shift in slope. Thus, it is important to compare the two data base to detect any systematic difference. In the following section we try to address this issue.

\begin{table*}
{\small
\caption{Sub-division of MCPS regions:}
\label{table:tab4}
\begin{tabular}{|c|c|c|c|c|c|c|c|}
\hline \hline
Sl. no. & No. of  & No. of  & No. of     & No. of     & No. of             & Area of        &  Number of       \\
        & stars   & regions & division   & division   & sub-divisions      & a sub-division &  subregions      \\
        &         & (a)     & along RA   & along Dec  & (d=b$\times$c)     & (arcmin sq.)   &  (a$\times$d)    \\
        &         &         & (b)        & (c)        &                    &                &                  \\
\hline\hline
1  & 0 $<$ N $\le$ 4000      & 192 & 1 & 1 & 1  & (10.53$\times$15.00) & 192 (black)     \\
2  & 4000  $<$ N $\le$ 7500  & 353 & 2 & 1 & 2  & (5.26$\times$15.00)  & 706 (brown)     \\
3  & 7500  $<$ N $\le$ 11000 & 293 & 3 & 1 & 3  & (3.51$\times$15.00)  & 879 (red)       \\
4  & 11000 $<$ N $\le$ 13800 & 233 & 2 & 2 & 4  & (5.26$\times$7.50)   & 932 (orange)    \\
5  & 13800 $<$ N $\le$ 16400 & 143 & 3 & 2 & 6  & (3.51$\times$7.50)   & 858 (yellow)    \\
6  & 16400 $<$ N $\le$ 19100 & 86  & 4 & 2 & 8  & (2.63$\times$7.50)   & 688 (dark green) \\
7  & N $>$ 19100           & 55  & 3 & 3 & 9  & (3.51$\times$5.00)   & 495 (blue)      \\
\hline
\end{tabular}
\begin{minipage} {180mm}
\vskip 1.0ex
{Note: The table describes the 7 binning criteria used to sub-divide MCPS regions. For each criteria, the second column denotes the limit on total number of stars (N) within a region. The third column gives the number of regions, having N within that specified limit. Column four and five specify the number by which a region is binned along RA and Dec respectively. Column six thus gives the total number of subregions, a single region is binned into. Whereas, the seventh column gives the area of each such subregion. The last (eighth) column denotes the total number of subregions corresponding to each of the 7 sub-division criteria. The colours adjacent to the numbers are used to denote Figure \ref{fig:mcps_np_vs_isum_div} and \ref{fig:mcps_np_vs_r_div}.}
\end{minipage}
}
\end{table*} 

\subsection{Comparison of OGLE III and MCPS data}

In order to find systematics, if any, between the OGLE III and MCPS data, we compared the V magnitude and (V$-$I) colour derived by both the catalogues and checked for systematic variation. We selected a few MCPS subregions for this purpose. The regions are selected based on the criteria that they have sufficient number of RGB stars ($N_p$ $\ge$ 10), located away from the central region to have relatively less crowding effect (radial distance $\ge$ 2.5 degree), and have well estimated slope suggesting relatively less differential reddening ($r$ $\ge$ 0.85 and $\sigma_{slope}$ $\le$ 0.50). These criteria ensure that the regions do not lie in the outskirts of the LMC where the stellar density is sparse or close to the central regions where there are problems due to crowding and differential reddening. Also, if there exists a difference in the filter system, it is unlikely to be dependent on the metallicity or RGB slope of a region. However, to remain unbiased, we selected four regions with different values of slope. Then for each of these MCPS subregions, a region of similar dimension is extracted from the OGLE III data.

\begin{figure*}
\centering
\begin{minipage}[b]{0.45\linewidth}
\includegraphics[height=3.2in,width=3.2in]{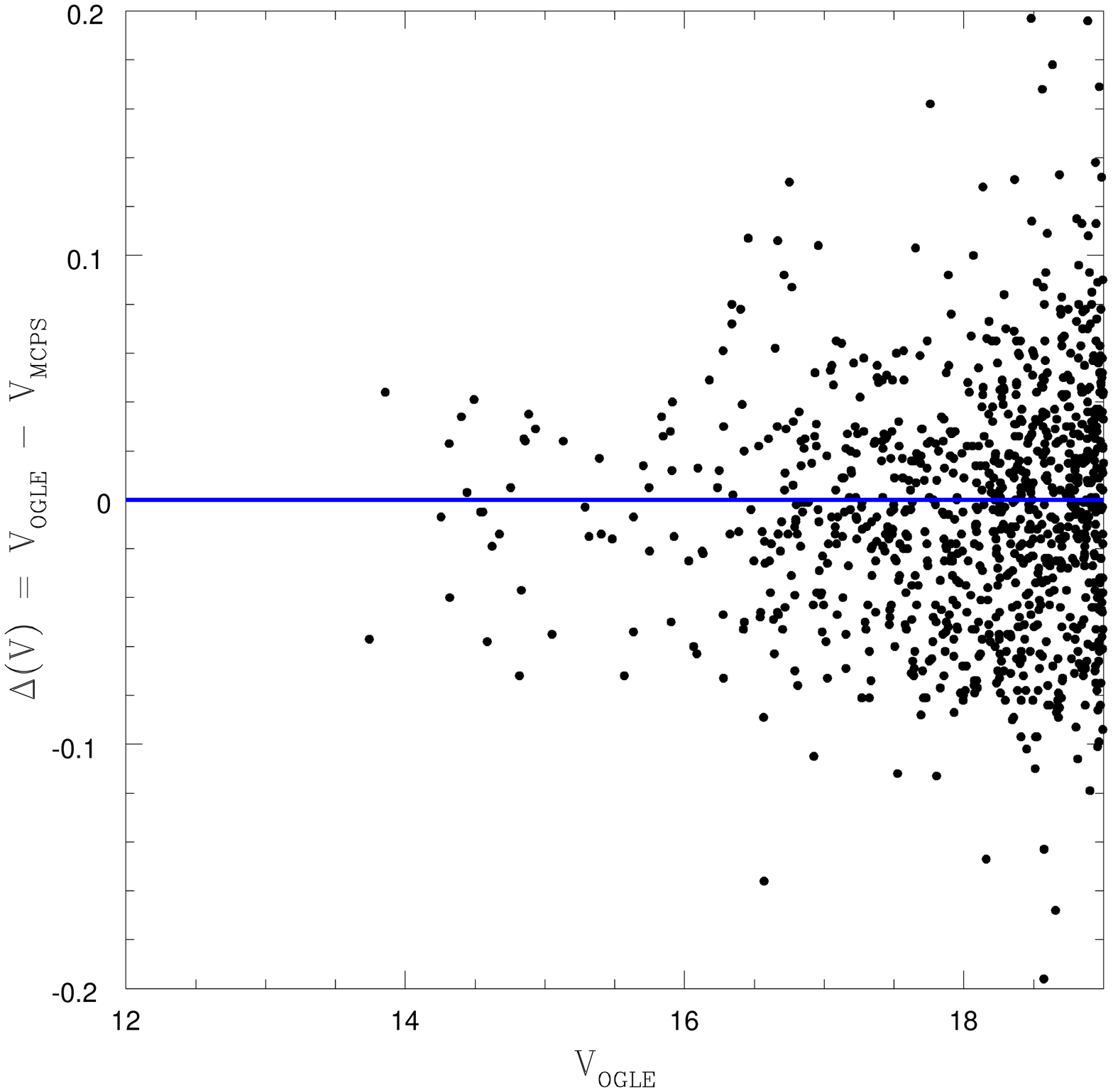}
\caption {Plot of $\Delta$(V) versus V$_{OGLE}$ for stars belonging to the four cross-correlated LMC subregions mentioned in Table \ref{table:tab5}.
\label{fig:cross1}}  
\end{minipage}
\quad
\begin{minipage}[b]{0.45\linewidth}
\includegraphics[height=3.2in,width=3.2in]{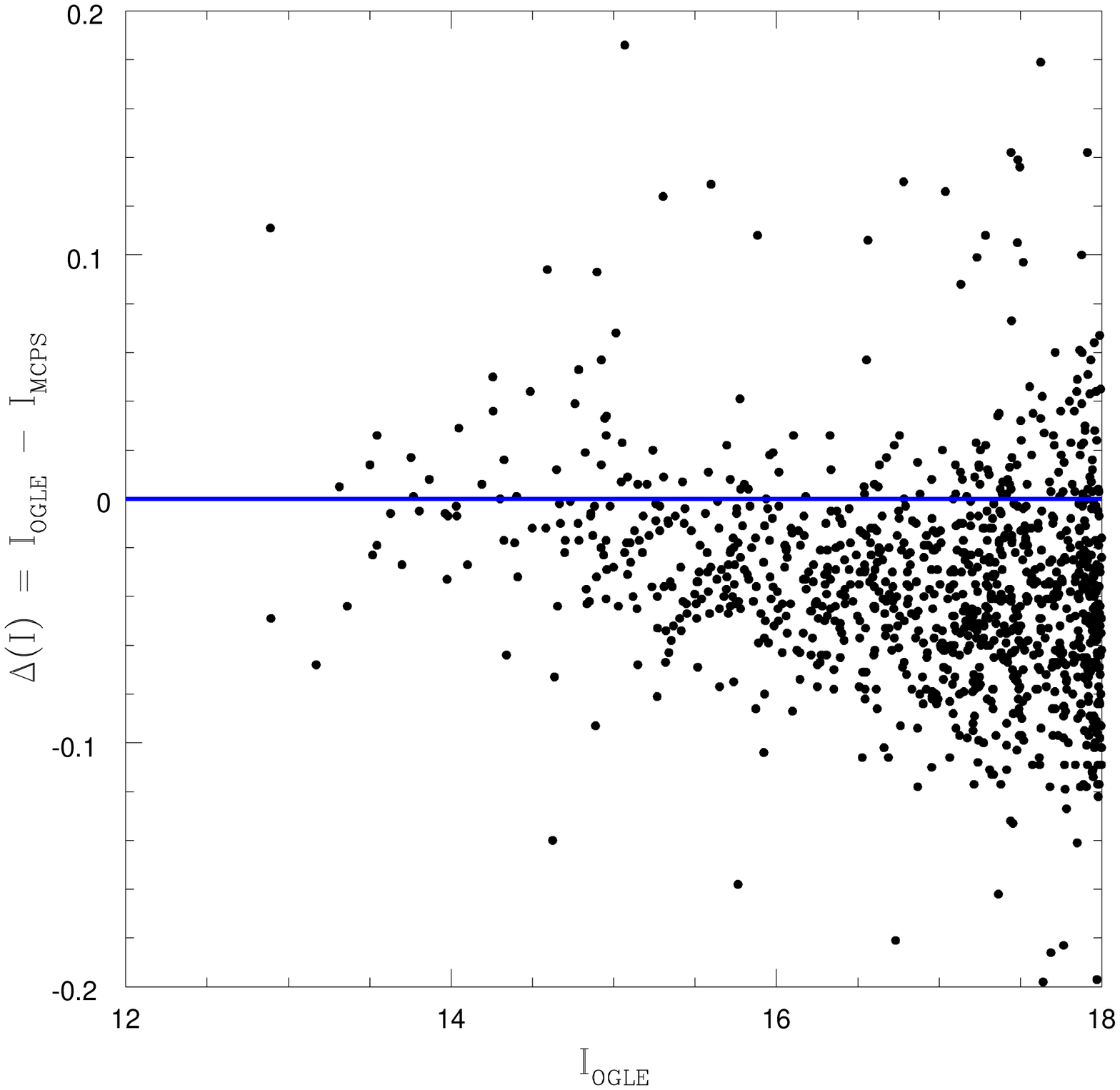}
\caption {Plot of $\Delta$(I) versus I$_{OGLE}$ for stars belonging to the four cross-correlated LMC subregions mentioned in Table \ref{table:tab5}.
\label{fig:cross2}}   
\end{minipage}
\end{figure*}

Table \ref{table:tab5} below lists the (RA, DEC), number of stars in the evolved part of the CMD, value of slopes, $r$, and $N_p$ corresponding to MCPS and OGLE III data for each of these subregions. To derive a cross correlation between the filters in the two different surveys, we  began by taking a star from MCPS data and identified its closest counterpart in the OGLE III data. For each star we then calculated the difference in V mag between OGLE III and MCPS ($\Delta$(V)= V$_{OGLE}$-V$_{MCPS}$) and plot it with respect to the V mag of OGLE III, in Figure \ref{fig:cross1}. The plot looks symmetric with respect to the zero value of $\Delta$(V). We can conclude that there is no significant difference in V magnitude between the two filter systems. Figure \ref{fig:cross2} shows a similar plot for $\Delta$(I) (i.e. I$_{OGLE}$-I$_{MCPS}$) versus I mag of OGLE III. The figure clearly shows a mild negative slope. It seems as one goes fainter (higher magnitude) in I band, $\Delta$(I) becomes more and more negative. This suggests that there exists a systematic trend in $\Delta$I as a function of I magnitude, which can cause a mild change in the slope of the RGB. Any such shift is not found in the case of V magnitude. Both $\Delta$(V) and $\Delta$(I) are invariant with respect to colour: any systematic variation is seen to be below 0.03 mag for $\Delta$(V) and 0.01 mag for $\Delta$(I) over (V$-$I)$_{OGLE}$ ranging from 0.5 to 2.5 mag.  

\begin{figure*} 
\begin{center} 
\includegraphics[height=4in,width=4in]{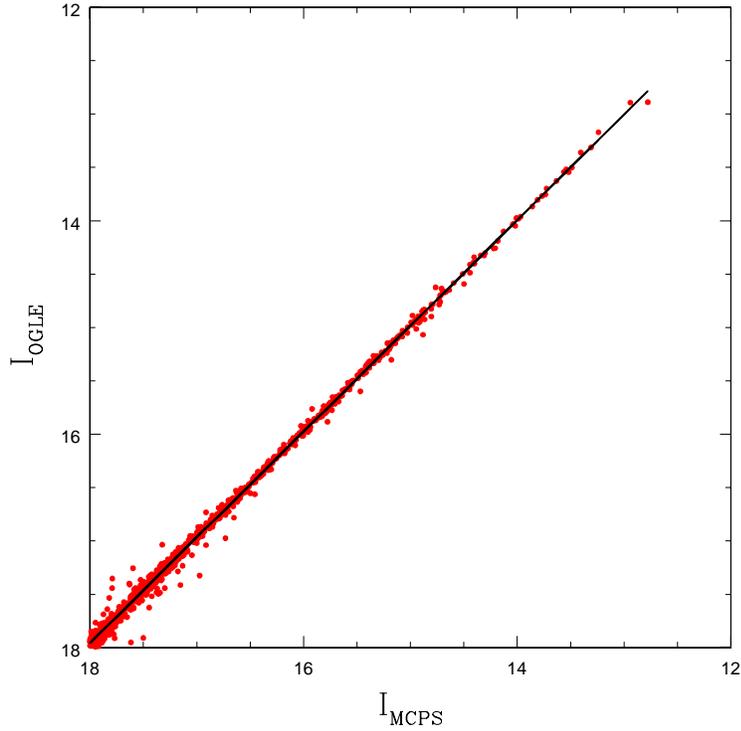}
\caption{\small Plot of I$_{OGLE}$ versus I$_{MCPS}$ for stars belonging to the four cross-correlated LMC subregions (red filled circles) mentioned in Table \ref{table:tab5}. The estimated slope is shown as a black solid line.
\label{fig:cross3}}
\end{center} 
\end{figure*}

Having found a systematic shift in I magnitude, we make a plot of I$_{OGLE}$ versus I$_{MCPS}$ and fit a relation, which is found to be linear. We have fitted a straight line (Figure \ref{fig:cross3} ) with a 3-sigma cut. We consider only those stars that are brighter than 18.0 mag as the RGB for both OGLE III and MCPS data, and exclude fainter stars. The relation followed by the two photometric systems can be then described as a function:
\begin{equation} \label{eq:2}
I_{OGLE}= (0.990 \pm 0.001) \times I_{MCPS}+(0.124 \pm 0.028); 
\end{equation}
with $r$=0.99. We can now transform the I magnitude of stars in MCPS system to their corresponding I magnitude in OGLE III system using Equation \ref{eq:2}, whereas the V magnitude in both the data sets remain the same. This transformation is carried out for all the stars in the MCPS data and the slopes are re-estimated for all the MCPS subregions. Thus, we converted the MCPS data to the OGLE III system and estimated the slope of the RGB. Now these slope values can be directly compared with the slopes estimated from the OGLE III data.

\begin{figure*}
\centering
\begin{minipage}[b]{0.45\linewidth}
\includegraphics[height=3.0in,width=3.0in]{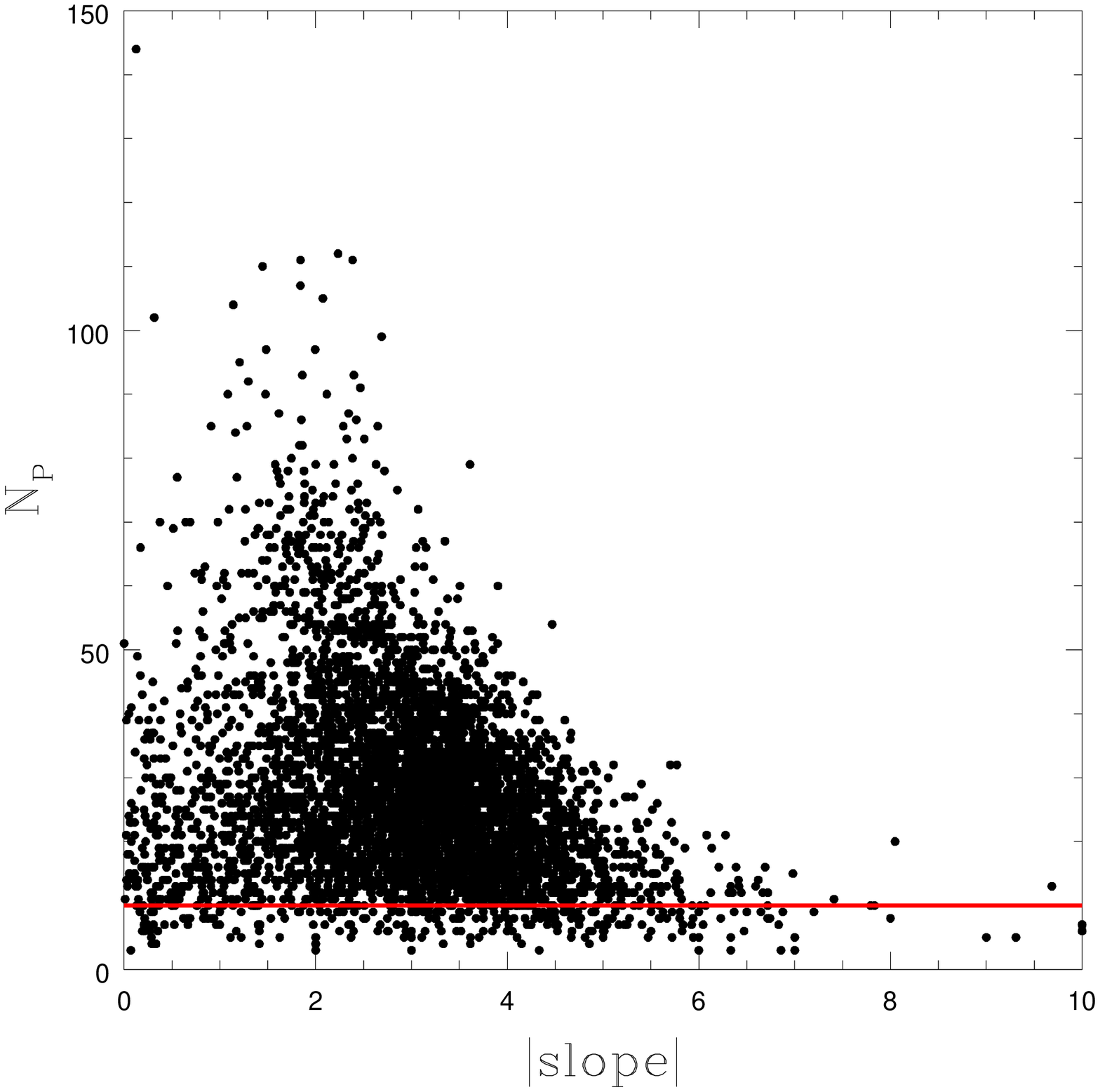}
\caption{\small Plot of $N_p$ versus $|$slope$|$ after I magnitude transformation for MCPS subregions. The red line at $N_p$ = 10 denotes the cut-off decided to exclude regions with poorly populated RGB.
\label{fig:mcpso3_np_vs_slope} } 
\end{minipage}
\quad
\begin{minipage}[b]{0.45\linewidth}
\includegraphics[height=3.0in,width=3.0in]{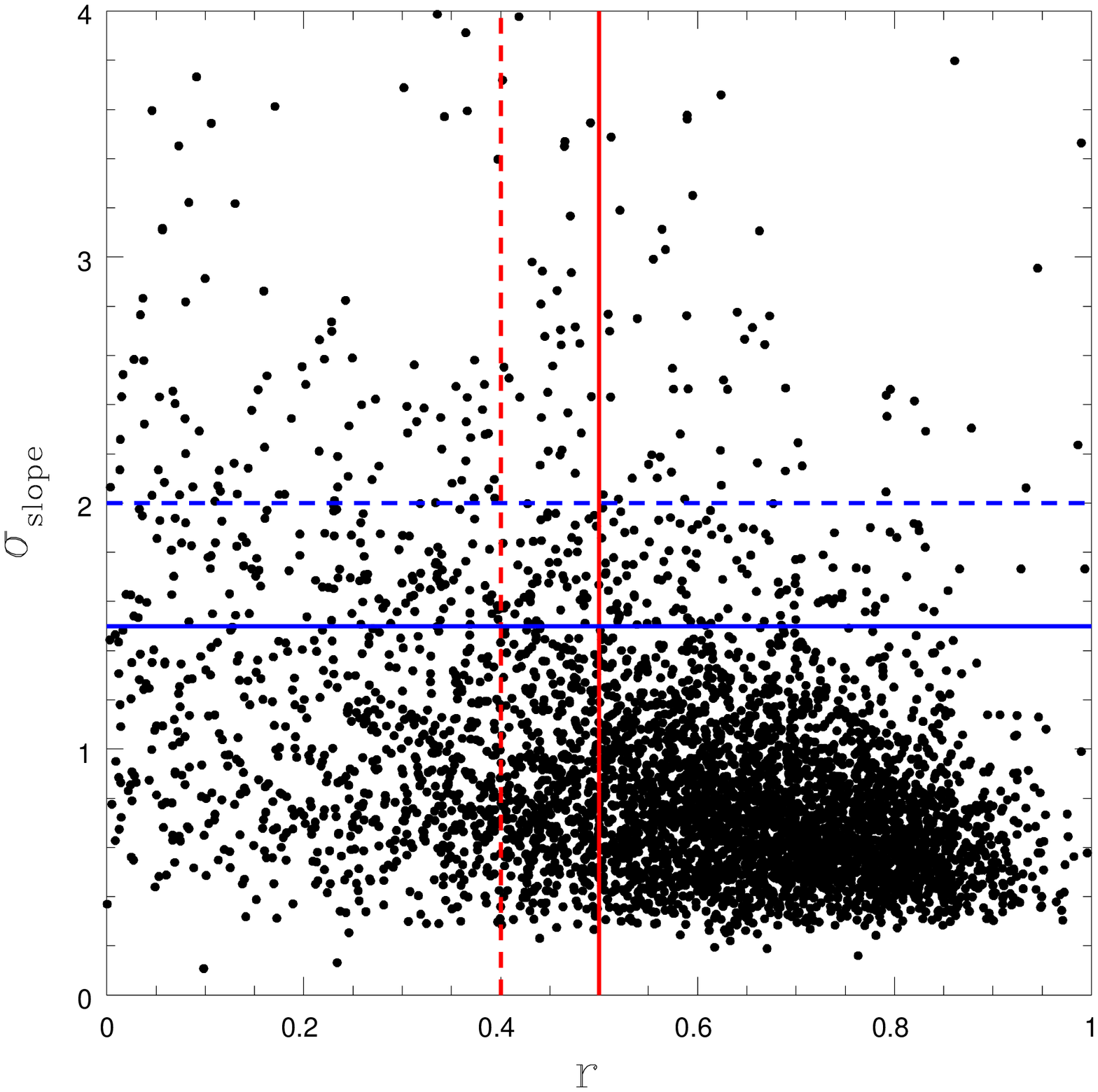}
\caption{\small Plot of $\sigma_{slope}$ versus $r$ after I magnitude transformation for MCPS subregions. The blue dashed and solid lines correspond to the cut-off criteria on $\sigma_{slope}$ at 2.0 and 1.5 respectively. The red dashed and solid lines denote the cut-off corresponding to $r$ at 0.4 and 0.5 respectively. 
\label{fig:mcpso3_errslope_vs_r} }  
\end{minipage}
\end{figure*}

\begin{figure*}
\centering
\begin{minipage}[b]{0.45\linewidth}
\includegraphics[height=3.0in,width=3.0in]{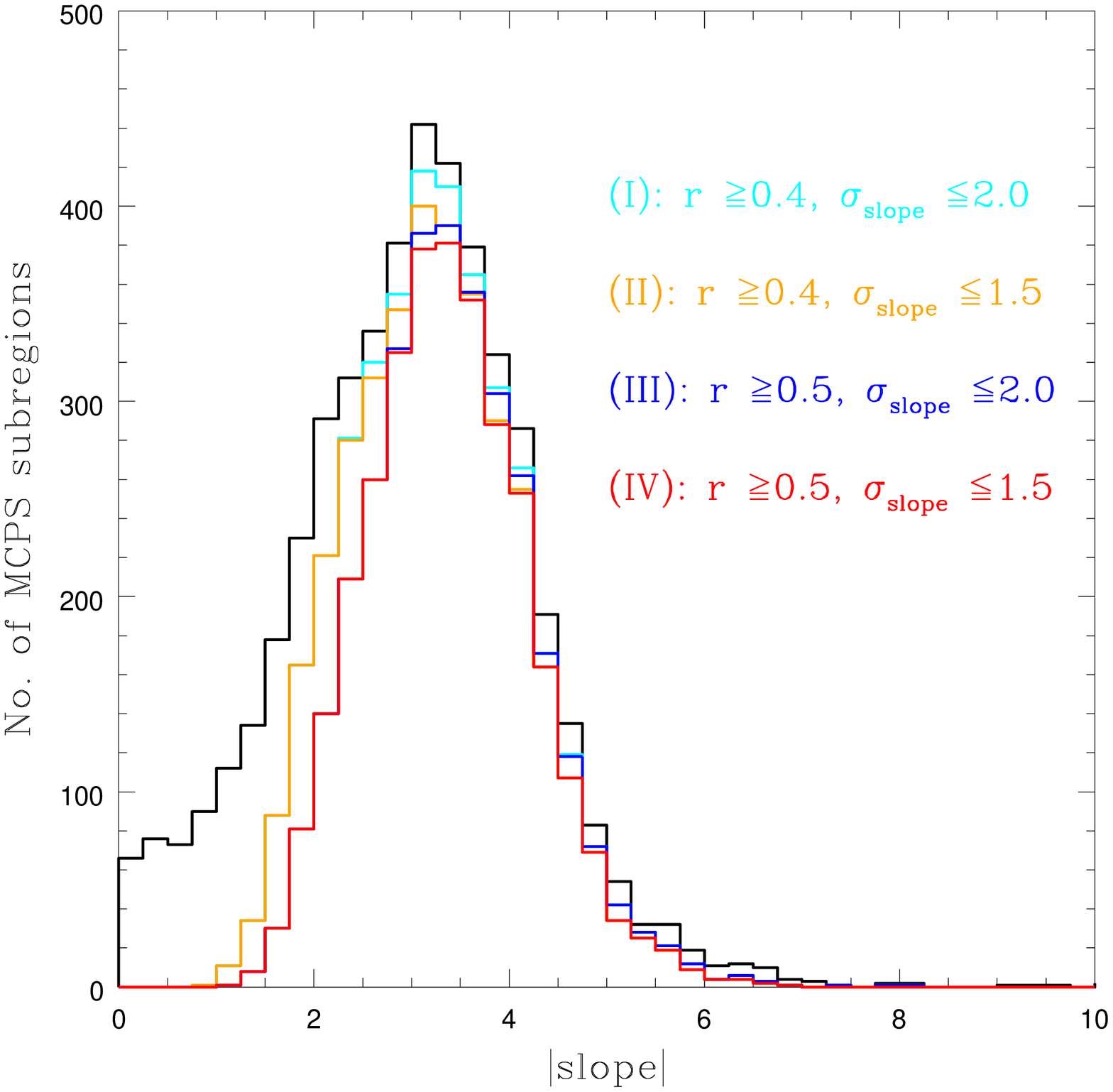}
\caption{\small Histogram of MCPS $|$slope$|$ values after I magnitude transformation, estimated for all the four cut-off citeria ($(I)$ in cyan, $(II)$ in orange, $(III)$ in blue, and $(IV)$ in red). $N_p$ $\ge$ 10 for all these four cases. The black solid line shows the distribution of $|$slope$|$ with no cut-offs, for all subregions. 
\label{fig:mcpso3_histslope}}  
\end{minipage}
\quad
\begin{minipage}[b]{0.45\linewidth}
\includegraphics[height=3.0in,width=3.0in]{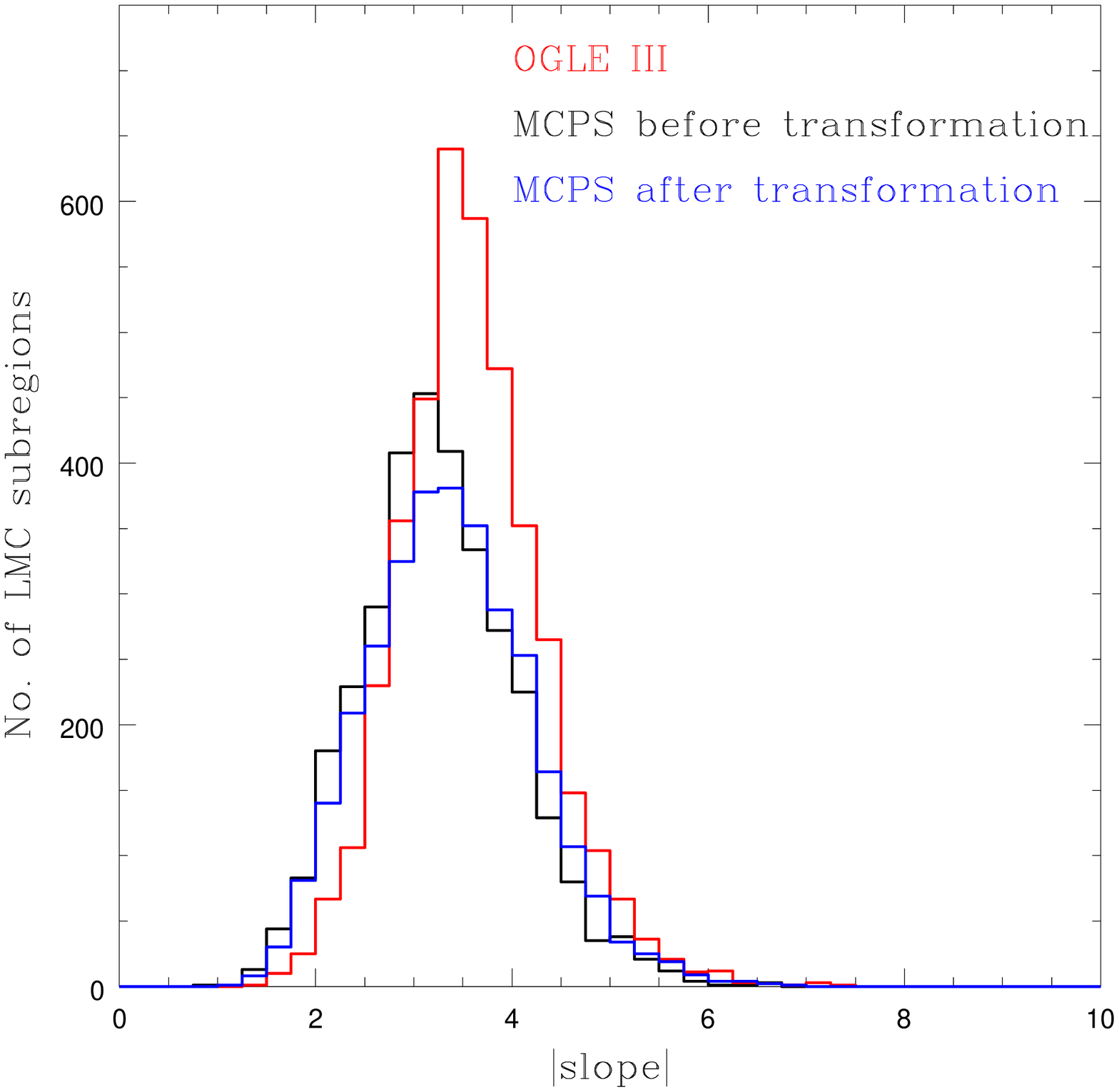}
\caption{\small Comparison of histogram for MCPS $|$slope$|$ values before (black) and after (blue) I magnitude transformation with respect to the OGLE III $|$slope$|$ values (red), where all estimations are with cut-off criteria $(IV)$: $N_p$ $\ge$ 10, $r$ $\ge$ 0.5 and $\sigma_{slope}$ $\le$ 1.5.
\label {fig:histslope_mcpso3_ogle}}
\end{minipage}
\end{figure*}

In Figure~\ref{fig:mcpso3_np_vs_slope} $N_p$ is plotted against the re-estimated slopes for the MCPS subregions. The plot of $\sigma_{slope}$ versus $r$ is shown in Figure \ref{fig:mcpso3_errslope_vs_r}. We considered 4 cut-off criteria, similar to OGLE III (Section 3) to select regions with reliable slope values, and show them in these figures. The slope distribution for the 4 cut-off criteria are shown in Figure \ref{fig:mcpso3_histslope}, along with the distribution with no cut-offs. The figure shows that, as the $r$ cut-off becomes more stringent, the width of the histogram reduces such that the lower slope values are removed. The effect of $\sigma_{slope}$ is insignificant on the distribution for same cut-off in $r$. The strictest cut-off, $N_p$ $\ge$ 10, $r$ $\ge$ 0.5, and $\sigma_{slope}$ $\le$ 1.5 shows that slope values ranges primarily from 2 to 6 for the regions studied here. 

We check whether the process of removing the systematics in the I magnitude has changed the slope distribution in the case of MCPS data, in Figure~\ref{fig:histslope_mcpso3_ogle}. It can be seen that the change of photometric system has not changed the slope distribution significantly, as there is hardly any shift in the peak value. Therefore, the observed shift between the OGLE III and MCPS slope distribution is not due to the photometric system. Thus, we consider the shift to be possibly a genuine one, likely to be arising due to difference in the area covered by these two data sets. 

\begin{table*}
{\small
\caption{Subregions to cross-correlate MPCS and OGLE III filter systems:}
\label{table:tab5}
\begin{tabular}{|c|c|c|c|c|c|c|}
\hline \hline
Serial no. & (RA$^{\circ}$, Dec$^{\circ}$) & Data & No. of stars           & $|$slope$|$ $\pm$ $\sigma_{slope}$ & $r$ & $N_p$ \\
           &                               &      & in evolved part of CMD &                                      &     &       \\
\hline\hline
$i$    & (90.12, $-$70.62) & MCPS & 714 & 3.14$\pm$0.38 & 0.89 & 19 \\      
       &                   & OGLE & 749 & 4.06$\pm$0.64 & 0.81 & 23 \\ 
\hline    
$ii$   & (71.59, $-$70.12) & MCPS & 759 & 3.53$\pm$0.33 & 0.92 & 22 \\      
       &                   & OGLE & 850 & 3.96$\pm$0.40 & 0.89 & 28 \\ 
\hline     
$iii$  & (70.12, $-$70.12) & MCPS & 823 & 4.01$\pm$0.41 & 0.91 & 22 \\     
       &                   & OGLE & 940 & 4.53$\pm$0.45 & 0.89 & 26 \\ 
\hline      
$iv$   & (70.12, $-$69.87) & MCPS & 781 & 4.68$\pm$0.48 & 0.91 & 20 \\      
       &                   & OGLE & 817 & 4.44$\pm$0.92 & 0.73 & 22 \\ 
\hline     
\end{tabular}
\begin{minipage} {180mm}
\vskip 1.0ex
{Note: The table lists out four subregions used to estimate cross-correlation between the MCPS and OGLE III filter systems. The central (RA, Dec) for each subregion is mentioned in column number two. The survey data used for each subregion is mentioned in the third column, and the corresponding number of stars in evolved part of the CMD (0.5 $<$ (V$-$I) $\le$ 2.5 mag and 12.0 $\le$ V $<$ 20.0 mag), is specified in the fourth column. In fifth column the estimated slope along with its associated error, for each data set is listed. The corresponding correlation coefficient ($r$) and number of density bins on RGB fitted with straight line ($N_p$), for each case, is mentioned in column number six and seven respectively.}
\end{minipage}
}
\end{table*} 



\begin{figure*} 
\begin{center} 
\includegraphics[height=4.5in,width=6in]{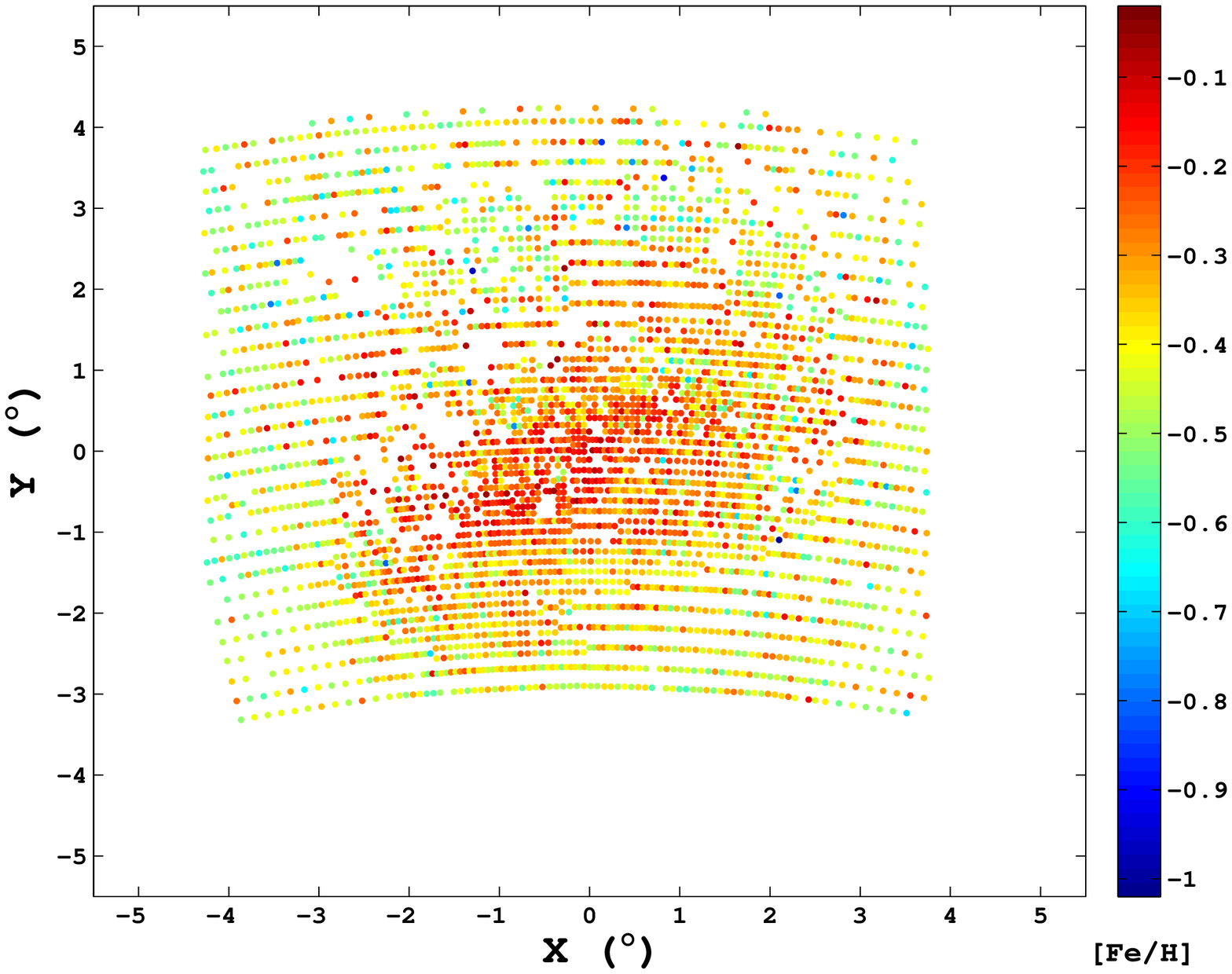}
\caption{\small MCPS metallicity map with cut-off criteria $(I)$: $N_p$ $\ge$ 10, $r$ $\ge$ 0.4 and $\sigma_{slope}$ $\le$ 2.0.
\label{fig:mcpsmap1}} 
\end{center} 
\end{figure*}

\begin{figure*} 
\begin{center} 
\includegraphics[height=4.5in,width=6in]{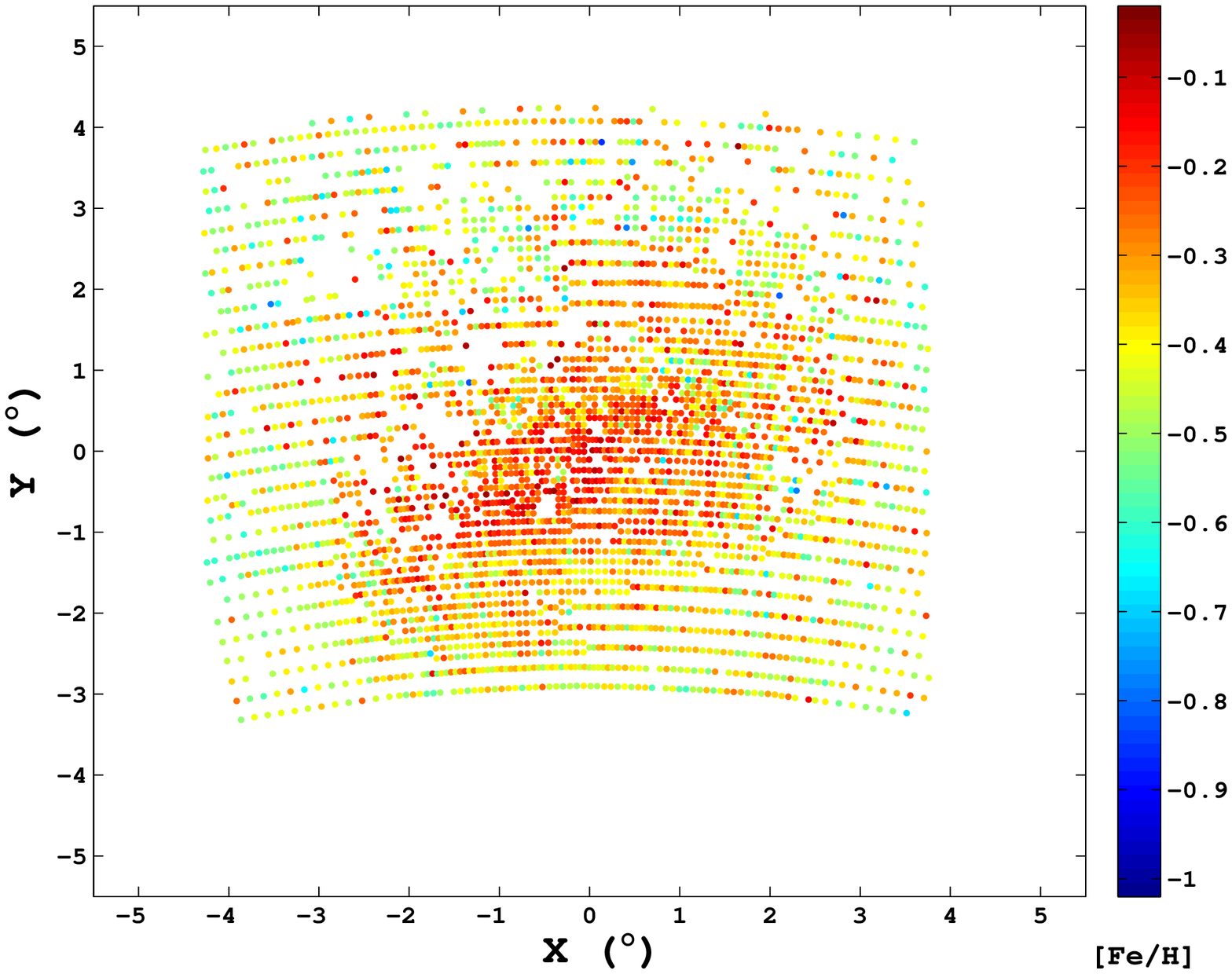}
\caption{\small MCPS metallicity map with cut-off criteria $(II)$: $N_p$ $\ge$ 10, $r$ $\ge$ 0.4 and $\sigma_{slope}$ $\le$ 1.5.
\label{fig:mcpsmap2}} 
\end{center} 
\end{figure*}

\begin{figure*} 
\begin{center} 
\includegraphics[height=4.5in,width=6in]{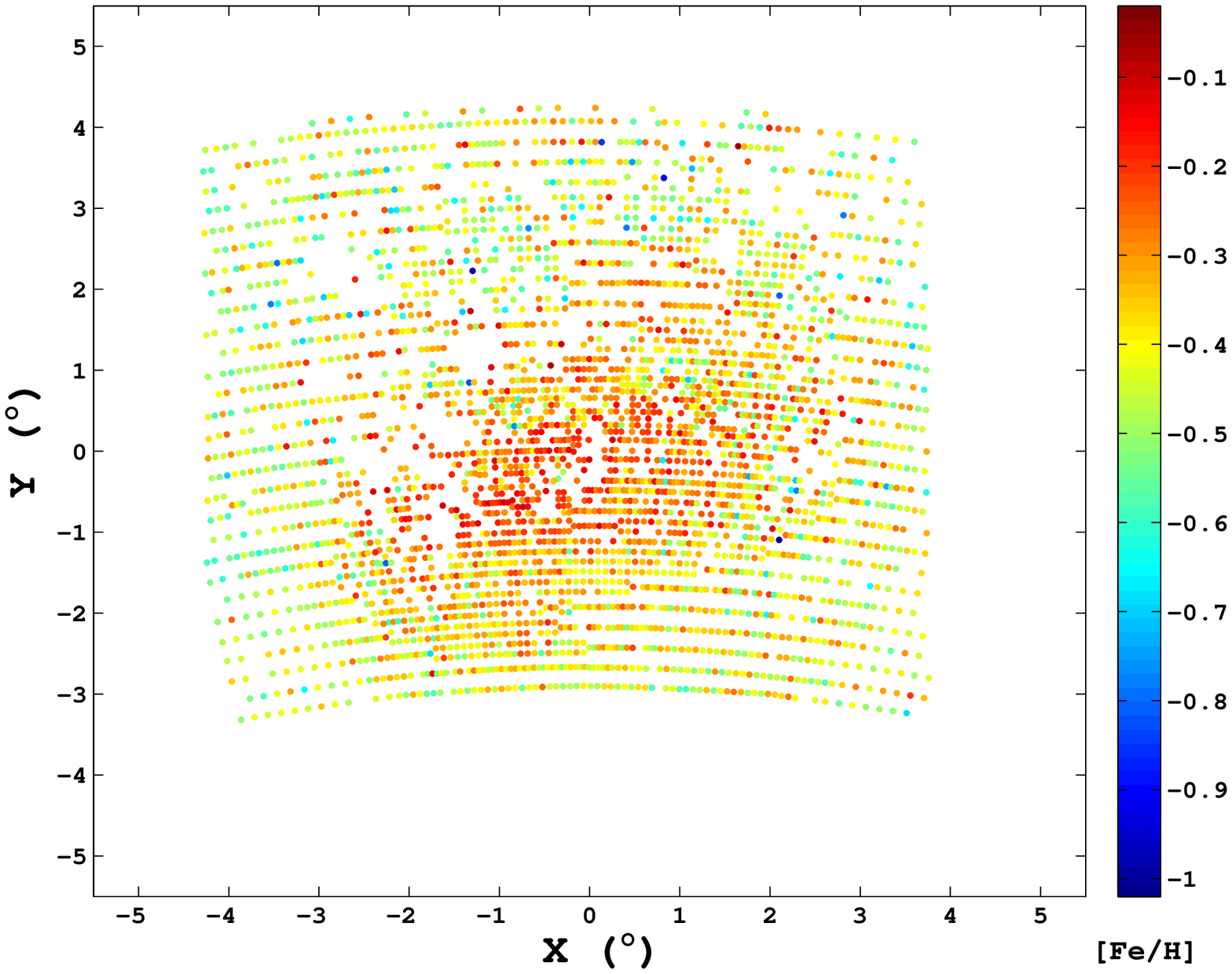}
\caption{\small MCPS metallicity map with cut-off criteria $(III)$: $N_p$ $\ge$ 10, $r$ $\ge$ 0.5 and $\sigma_{slope}$ $\le$ 2.0.
\label{fig:mcpsmap3}} 
\end{center} 
\end{figure*}

\begin{figure*} 
\begin{center} 
\includegraphics[height=4.5in,width=6in]{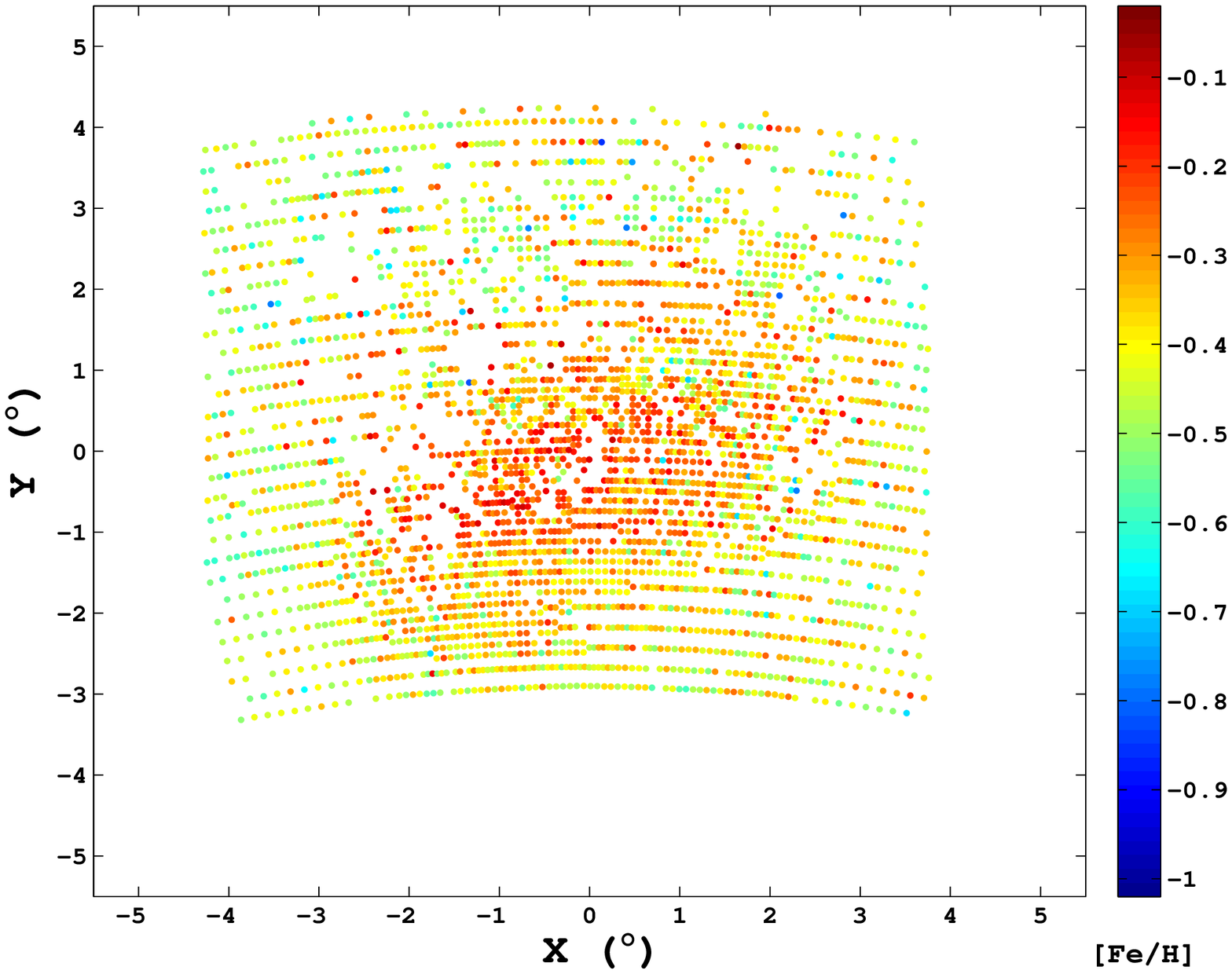}
\caption{\small MCPS metallicity map with cut-off criteria $(IV)$: $N_p$ $\ge$ 10, $r$ $\ge$ 0.5 and $\sigma_{slope}$ $\le$ 1.5.
\label{fig:mcpsmap4}} 
\end{center} 
\end{figure*}

\subsection{Metallicity map using MCPS data}

It is ideal that we formulate a slope-metallicity relation similar to the one derived for the OGLE III data. In the case of MCPS data, many regions in the central LMC were eliminated due to poor slope estimation, and the outer regions are found to have poorly populated RGBs. Thus, we could not estimate slopes of regions close to the spectroscopically studied regions. Due to above mentioned reasons, we were unable to derive an independent slope-metallicity relation for the MCPS data. An alternate way for calibration is to use the slope-metallicity relation derived for OGLE III, since both the data sets use V and I passband filters. Now, that the photometric system of MCPS is converted to the OGLE III system, the slopes estimated in the previous section are converted to metallicities using Equation \ref{eq:1}. Figures \ref{fig:mcpsmap1}, \ref{fig:mcpsmap2}, \ref{fig:mcpsmap3}, and \ref{fig:mcpsmap4} show the metallicity map derived using MCPS data, created using the 4 cut-off criteria. It can be seen that, similar to the OGLE III data, there is not much difference between the maps created using 4 different criteria. Similar to the OGLE III analysis, we shall use the most stringent criteria, with the minimum number of subregions for further analysis. 

The gaps in the map are due to regions with poorly estimated slopes. The regions which get missed out due to this effect are those located near the 30 Dor region (similar to OGLE III map), along with some regions to the north-east and north of the LMC centre. More gaps appear around these regions, as well as within the bar region, when the selection criteria become more stringent. As most of these regions are near regions of star formation, the poor estimate of slopes might be due to the presence of small scale variation in reddening. We notice that the central region has more or less uniform metallicity distribution, which is relatively high when compared to the outer regions. Even though there is a radial variation of metallicity, the maps also suggest that the variation is likely to be very small. We also notice that regions with metallicity less than $-$0.60 are located mostly in the outer regions, with a few in the metal rich central region. In the MCPS map, a large number of regions in the bar region get removed due to poor slope estimation; due to this, the appearance of the bar as a homogeneous metal rich region is not present in the map. Also, we observe the metallicity distribution of the northern and southern regions here. These regions are not as metal poor as the eastern and western disks, but rather moderately metal rich. That is, we do see a change in the metallicity as we we move from the bar to the north or south. Even though we can visually identify these variations, quantitatively these may not be very significant. 

\begin{figure*}
\centering
\begin{minipage}[b]{0.45\linewidth}
\includegraphics[height=3.0in,width=3.0in]{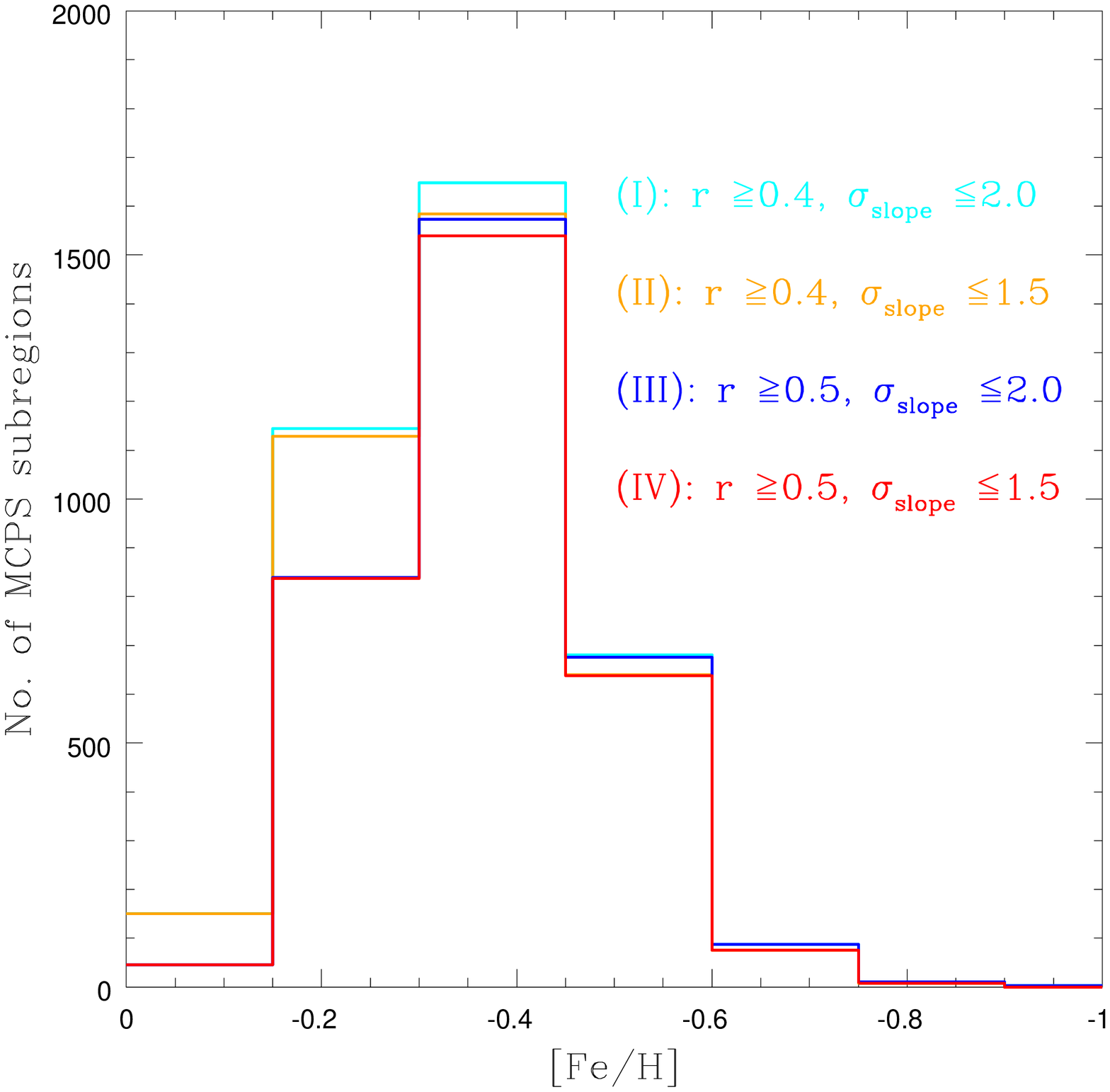}
\caption{\small Histogram of metallicity ($[Fe/H$]) for MCPS data, estimated for all the four cut-off criteria ($(I)$ in cyan, $(II)$ in orange, $(III)$ in blue, and $(IV)$ in red). $N_p$ $\ge$ 10 for all these four cases.
\label{fig:mcpso3_histabun}} 
\end{minipage}
\quad
\begin{minipage}[b]{0.45\linewidth}
\includegraphics[height=3.0in,width=3.0in]{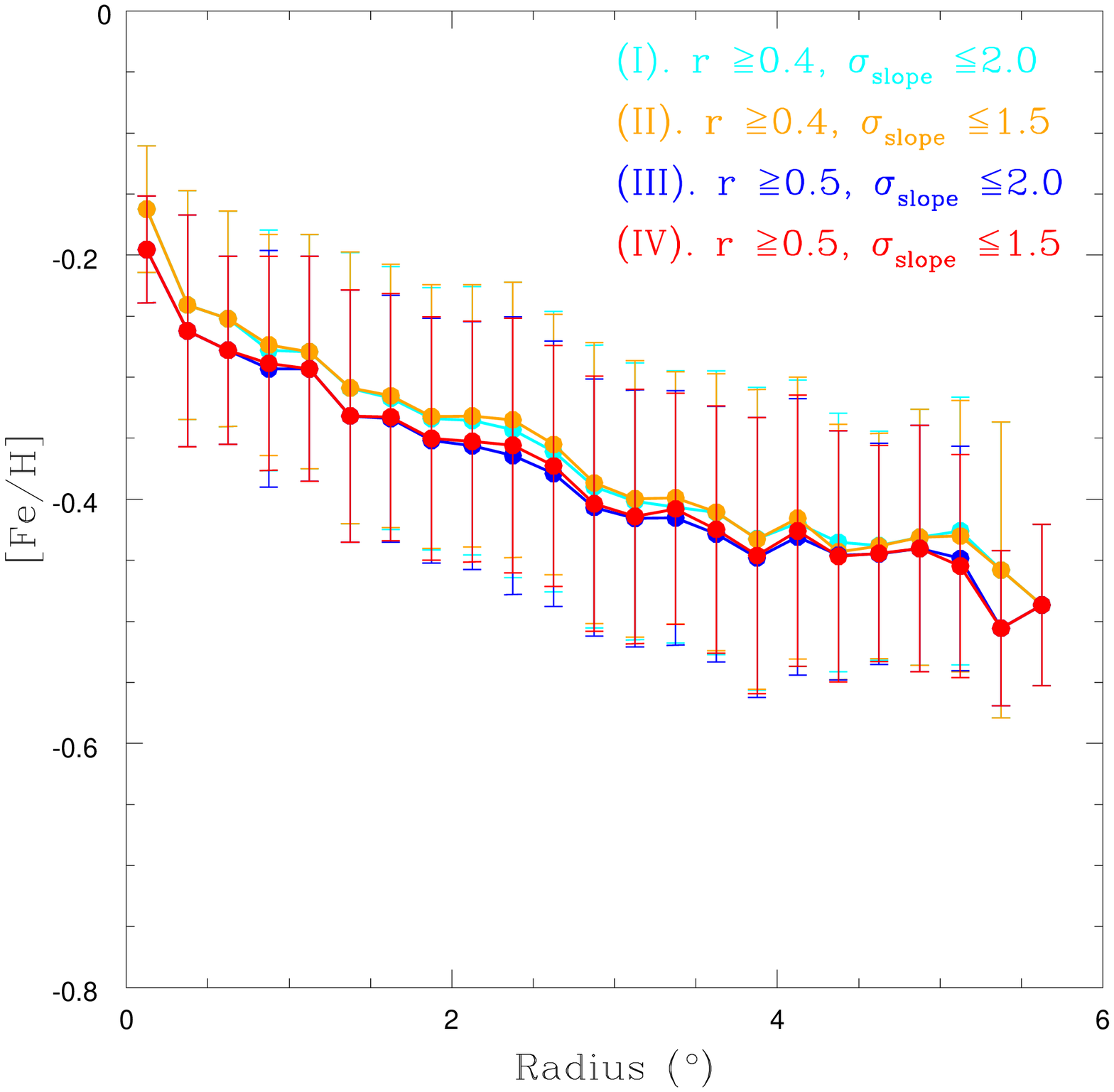}
\caption{\small Radial variation of metallicity ($[Fe/H$]) for MCPS data, estimated for all the four cut-off criteria ($(I)$ in cyan, $(II)$ in orange, $(III)$ in blue, and $(IV)$ in red). $N_p$ $\ge$ 10 for all these four cases.
\label{fig:mcpso3_rad}}
\end{minipage}
\end{figure*}

The metallicity distribution of the LMC covered by the MCPS data is shown in Figure~\ref{fig:mcpso3_histabun}, for all the 4 cut-off criteria. The data is binned with a width of 0.15 dex, which is similar to the error in the average metallicity. The histogram suggests that the peak of the distribution lies in the $-$0.30 to $-$0.45 dex bin. There is not much difference between the four criteria used, though the fourth criteria has the smallest number of regions and fewer regions with high metallicity values. Also, there are almost no regions with metallicity less than $-$0.75 dex. The range in metallicity is thus found to be from $-$0.15 to $-$0.75 dex. The histogram is very similar to the OGLE III distribution, the striking difference being, in this distribution, apart from having a reduced peak height, the distribution is skewed to positive values, which is opposite to the OGLE III distribution. This difference may be due to the inclusion of more northern and southern regions and fewer eastern and western regions, and the exclusion of the relatively metal-rich bar due to crowding.

As the MCPS data covers a larger area of the LMC, the MCPS metallicity distribution can be used to estimate the average of the metallicity in the bar region with respect to the outer regions of the LMC, as well the radial variation of the metallicity. The average values of metallicity of the complete LMC, the bar region, and the outer region of LMC are listed in Table~\ref{table:tab6}, for all the four cut-off criteria. The table also shows the number of regions used to estimate the average. The estimated average value, listed in the last column has an error, which reflects the standard deviation of the average and does not include the error in metallicity estimation of each region. The table shows that the bar region is relatively metal rich when compared to the outer regions, but the variation is quite small. As we consider the fourth criteria, the averages shown in the last row is used for further discussion. The bar region is found to have an average metallicity of $-$0.28$\pm$0.10 whereas the outer LMC has an average metallicity of $-$0.41$\pm$0.10. The average metallicity of the LMC is found to be $-$0.37$\pm$0.11 dex. The average value of metallicity estimated from OGLE III, as given in Table~\ref{table:tab3}, is very similar to the value estimated here, whereas the metallicity estimate for the bar region and the outer region are slightly different. The OGLE III estimate of average metallicity for the outer LMC is relatively less, probably due to the fact that the OGLE III covers more of the eastern and western regions which are found to be metal poor. The OGLE III data has more regions (1189) in the bar, when compared to the MCPS data (420); the difference in metallicity may be because of this effect. The metallicity maps in fact show that though the bar region is relatively metal rich, there is a lot of small scale variation. 

The metallicity gradient estimated using the MCPS data is presented in Figure \ref{fig:mcpso3_rad}. The radial average is estimated for 0.25 degree annuli. The radial variation for all selection criteria are shown and the error bar indicates the standard deviation about the average for the radial bin. There are a good number of regions within a radial distance of 4 degrees, beyond which the number of regions reduce and hence the metallicity values may be biased. The radial profile shows a peak near the central region, which was absent in the plot derived from OGLE III data. The peak might be an artifact, as it arises due to the higher metallicity of the inner most bin, which originates with a smaller number of regions. The profile from the second bin, up to a radial distance of 4 degrees, shows a smooth variation from the inner to the outer region. Thus, this once again confirms the smooth and gradual decrease in metallicity from the inner to outer regions. Also, the radial gradient does not show the bar effect, similar to the metallicity map. The profile shows a rather gradual gradient, which is probably the effect of inclusion of the northern and southern regions. As the MCPS data covers more of the disk of the LMC and less of the bar, the profile could be suggestive of the metallicity distribution of the LMC disk.   

\begin{table*}
{\small
\caption{Mean metallicity for different regions of the LMC using MCPS data:}
\label{table:tab6}
\begin{tabular}{|c|c|c|c|c|c|c|}
\hline \hline
Cut-off criteria & $r$ & $\sigma_{slope}$ & Region of the LMC & Number of subregions & Mean [Fe/H] (dex) \\
\hline\hline
I   & $\ge$ 0.40 & $\le$ 2.0 & COMPLETE & 3724 & $-$0.35$\pm$0.12\\      
    &            &           & BAR      & 544  & $-$0.26$\pm$0.10\\      
    &            &           & OUTER    & 1792 & $-$0.40$\pm$0.12\\         
\hline    
II  & $\ge$ 0.40 & $\le$ 1.5 & COMPLETE & 3587 & $-$0.35$\pm$0.12\\      
    &            &           & BAR      & 544  & $-$0.26$\pm$0.10\\      
    &            &           & OUTER    & 1694 & $-$0.40$\pm$0.11\\         
\hline    
III & $\ge$ 0.50 & $\le$ 2.0 & COMPLETE & 3235 & $-$0.37$\pm$0.12\\      
    &            &           & BAR      & 420  & $-$0.28$\pm$0.10\\      
    &            &           & OUTER    & 1637 & $-$0.42$\pm$0.11\\         
\hline    
IV  & $\ge$ 0.50 & $\le$ 1.5 & COMPLETE & 3144 & $-$0.37$\pm$0.11\\      
    &            &           & BAR      & 420  & $-$0.28$\pm$0.10\\      
    &            &           & OUTER    & 1569 & $-$0.41$\pm$0.10\\         
\hline       
\end{tabular}
\begin{minipage} {180mm}
\vskip 1.0ex
{Note: The first column denotes the four different cut-off criteria considered to filter out the LMC subregions. It is to be noted that we considered $N_p$ $\ge$ 10 for all four cut-off criteria. The second and third column specify the constraint on correlation coefficient ($r$) and $\sigma_{slope}$ respectively, corresponding to each cut-off criteria. The fourth column mentions three specific regions of the LMC: the complete coverage (for MCPS data), bar region and outer region (as defined in Section 3.2). The number of subregions that satisfy the cut-off for each of these three specific regions, are mentioned in the fifth column. The mean metallicity and standard deviation for these three specific LMC regions, are mentioned in the last (sixth) column.}
\end{minipage}
}
\end{table*}


\section{Error analysis}

\begin{itemize}
\item Estimation of error in slope and metallicity for OGLE III data: The factors that can contribute to the error in slope estimation are the photometric error associated with individual points in the CMD (error in V and I magnitude are $\le$ 0.15), and error due to fine binning of the CMD (dimension of each bin is 0.10 in  colour, and 0.05 in magnitude). In this scenario, we are not looking into the strength of individual CMD bins, and concentrating only on the overall feature of the RGB. After we identify the most populated bins as a part of RGB, the spread in the distribution of RGB bins is such that it more or less makes a good representation of the RGB. These bins are then fitted with the standard least square fitting technique, to estimate the slope ($|$slope$|$) and its corresponding error ($\sigma_{slope}$). In Figure \ref{fig:ogle3_errslope_vs_r} for OGLE III, we can see that for most of the regions, $\sigma_{slope}$ shows a clumpy distribution below 2. The values of $\sigma_{slope}$ estimated in our study are relatively larger. This is caused by the natural spread of the populated RGB bins. Thus in this study, we have not considered the contribution of errors associated with individual stars and colour-magnitude binning, while estimating error in slope. We expect these contributions to be negligible.

We have used three spectroscopic studies to formulate a slope-metallicity relation, given in Equation \ref{eq:1}. The error associated with the metallicity estimation of individual RGs, in all these are of the order of 0.15 dex. We can express Equation \ref{eq:1} as:
\begin{equation} \label{eq:3}
[Fe/H]= b \times |slope|+ a;
\end{equation}
where $b$ (=$-$0.137) is the slope and $a$ (=0.092) is the y-intercept of the slope-metallicity calibration relation. If $\sigma_b$ (=0.024) and $\sigma_a$ (=0.091) are the errors associated with $b$ and $a$ respectively, the error associated with metallicity ($error_{[Fe/H]}$) for each subregion can be calculated by propagation of error, as:
{\small
\begin{equation} \label{eq:4}
error_{[Fe/H]}=\sqrt{{(b \times |slope|)}^2\times\left(\left(\frac{\sigma_b}{b}\right)^2+\left(\frac{\sigma_{slope}}{|slope|}\right)^2\right)+{\sigma_a}^2}.
\end{equation} 
}
It is to be noted that while calculating $error_{[Fe/H]}$, we have not considered the error associated with individual calibration points.

\item Estimation in error in slope and metallicity for MCPS data: We have corrected for the systematic difference in I bands between the OGLE III and MCPS filter systems, that is given by Equation \ref{eq:2}. The errors associated with the slope and y-intercept of Equation \ref{eq:2} are smaller than the photometric error for individual stars, as well as the error associated with the fine binning of CMD. After transforming the I magnitudes for all MCPS subregions, their RGB slope and corresponding error in slope were estimated, by technique similar to OGLE III.
The value of $\sigma_{slope}$ is found to be relatively high, similar to those derived from OGLE III data. The $error_{[Fe/H]}$ is then derived in a similar way as OGLE III i.e. using propagation of error in Equation~\ref{eq:3}.

\end{itemize}

\begin{figure*}
\centering
\begin{minipage}[b]{0.45\linewidth}
\includegraphics[height=3.0in,width=3.0in]{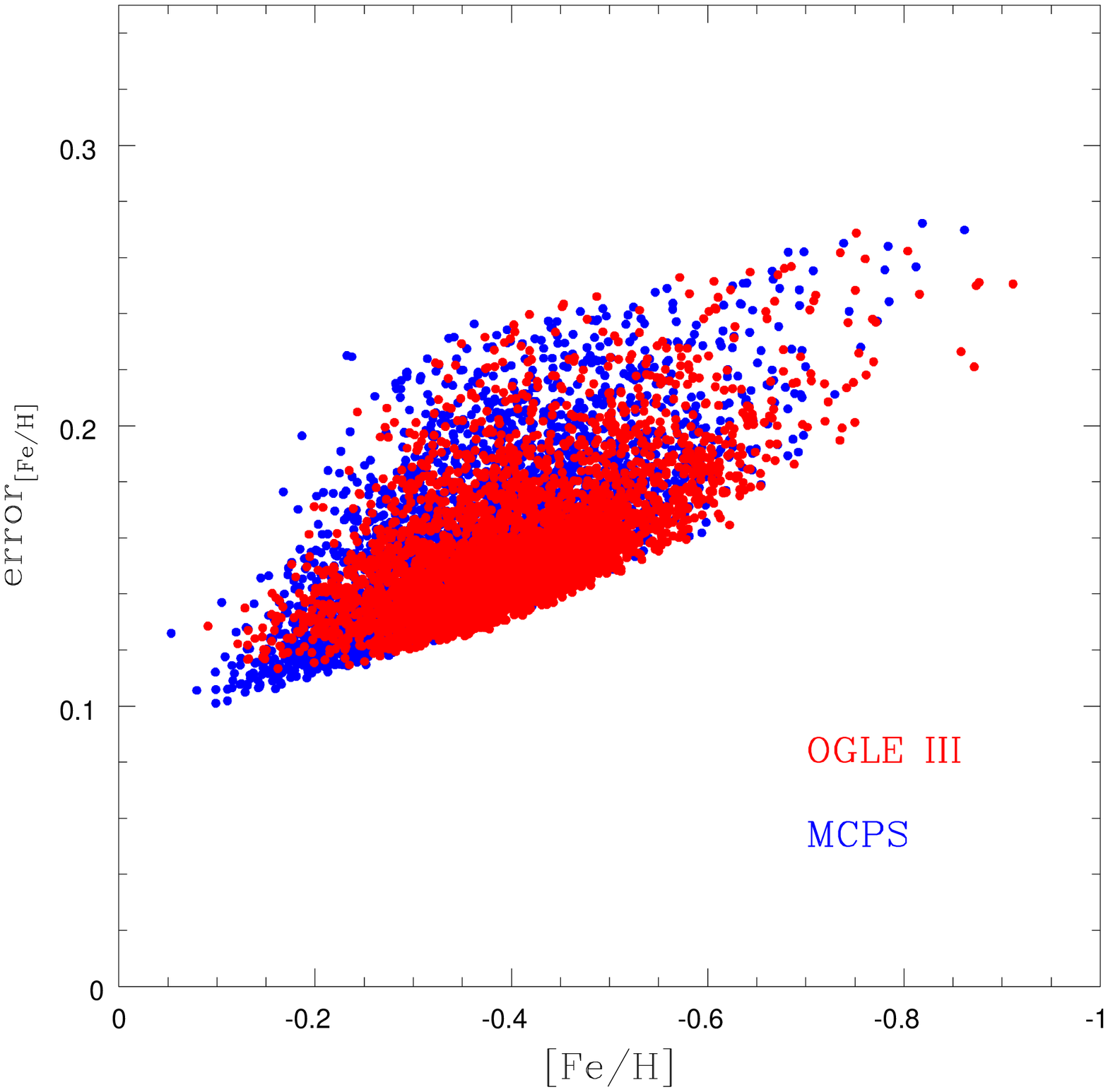}
\caption{\small Plot of $error_{[Fe/H]}$ versus [Fe/H] for OGLE III (red filled circles) and MCPS (blue filled circles), for cut-off criteria $(IV)$: N$_p$ $\ge$ 10, $r$ $\ge$ 0.5, and $\sigma_{slope}$ $\le$ 1.5.
\label{fig:mcpso3_ogle_errabun}} 
\end{minipage}
\quad
\begin{minipage}[b]{0.45\linewidth}
\includegraphics[height=3.0in,width=3.0in]{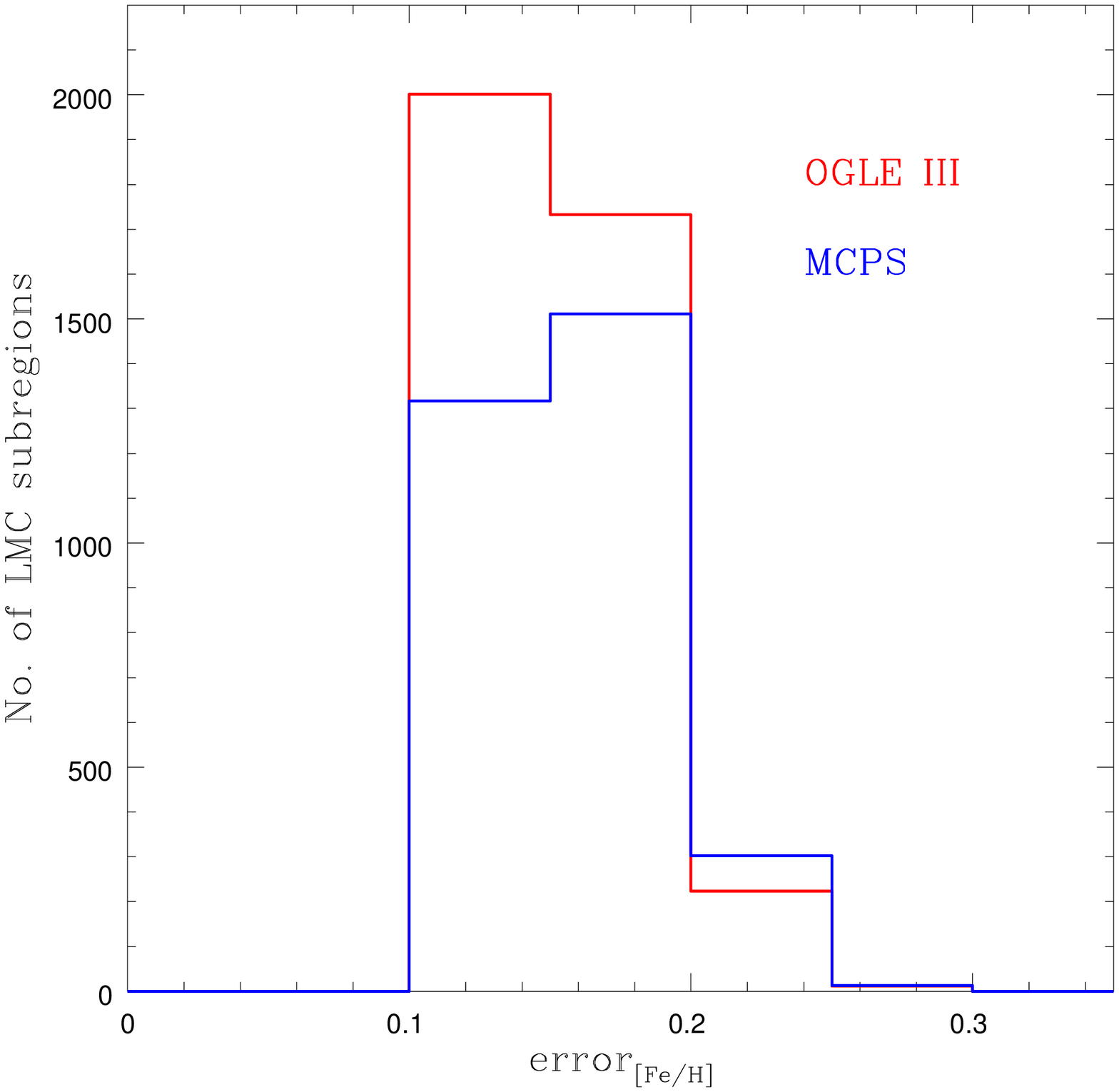}
\caption{\small Plot of histogram for $error_{[Fe/H]}$ for OGLE III (red) and MCPS (blue), for cut-off criteria $(IV)$: N$_p$ $\ge$ 10, $r$ $\ge$ 0.5, and $\sigma_{slope}$ $\le$ 1.5.
\label{fig:histerrabun_mcpso3_ogle}}
\end{minipage}
\end{figure*}

Figure \ref{fig:mcpso3_ogle_errabun} shows a plot between $error_{[Fe/H]}$ and [Fe/H], estimated for the most stringent cut-off criteria $({IV)}$, for both OGLE III and MCPS. It is very well seen that $error_{[Fe/H]}$ falls primarily in the range of 0.10--0.25 for both OGLE III amd MCPS. Thus, the values of $error_{[Fe/H]}$ are of the similar order as the error in spectroscopic values of metallicity of RGs ($\sim$ 0.15 dex). A slight trend is observed in the figure, where the error in abundance is seen to increase with decrease in metallicity. Figure \ref{fig:histerrabun_mcpso3_ogle} shows the histogram of $error_{[Fe/H]}$ for OGLE III and MCPS. The OGLE III data has more regions with error in the range 0.10--0.15 dex, whereas, the MCPS data has more regions with error in the range 0.15--0.20 dex. Both data sets have similar number of regions within the error range 0.20--0.25 dex. We conclude that the error associated with [Fe/H]  estimation is similar in both the data sets.


\section{Discussion}

We have estimated photometric metallicity maps of the LMC using OGLE III and MCPS data. The two data sets are found to be complementary in terms of the area coverage of the LMC. We estimated the slope of the RGB of several subregions in the LMC using a method which takes care of reddening and density variation between regions. The slope is converted to metallicity using spectroscopic measures which cover nearly the full range. The following points are summarized to bring out the results presented in the paper to the right perspective.

\begin{itemize}
\item The aim of this study is to  map the average metallicity distribution of the LMC. The method traces the densest part of the RGB and the slope is estimated using a straight line fit. We thus estimate only the metallicity of the population which has the largest number of stars in the RGB phase. The analysis does not bring out any information regarding the contribution from the less dominant populations.
\item If there are multiple populations with similar number of stars, then the RGB tends to be broad and the slope may be poorly estimated. If there are small scale variations in the reddening, then also the RGB tends to be broad resulting in a poor estimation of the slope. These regions will not get considered in our analysis, due the removal of regions with poor estimation of slope.
\item A metallicity estimation based on photometry can cover a large area of the LMC, unlike the spectroscopic method, which can over only relatively small areas. The photometric method can bring out the overall distribution, variation, and global average of metallicity. It also can identify regions which might show large deviation with respect to the mean value. These regions can then be ear marked for further detailed studies. Thus, photometric metallicity map, though not very accurate, has its own advantages.
\item The calibration of RGB slope to metallicity rests on the assumption that the spectroscopic targets are drawn from the dominant population of the subregion. The assumption is fair enough, as the spectroscopic targets are also chosen from the RGB and are likely to be picked from the population with most number of stars. The tail of metal-poor stars that is observed in spectroscopic studies is not detected in our sample because it does not strongly influence the mean metallicity of regions.
\item Across the inner 4 degrees of the LMC, the RGB slope has a relatively large range; the calibration relation between slope and metallicity should hold good for the full range. We combined results from three different spectroscopic studies to achieve this calibration. All the three studies have used similar spectral lines as well as technique to estimate metallicity, thus we expect no systematic differences between them. We have used field giants and two star clusters for the calibration. Thus, the metallicity values presented in this study are tied to this choice of metallicity calibration.
\end{itemize}

\subsection{Assumptions and their impact}
In the present study, we have used two large photometric datasets to create a metallicity map. The steps taken to do this process might have affected the outcome of the study and below we discuss each step and its impact on the estimated value of RGB slope and metallicity. 

\subsubsection{Effect of sub-division}
The two data sets are sub-divided to create regions with smaller area. We have also considered different sizes for sub-division, in the outer and inner LMC, in both the data sets. As the depth and resolution of MCPS and OGLE III are different, we were unable to make subregions of same area in the data sets. The MCPS subregions are larger than the OGLE III subregions, in all parts of the LMC. Due to this, even though we eventually converted the MCPS data to the OGLE III system, we are unable to make one-to one correlation between the metallicity estimated using the two data sets. As the area changes, we notice that the estimated slope changes mildly, though within the errors. Thus, the estimated value of slope mildly depends on the area considered, but the change is found to be within the error of slope estimation. This may be due to the fact that the dominant population as well as differential reddening can change with area. We note that the important result of this study is not the average metallicity of region by region, but the global average and its variation across the LMC.

\subsubsection{Effect of reddening and differential reddening}
The LMC is known to have variation in reddening with respect to population \citep{Zaritsky+2004AJmcps}. The reddening variation can shift the location of the RGB in the CMD. This effect is taken care in the analysis by anchoring the RGB to the densest part of the RC. But, the effect of differential reddening will remain and make the RGB broad. Large scale variation of reddening can broaden the RGB, resulting in a poorly estimated slope. These regions are eliminated from the analysis. Most of the regions which get eliminated due to poor slope estimation, are found to be located near star forming regions. As we are interested in the statistical average estimated using large number of regions, their impact is likely to be negligible.

\subsubsection{Systematic effects}
There can be two systematic effects in our study. The first one is due to the photometric systems of the two data sets used and the second one is in the conversion of slope to metallicity. We have tried to minimise the first one by converting the MCPS system to the OGLE III system. 
For slope-metallicity calibration, we used three references from the same group and hence there is minimal systematic shift among the three calibrations. Thus, we have tried to make the second systematics as minimum as possible. If the photometric shift is not corrected completely, then the shift in metallicity between the MCPS and OGLE III could be there such that the higher metallicity observed in the MCPS regions could be an artefact. On the other hand, the radial variation as well as the global averages have less impact.  

\subsection{Comparison of metallicity distribution}
After we have addressed the impact of various assumptions on the estimates, we now compare the results estimated from the two data sets. Using the OGLE III data, we find the bar region to be a more or less homogeneous metal rich region. We do notice some small scale variation in [Fe/H], but are within errors. The eastern and the western disks are found to be relatively metal poor. The MCPS map shows a gradual gradient from near the central region to the outer regions. The northern and the southern regions are only marginally metal poor. If the OGLE III data is considered to describe the bar and the MCPS data to describe the disk, then our analysis suggests that the LMC has a metal rich bar and a disk with a very shallow metallicity gradient. The analysis is also indicative that the east and the west LMC disk may be metal poor than the north and south disk - this needs to be verified with more spectroscopic data in these regions.

\begin{figure*}
\centering
\begin{minipage}[b]{0.45\linewidth}
\includegraphics[height=3.0in,width=3.0in]{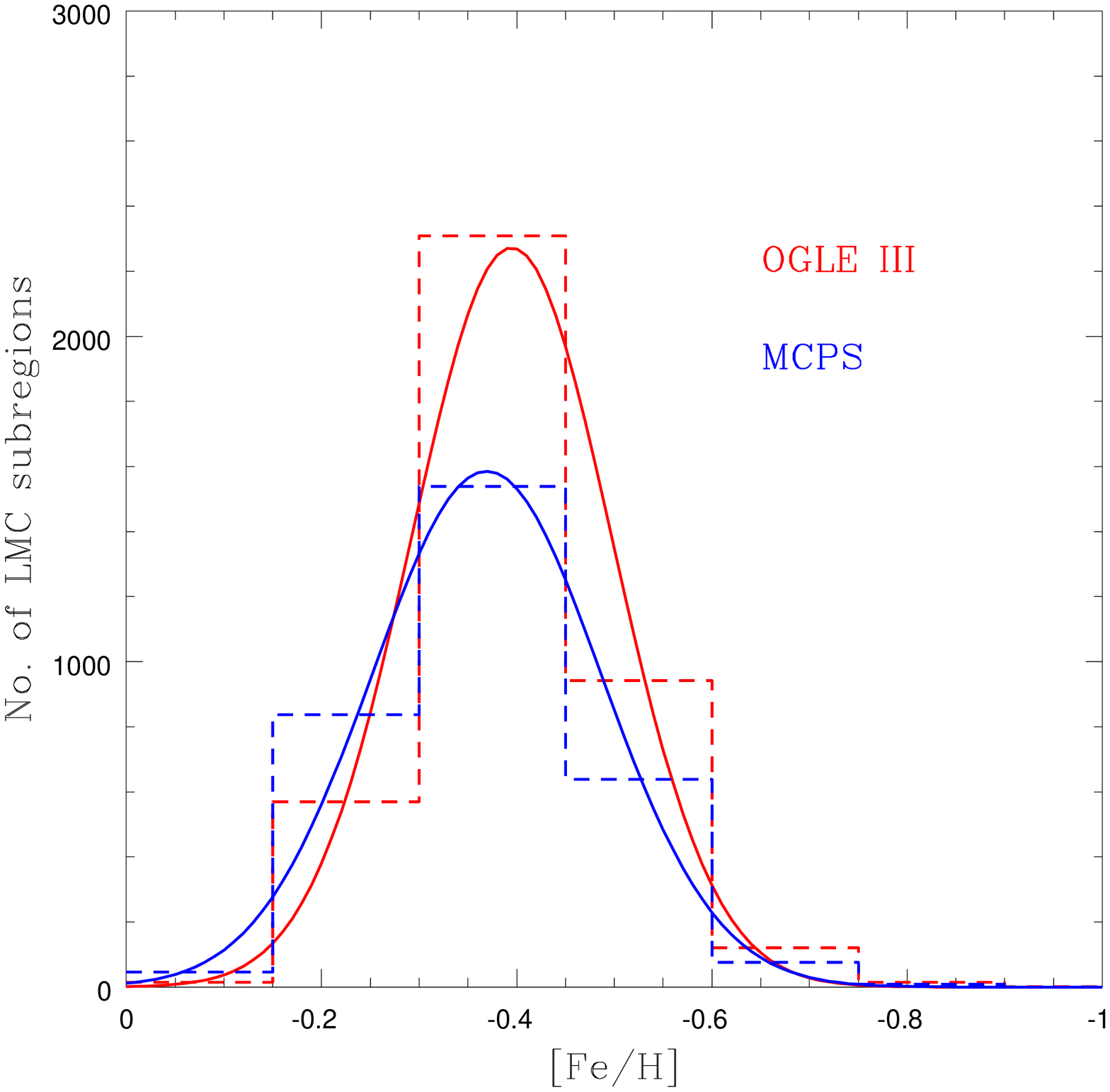}
\caption{\small Histogram of metallicity (dashed line) for the complete LMC fitted with a Gaussian function (solid line), for OGLE III (red) and MCPS data (blue). 
\label{fig:gauss1}} 
\end{minipage}
\quad
\begin{minipage}[b]{0.45\linewidth}
\includegraphics[height=3.0in,width=3.0in]{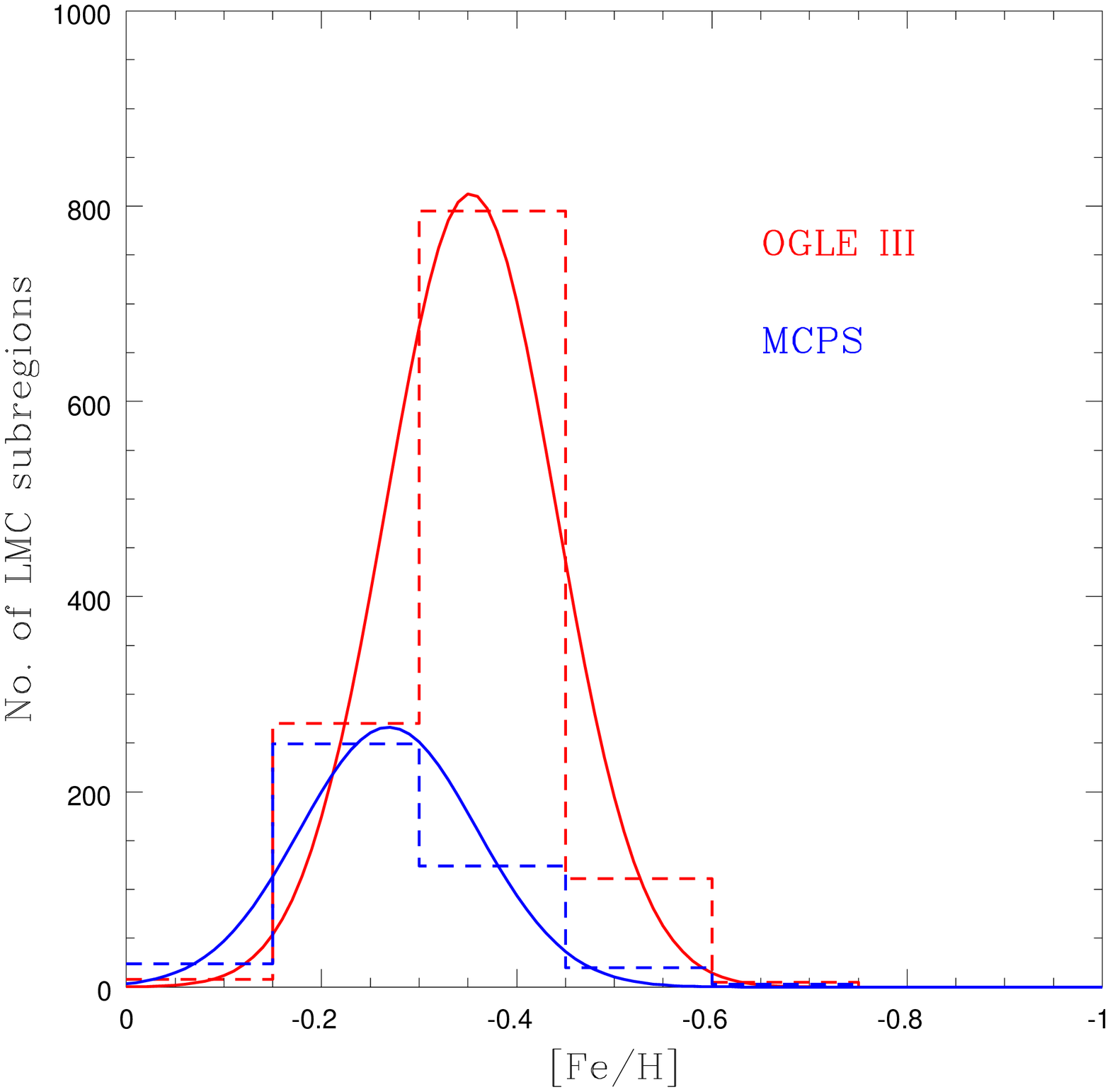}
\caption{\small Histogram of metallicity (dashed line) for the LMC bar fitted with a Gaussian function (solid line), for OGLE III (red) and MCPS data (blue). 
\label{fig:gauss2}}
\end{minipage}
\end{figure*}

In Figure \ref{fig:gauss1}, we compare the distribution of metallicity for the complete LMC as estimated from the two data sets. It can be seen that, in general the distributions are similar. The peaks of the distributions are almost identical but the two surveys differ in their secondary peaks, as OGLE III data shows a metal poor second peak, whereas the MCPS data shows a metal rich second peak. This is likely to be due to the difference in the area coverage. We have fitted Gaussian to the distribution. For the OGLE III data, the peak of the distribution is estimated to be at [Fe/H] = $-$0.395$\pm$0.002 dex and the width of the distribution to be $\sigma$[Fe/H] = 0.103$\pm$0.001 dex. In the case of MCPS data, the peak of the distribution is found to be [Fe/H] = $-$0.369$\pm$0.002 dex and the width of the distribution to be $\sigma$[Fe/H] = 0.117$\pm$0.001 dex. 

In Figure \ref{fig:gauss2}, we have shown the distribution of metallicity for the bar region. The distribution clearly shows that the OGLE III data has a large number of data points where most of them are in the $-$0.3 to $-$0.45 dex bin, whereas the distribution using the MCPS data with relatively much less number of points have the peak in the $-$0.15 to $-$0.30 dex bin. For OGLE III the peak of the distribution is estimated to be at [Fe/H] = $-$0.353$\pm$0.002 dex and the width of the distribution to be $\sigma$[Fe/H] = 0.087$\pm$0.002 dex. On the other hand, the distribution using the MCPS data estimated [Fe/H] = $-$0.269$\pm$0.004 dex and the width of the distribution to be $\sigma$[Fe/H] = 0.091$\pm$0.004 dex. Thus, the bar region is found to be metal rich by both OGLE III and MCPS data. We also notice that the distributions have a relatively less $\sigma$[Fe/H] value, suggesting that the range of [Fe/H] is relatively narrow in the bar region. 

\begin{figure*}
\centering
\begin{minipage}[b]{0.45\linewidth}
\includegraphics[height=3.0in,width=3.0in]{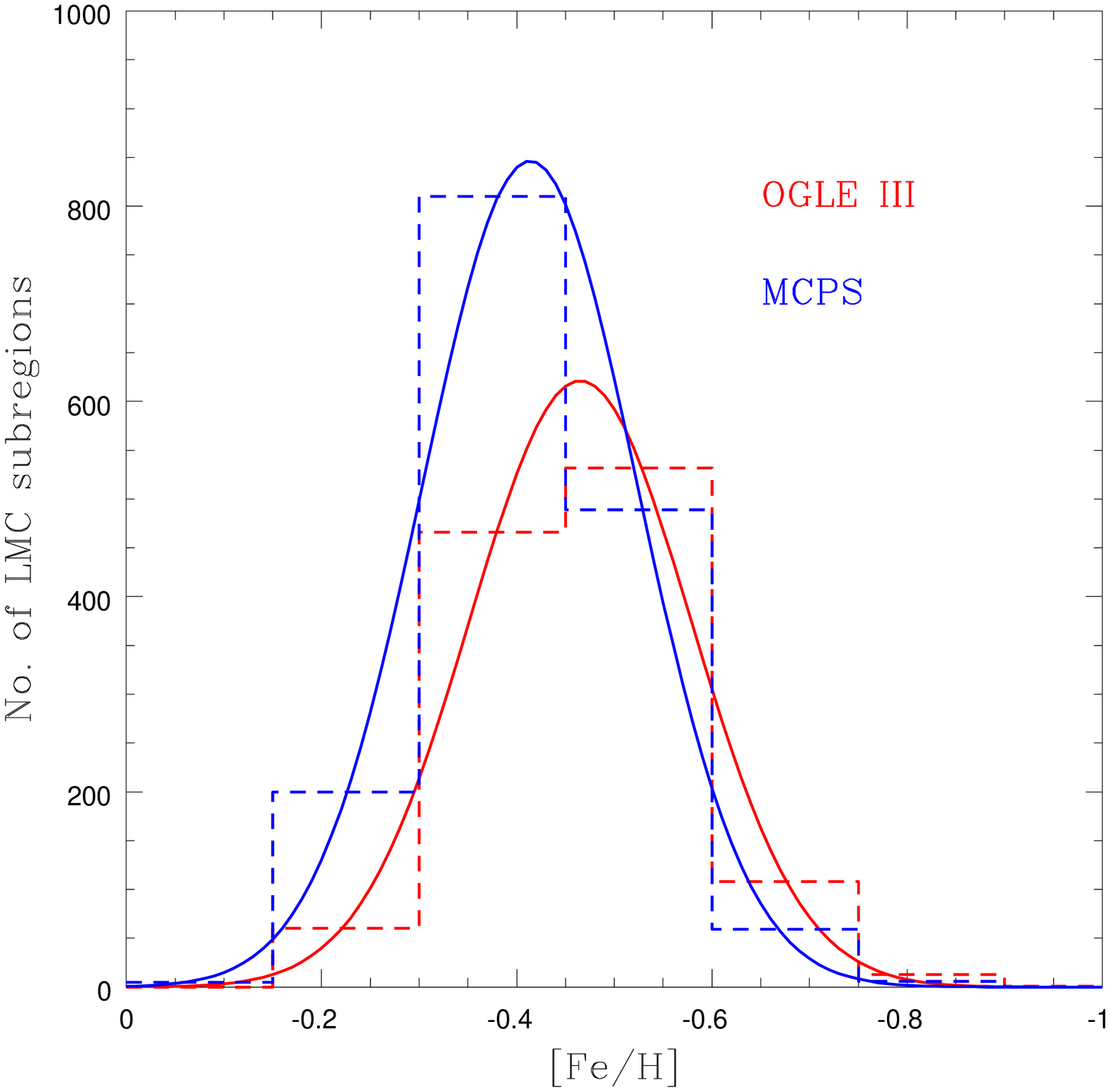}
\caption{\small Histogram of metallicity (dashed line) for the outer LMC fitted with a Gaussian function (solid line), for OGLE III (red) and MCPS data (blue).
\label{fig:gauss3}} 
\end{minipage}
\quad
\begin{minipage}[b]{0.45\linewidth}
\includegraphics[height=3.0in,width=3.0in]{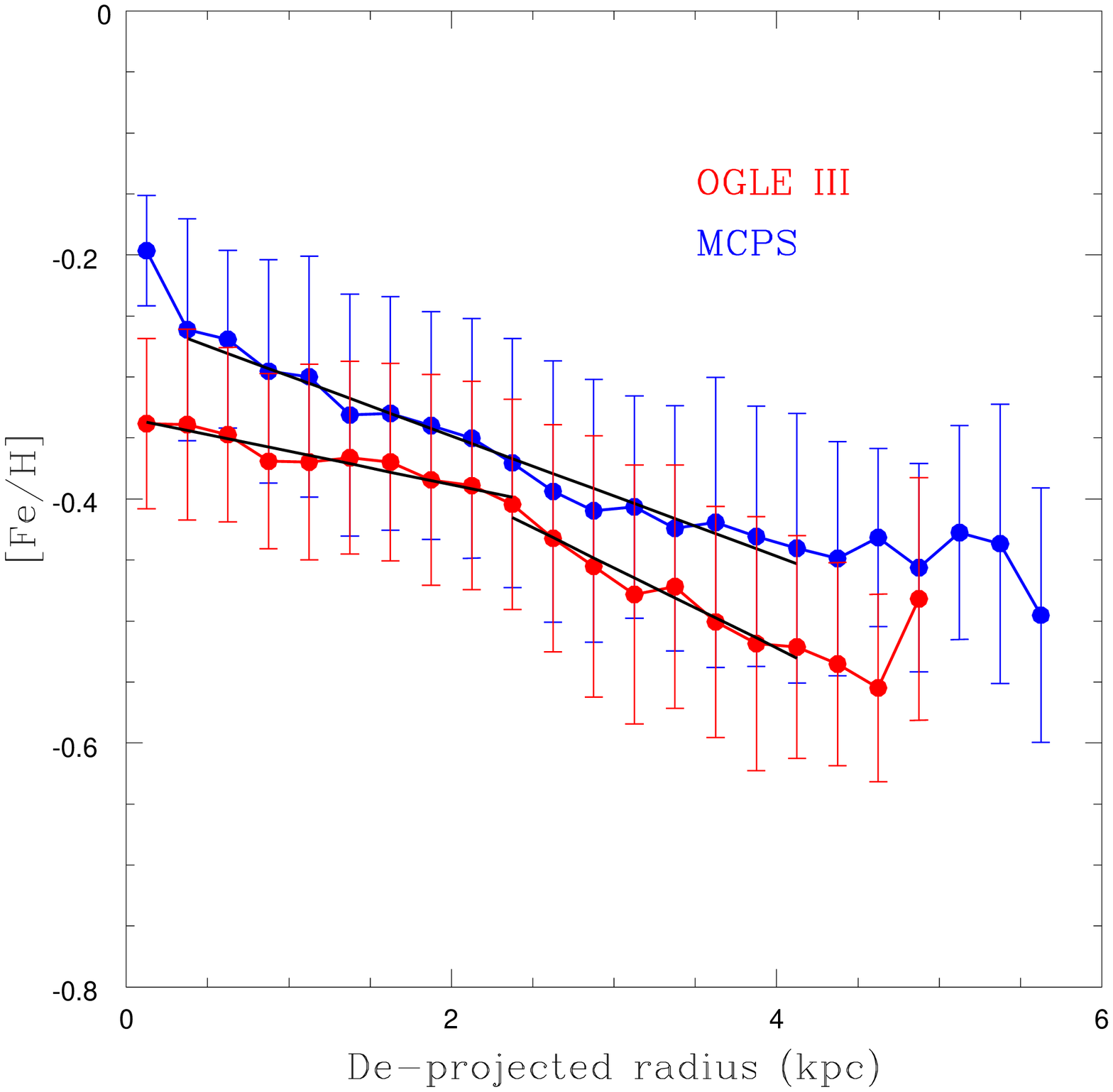}
\caption{\small Variation of metallicity with radius in the de-projected plane of the LMC for OGLE III (red solid line) and MCPS (blue solid line) data, estimated for cut-off criteria $(IV)$: N$_p$ $\ge$ 10, $r$ $\ge$ 0.5, and $\sigma_{slope}$ $\le$ 1.5. The gradients estimated till a radius of about 4 kpc are shown as black solid lines for both OGLE III and MCPS.
\label{fig:mcpso3_ogle_rad}}
\end{minipage}
\end{figure*}

In Figure \ref{fig:gauss3}, we have shown the distribution of [Fe/H] for the outer LMC. The MCPS distribution shows the peak at $-$0.3 to $-$0.45 bin, whereas the OGLE III distribution has the peak at $-$0.45 to $-$0.60, though one can in general consider a broad peak from $-$0.30 to $-$0.60 dex. The MCPS distribution has a relatively more number of metal rich regions and OGLE III has marginally more metal poor regions. The peak of the MCPS data is also marginally metal rich. The above differences, as indicated before, is likely to be due to the difference in area coverage. The MCPS distribution has a peak at [Fe/H] = $-$0.414$\pm$0.003 dex and the width of the distribution to be $\sigma$[Fe/H] = 0.110$\pm$0.002 dex. The OGLE III distribution has a peak at [Fe/H] = $-$0.465$\pm$0.003 dex and the width of the distribution to be $\sigma$[Fe/H] = 0.113$\pm$0.002 dex. Thus, we find that there is a statistically significant difference between the metallicity of the bar region and the outer region. We also suggest that the northern and southern regions of the LMC could be marginally more metal rich than the eastern and western regions.

In Figure \ref{fig:mcpso3_ogle_rad}, we compare the radial gradient of metallicity from the two data sets. The LMC plane is inclined with respect to the sky plane by an angle $i$, and the position angle of the line of nodes is given by $\Theta$. If we know the mean distance to the LMC centre $D_0$, we can apply the correction for the position angle and $i$, using the conversion equations from \cite{vander2001AJ-MCstructure}. For our purpose we have used the value of $i$ to be $37^{\circ}.4$, and $\Theta$ to be $141^{\circ}.2$, from \cite{Smitha&Purni2010A&Aanestimate}. The sky-plane ($X$, $Y$) is de-projected to the LMC plane ($X^{'}$, $Y^{'}$) assuming $D_0$ = 50 kpc. We constructed a radial metallicity gradient, by assuming a bin width of 0.25 kpc. We have quantified the variation of metallicity with the de-projected radius from the LMC centre ($\rho^{'}$ in kpc), for both OGLE III and MCPS, by fitting a straight line using least square fit as shown in the Figure \ref{fig:mcpso3_ogle_rad}. For MCPS data, to avoid issues due to sampling, we exclude the region nearest to the LMC centre and limit our estimation from $\rho^{'}$ $\sim$ 0.5 to 4~kpc. The metallicity is found to decrease with increase in $\rho^{'}$ as: [Fe/H]=($-$0.049$\pm$0.002) $\times$ $\rho^{'}$+ ($-$0.250$\pm$0.005); with a correlation coefficient of $r$ =0.99. The unit for the slope is dex kpc$^{-1}$. For OGLE III, we have estimated the slope for the central region that contains the bar, where the metallicity seems to remain almost constant (up to $\rho^{'}$ $\sim$ 2.5 kpc). This is found to be: [Fe/H]=($-$0.027$\pm$0.003) $\times$ $\rho^{'}$ + ($-$0.333$\pm$0.004); with $r$= 0.96. For outside the central LMC ($\rho^{'}$ $\sim$ 2.5 to 4 kpc), the variation of metallicity is found to be: [Fe/H]=($-$0.066$\pm$0.006) $\times$ $\rho^{'}$+ ($-$0.259$\pm$0.020); with $r$ =0.98. The comparison of these estimated slopes suggests that, the metallicity variation in MCPS is steeper as compared OGLE III till a distance of $\rho^{'}$ $\sim$ 2.5 kpc. Beyond this distance, the metallicity decreases more rapidly for OGLE III regions than for the MCPS regions. Thus, it is suggestive of a marginally different radial gradient for the northern and southern LMC, when compared to the eastern and western regions. 

\cite{Cioni2009A&Athemetallicity}, using the C/M ratio of the field AGB population as an indicator of metallicity, found that the [Fe/H] of the LMC decreases linearly with distance from the centre up to a distance of 8 kpc, following the relation: [Fe/H]=$-0.047\pm0.003\times\rho^{'}-1.04\pm0.01$. This variation is shown in their Figure 2. In this figure they also plotted the metallicity of RR Lyrae stars  within a radius of about 3 kpc, from  \cite{Borissova+2006A&Apropert}, which have a steeper metallicity gradient ($-$0.078$\pm$0.007 dex kpc$^{-1}$)  compared to that of the AGBs. There is no requirement that the two populations share a common distribution, as the RR Lyrae stars are older than 10~Gyr, while the AGB stars are strongly biased towards ages $\approx$1--2~Gyr. Their Figure 2 also showed the radial variation of metallicity of field RGBs (from \citealt{Cole+2005AJspectroOfRGs, Pompeia+2008A&Achemi, Carrera+2008AJ-CEH-LMC}), and star clusters (from \citealt{Grocholski+2006AJCaIItriplet,Grocholski+2007AJdist}), which is almost constant within the inner LMC.  We note that the metallicity gradient estimated in this study using the MCPS data, matches very well with that estimated by \cite{Cioni2009A&Athemetallicity}, though the y-intercept, which is the metallicity at the centre, are very different. The origin of this discrepancy requires further investigation, and may be due to the difference in mean age between the general RGB field discussed here and AGB stars previously studied. The results derived in this section are tabulated in Table \ref{table:tab7}.

The mean [Fe/H] as well as a shallow gradient in the bar region suggests that the bar of the LMC could be considered to show the expected enhancement seen in the barred galaxies. While many observations support the basic scenario of bar and galaxy evolution, several recent studies of barred galaxies using large samples reveal a dependence on mass of the galaxy. A study of 294 galaxies with strong bars from the Sloan Digital Sky Survey (SDSS) by \cite{Ellison+2011MNRAStheimpact} showed that barred galaxies  with stellar masses M $<$ 10$^{10}$ M$_\odot$ also show an increase in central metallicity, but without a corresponding increase in central SFR. One possible explanation is that star formation in the centre of low-mass barred galaxies, which is presumably responsible for the observed higher metallicities, has now ceased, while stars are still forming in the centre of high-mass galaxies \citep{Martel+2013MNRAStheconnect}. Thus, the LMC might also be belonging to this class of objects, where we do detect higher metallicity, but no enhanced star formation.

\subsection{Map of metallicity outliers}

\begin{figure*} 
\begin{center} 
\includegraphics[height=4in,width=7in]{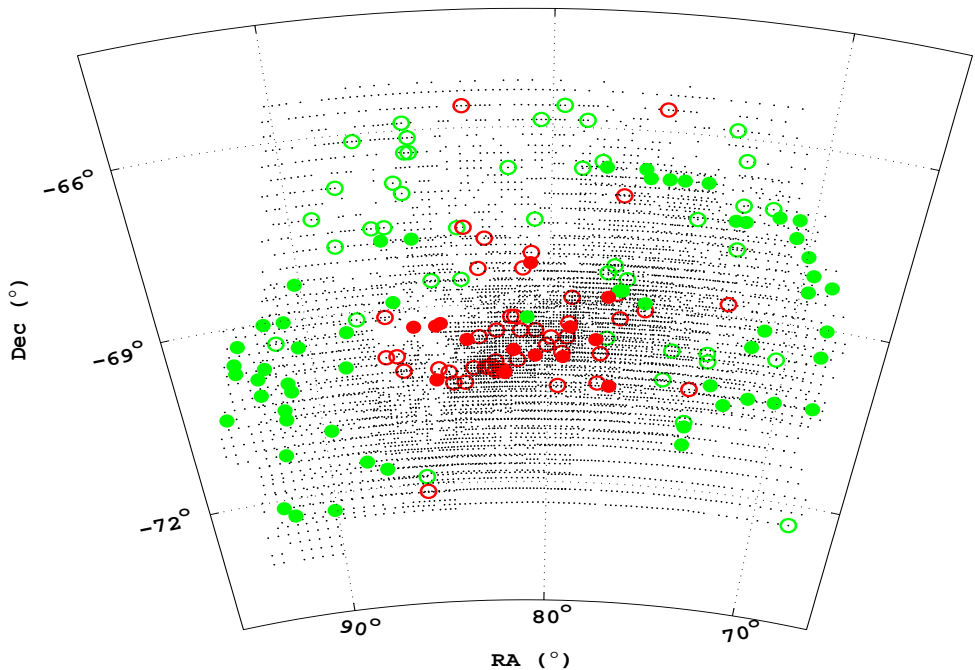}
\caption{\small A combined metallicity map for OGLE III and MCPS data (cut-off criteria $({IV})$: $N_p$ $\ge$ 10, $r$ $\ge$ 0.5 and $\sigma_{slope}$ $\le$ 1.5) displaying subregions with large deviation from mean metallicity. The significantly metal rich regions ([Fe/H]$\ge$$-$0.15 dex) are shown in red colour (filled circles for OGLE III, and open circles for MCPS), whereas the significantly metal poor regions ([Fe/H]$\le$$-$0.65) are shown in green colour (filled circles for OGLE III, and open circles for MCPS).
\label {fig:combinemap}
}
\end{center} 
\end{figure*}

One of the aims of the study was to identify regions which have metallicity significantly deviating from the mean value, or with respect to the surrounding regions. As we have the average metallicity as $-$0.40 dex, we considered regions with metallicity less than $-$0.65 dex and more than $-$0.15 dex as deviations, which corresponds to deviations beyond 2.5$\sigma$ about the mean. In Figure \ref{fig:combinemap}, we show the combined OGLE III and the MCPS data. All regions with metallicity between $-$0.65 dex to $-$0.15 dex are shown as small dots, which covers most of the region. Regions which have metallicity higher than $-$0.15 dex are shown as red filled circles (OGLE III data) and red open circles (MCPS data). It can be seen from the figure that the red points, both from OGLE III and MCPS, are mostly located in the central region coincident with the bar. We stress that co-location of points from both the data sets are seen here and re-emphasises that the bar is indeed the most metal rich part of the LMC. We also find a location to the north of the bar (RA $\sim$ 82$^{\circ}$ and Dec $\sim$ $-$68.5$^{\circ}$) to be metal rich, as identified by both the data sets. A few additional regions in the north of the LMC are seen as metal-rich by the MCPS data. The southern and eastern LMC are not found to have any metal rich regions, apart from a couple of regions close to the bar.

Regions with metallicity less than $-$0.65 dex are shown as filled green circles (OGLE III data) and open green circles (MCPS data). Most of the metal poor regions are in the outer LMC, with a few points in the bar region as well.  The eastern and the western disk of the LMC has a large number of metal poor regions from both the data sets. The northern disk has such regions identified from the MCPS data. In general, the southern LMC does not show either metal rich or metal poor regions. Thus there is a clear demarcation between the bar and the outer regions of the LMC. We also identify a few locations which are either metal rich or poor with respect to the surrounding. These regions are candidates for detailed spectroscopic study in order to understand the source of deviation. There does not seem to be any correlation between the regions identified as metal-poor and the locations in which \cite{Olsen+2011ApJapopulation} discovered a population of kinematically distinct, metal-poor giants which they attributed to an accreted SMC population. This suggests that at no location in the inner LMC is the kinematically distinct population strong enough to significantly draw down the metallicity of the field in general, relative to neighbouring fields; either the total accreted population is very small, or well-mixed into the disk as a whole.

 
\begin{table*}
{\small
\caption{Summary of average [Fe/H] and gradient for different regions of the LMC}
\label{table:tab7}
\begin{tabular}{|c|c|c|c|c|c|c|}
\hline \hline
Data & LMC region & [Fe/H] (dex) & $\sigma$[Fe/H] (dex) & Gradient (dex kpc$^{-1}$)\\
\hline\hline   
OGLE III    & COMPLETE & $-$0.39 & 0.10 &                    \\      
            & BAR      & $-$0.35 & 0.09 & $-$0.027$\pm$0.003 \\      
            & OUTER    & $-$0.46 & 0.11 & $-$0.066$\pm$0.006 \\            
\hline    
MCPS        & COMPLETE & $-$0.37 & 0.12 & $-$0.049$\pm$0.002 \\      
            & BAR      & $-$0.27 & 0.09 &                    \\      
            & OUTER    & $-$0.41 & 0.11 &                    \\         
\hline    
   
\end{tabular}
\begin{minipage} {180mm}
\vskip 1.0ex
{Note: The table presents a summary of results derived from OGLE III and MCPS data. The mean [Fe/H] and the corresponding $\sigma$[Fe/H], estimated using these two data sets, for different regions within the LMC, are listed in the third and fourth column respectively. The fifth column gives the radial metallicity gradient derived corresponding to each case.}
\end{minipage}
}
\end{table*}

\section{Summary}
This paper presents an estimate of the average and radial variation of metallicity ([Fe/H]) in the LMC based on photometric data. The results can be summarised as follows:
\begin{enumerate}
\item We present a metallicity map of the LMC derived from photometric data and calibrated using spectroscopic data. This is a first of its kind map derived using RGB stars from the OGLE III and MCPS data sets.
\item We estimate the RGB slope of several subregions in the LMC and convert the slope to metallicity using spectroscopic data of giants in the field as well as cluster.
\item The average metallicity of the LMC is found to be $-$0.37 dex ($\sigma$[Fe/H] = 0.12) from MCPS data and $-$0.39 dex ($\sigma$[Fe/H] = 0.10) from OGLE III data, within a radius of 4 degrees.
\item The bar region of the LMC is found to have an average metallicity of $-$0.35 dex ($\sigma$[Fe/H] = 0.9), and a shallow gradient in metallicity.
\item The outer regions have metallicity ranging from $-$0.41 dex ($\sigma$[Fe/H] = 0.11) from the MCPS data to $-$0.46 dex ($\sigma$[Fe/H] = 0.11) from OGLE III data.
\item Both the data sets suggest a shallow radial metallicity gradient for the LMC disk, up to a radius of 4~kpc (from $-$0.049$\pm$0.002 dex kpc$^{-1}$ for MCPS to $-$0.066$\pm$0.006 dex kpc$^{-1}$ for OGLE III) which agree with \cite{Cioni2009A&Athemetallicity}.
\item We identify a few areas where the metallicity is found to be significantly different from the surrounding regions, which need to be studied in detail using spectroscopic studies.
\end{enumerate}
\section*{Acknowledgments}
Samyaday Choudhury acknowledges Smitha Subramanian for a critical reading of the original version of the manuscript and comments, and Indu G. for the interesting discussions. The authors thank the anonymous referee for the encouraging comments which improved the presentation of the paper. The authors also thank the OGLE and the MCPS team for making the data available in public domain.
\bibliographystyle{mn2e}
\bibliography{bibliography}

\label{lastpage}

\end{document}